\documentclass[fleqn,usenatbib]{mnras}
\usepackage{amsmath,amssymb}
\usepackage{newtxtext,newtxmath}
\usepackage[T1]{fontenc}
\usepackage{ae,aecompl}

\usepackage{graphicx}
\usepackage{gensymb}
\usepackage{hyperref}

\DeclareRobustCommand{\ion}[2]{\relax\ifmmode\ifx\testbx\f@series{\mathbf{#1\,\mathsc{#2}}}\else{\mathrm{#1\,\mathsc{#2}}}\fi\else\textup{#1\,{\mdseries\textsc{#2}}}\fi}

\newcommand{\cb}{C\&B}

\newcommand{\civ}{\ion{C}{iv}}
\newcommand{\heii}{\ion{He}{ii}}

\newcommand{\heplus}{\ensuremath{\rm He^+}}

\providecommand{\e}[1]{\ensuremath{\times 10^{#1}}}
\newcommand{\unit}[1]{\ensuremath{\, \mathrm{#1}}}

\newcommand{\hst}{\textit{HST}}
\newcommand{\hstcos}{\textit{HST}/COS}

\newcommand{\jwst}{\textit{JWST}}
\newcommand{\chandra}{\textit{Chandra}}
\newcommand{\spitzer}{\textit{Spitzer}}

\newcommand{\ott}{\ensuremath{\mathrm{O}_{32}}}
\newcommand{\rtt}{\ensuremath{\mathrm{R}_{23}}}

\newcommand{\cloudy}{\textsc{cloudy}}

\newcommand{\pyneb}{\textsc{pyneb}}
\newcommand{\beagle}{\textsc{beagle}}
\newcommand{\parsec}{\textsc{parsec}}
\newcommand{\tauV}{\hbox{$\hat{\tau}_V$}}

\newcommand{\Hii}{\mbox{H\,{\sc ii}}}

\defcitealias{senchynaUltravioletSpectraExtreme2017}{S17}
\defcitealias{senchynaExtremelyMetalpoorGalaxies2019}{S19}
\defcitealias{shiraziStronglyStarForming2012}{SB12}
\defcitealias{brinchmannGalaxiesWolfRayetSignatures2008}{B08}

\title[Massive star populations in nearby WR galaxies]{Ultraviolet spectra of extreme nearby star-forming regions: evidence for an overabundance of very massive stars}
\author[P. Senchyna et al.]{
    Peter Senchyna,$^{1}$\footnotemark[1]
    Daniel P. Stark,$^{1}$
    St\'{e}phane Charlot,$^{2}$
    Jacopo Chevallard,$^{2}$
    \newauthor
    Gustavo Bruzual,$^{3}$
    and
    Alba Vidal-Garc\'ia$^{2}$
    \vspace{0.1in}\\
    $^{1}$ Steward Observatory, University of Arizona, 933 N Cherry Ave, Tucson, AZ 85721 USA \\
    $^{2}$ Sorbonne Universit\'{e}, CNRS, UMR7095, Institut d'Astrophysique de Paris, F-75014, Paris, France \\
    $^{3}$ Instituto de Radioastronom{\'i}a y Astrof{\'i}sica, UNAM, Campus Morelia, Michoacan, M{\'e}xico, C.P. 58089, M{\'e}xico\\
}

\date{Submitted to MNRAS}

\pubyear{2020}

\begin{document}
\label{firstpage}
\pagerange{\pageref{firstpage}--\pageref{lastpage}}
\maketitle

\begin{abstract}
    As deep spectroscopic campaigns extend to higher redshifts and lower stellar masses, the interpretation of galaxy spectra depends increasingly upon models for very young stellar populations.
    Here we present new \hstcos{} ultraviolet spectroscopy of seven nearby ($<120$ Mpc) star-forming regions hosting very young stellar populations ($\sim 4$--20 Myr) with optical Wolf-Rayet stellar wind signatures, ideal laboratories in which to test these stellar models.
    We detect nebular \ion{C}{iii}] in all seven, but at equivalent widths uniformly $< 10$ \AA{}.
    This suggests that even for very young stellar populations, the highest equivalent width \ion{C}{iii}] emission at $\geq 15$ \AA{} is reserved for inefficiently-cooled gas at metallicities at or below that of the SMC.
    The spectra also reveal strong \civ{} P-Cygni profiles and broad \heii{} emission formed in the winds of massive stars, including some of the most prominent \heii{} stellar wind lines ever detected in integrated spectra.
    We find that the latest stellar population synthesis prescriptions with improved treatment of massive stars nearly reproduce the entire range of stellar \heii{} wind strengths observed here.
    However, we find that these models cannot simultaneously match the strongest wind features alongside the optical nebular line constraints.
    This discrepancy can be naturally explained by an overabundance of very massive stars produced by a high incidence of binary mass transfer and mergers occurring on short $\lesssim 10$ Myr timescales, suggesting these processes may be crucial for understanding the highest-sSFR galaxies in the early Universe.
    Reproducing both the stellar and nebular light of young systems such as these will be a crucial benchmark for the next generation of stellar population synthesis models.
\end{abstract}


\footnotetext[1]{E-mail: senchp@email.arizona.edu}

\section{Introduction}

Nebular line emission represents one of our only windows onto the nature of galaxies at cosmological redshifts.
The measurement of rest UV--optical emission lines via near-infrared spectroscopy with the {\it James Webb Space Telescope} (\jwst{}) and the next generation of extremely large ground-based telescopes (ELTs) promises to unveil the chemical abundances and star formation histories of large samples of galaxies into the reionization era at $z>6$ in the coming years \citep{starkGalaxiesFirstBillion2016,kewleyUnderstandingGalaxyEvolution2019}, approaching some of the earliest epochs of star formation in the Universe.
Successful extraction of constraints on all physical parameters of interest from these data will rely on models for the ionizing radiation output of populations of stars as a function of metallicity and age (stellar population synthesis) coupled to radiative transfer codes to predict the emergent nebular spectrum this radiation powers (photoionization modeling).
While many sophisticated stellar population synthesis models incorporating stellar physics canonically neglected \citep[such as fast rotation and binary mass transfer; e.g.][]{leithererEffectsStellarRotation2014,eldridgeBinaryPopulationSpectral2017,gotbergImpactStarsStripped2019} and new frameworks for performing inference using these predictions \citep[e.g.][]{chevallardModellingInterpretingSpectral2016,lejaDerivingPhysicalProperties2017} have emerged in recent years, these predictions remain largely untested under the extreme star formation conditions we expect to encounter routinely in reionization era galaxies.

Broadband SED fitting has provided an early glimpse of the nebular properties of star-forming systems at $z>6$, revealing a population significantly different from typical galaxies at lower redshifts.
Leveraging the fact that the strong rest-optical lines contaminate the \spitzer{}/IRAC [3.6] and [4.5] micron bands at $z\sim 6$--$9$, several authors have robustly demonstrated that the specific star formation rates (sSFRs) of typical systems at these redshifts are well in-excess of galaxies at lower redshift \citep{labbeSpectralEnergyDistributions2013,smitEvidenceUbiquitousHighequivalentwidth2014}.
Indeed, \citet{endsleyOIIIBetaEquivalent2020} have recently shown that 20\% of the most luminous systems at $z\sim 7$ reveal evidence for extremely high equivalent width optical nebular emission corresponding to sSFRs $\gtrsim 50 \unit{Gyr^{-1}}$, indicative of a substantial population of galaxies dominated entirely by massive stars formed within the last $<$20 Myr.
Accurately interpreting the nebular line emission detected in such extreme systems will rely sensitively on the accuracy of models for the ionizing radiation fields powered by very young populations of massive stars.

Deep pilot near-infrared spectroscopy with large aperture ground-based telescopes has provided an early preview of the spectra that \jwst{} will make commonplace.
These data have revealed surprisingly prominent emission from doubly-\ and triply-ionized carbon in the rest-UV of several intrinsically-luminous or gravitationally lensed systems at $z>6$ \citep{starkSpectroscopicDetectionsIII2015,starkSpectroscopicDetectionIV2015,mainaliEvidenceHardIonizing2017,schmidtGrismLensAmplifiedSurvey2017,hutchisonNearinfraredSpectroscopyGalaxies2019}.
While emission in \ion{C}{iii}] $\lambda \lambda 1907,1909$ (hereafter \ion{C}{iii}]) at such high equivalent widths ($\gtrsim 15$~\AA{}) is rare in typical massive galaxies at lower redshifts \citep{duIIIEmissionStarforming2017}, a growing body of work at $z\sim 2$ with VIMOS and MUSE and in the local Universe ($\lesssim 100$ Mpc) with the Cosmic Origins Spectrograph onboard the {\it Hubble} Space Telescope (\hstcos{}) has revealed that this semi-forbidden doublet is routinely detected in subsolar metallicity galaxies at very high sSFR \citep[][]{rigbyIIIEmissionStarforming2015,masedaMUSEHubbleUltra2017,senchynaUltravioletSpectraExtreme2017,nakajimaVIMOSUltraDeep2018,duSearchingAnalogsPeak2020}.
However, even the latest stellar population synthesis models struggle to reproduce the highest equivalent width emission in \ion{C}{iii}] detected at $z\sim 2$--6.
Though AGN may contribute to some of this extreme emission at intermediate redshifts \citep{lefevreVIMOSUltraDeepSurvey2019}, the multiple detections of \ion{C}{iii}] in-excess of 15 \AA{} at $z>6$ is suggestive of prominent massive stellar populations for which theoretical ionizing spectra predictions may not be accurate.

These results indicate that our ability to correctly interpret and extract constraints on underlying physics from nebular line emission will depend fundamentally on models for metal-poor stellar populations at very young ages.
The ionizing spectrum powered by massive stars is essentially entirely theoretical at all metallicities and ages since the earliest stars observed directly at the extreme ultraviolet (EUV) energies relevant for \Hii{} region photoionization are B giants \citep{bowyerExtremeUltravioletAstronomy2000}.
Our best hope of calibrating these models is through analysis of nearby star-forming regions that can be studied in great detail.
Much attention has traditionally been paid to the ability of population synthesis models to reproduce the distribution of strong-line ratios for typical star-forming systems, an important litmus test for their veracity \citep[e.g.][]{mcgaughIIRegionAbundances1991,bresolinIonizingStarsExtragalactic1999,charlotNebularEmissionStarforming2001,bylerNebularContinuumLine2017,xiaoEmissionlineDiagnosticsNearby2018}.
However, many of the uncertain effects of stellar modeling (binary mass transfer, fast rotation, IMF variation) are degenerate in this space, especially at very young ages where constraints from large spectroscopic surveys are relatively lacking.
More crucially, this does not test the core assumption underlying nebular emission line modeling: that model fits to the nebular emission lines alone can accurately constrain the properties of the responsible stellar population such as metallicity and age.

The difficulty in reproducing the first rest-UV nebular line detections at $z>6$ underscores the urgent need for model testing for very young stellar populations.
Modern stellar population synthesis results have diverged significantly in the past decade, with discrete improvements to the treatment of stellar winds and atmospheres, binary mass transfer, and rotation yielding predictions that diverge substantially at fixed physical parameters, especially for hard ionizing radiation \citep[e.g.][]{woffordComprehensiveComparativeTest2016}.
If models predict the wrong stellar content and ionizing radiation field for galaxies dominated by a burst in star formation at the very young ages and subsolar metallicities that many reionization-era systems reside at, physical inferences including on star formation histories and metallicities from even the highest-quality \jwst{} spectra will suffer from potentially dramatic systematic uncertainties.
Fortunately, spectroscopy of relatively bright star-forming regions in the local Universe can unveil far more detail than just nebular line strengths.
In particular, ultraviolet and deep optical observations especially at moderately sub-solar metallicities $Z/Z_\odot\gtrsim 0.2$ directly accesses continuum signatures of the massive stars producing the ionizing radiation and nebular emission, including very prominent wind features formed in the dense radiatively-driven outflows these stars power \citep[e.g.][]{allenDwarfEmissionGalaxy1976,brinchmannGalaxiesWolfRayetSignatures2008,leithererUltravioletSpectroscopicAtlas2011,woffordRareEncounterVery2014}.
With direct information about the massive stars present in these star-forming regions, far more stringent tests of stellar population models can be conducted.

In this paper, we present new \hstcos{} UV spectroscopy for seven local ($\lesssim 100$ Mpc) star-forming regions specifically selected to enable such a test of theoretical prescriptions for young stellar populations.
In particular, we focus on the subset of objects from the \citet{shiraziStronglyStarForming2012} sample of star-forming \heii{}-emitters in SDSS with evidence for broad Wolf-Rayet \citep[WR; see e.g.][]{crowtherPhysicalPropertiesWolfRayet2007} stellar wind emission at \heii{} $\lambda 4686$, produced by some combination of very massive and envelope-stripped or spun-up stars (Section~\ref{sec:sampleselect}).
These systems have properties similar by selection to the two prominent \ion{C}{iii}]-emitters at $12+\log\mathrm{O/H} \simeq 8.3$ presented in \citet[][hereafter \citetalias{senchynaUltravioletSpectraExtreme2017}]{senchynaUltravioletSpectraExtreme2017}, with optical nebular emission and inferred ages comparable to that inferred for the highest-sSFR systems in the reionization era (Section~\ref{sec:observations}).
Besides ensuring that direct constraints can be placed on the stellar populations present, the presence of prominent WR wind emission evinces a significant population of very hot stars.
The UV data provide access both to nebular emission in \ion{C}{iii}] and \ion{O}{iii}] $\lambda\lambda 1661,1666$, as well as the \heii{} $\lambda 1640$ recombination line and \civ{} $\lambda 1548,1550$ resonant doublet.
At these metallicities and ages, both \civ{} and \heii{} will be dominated by strong stellar wind features (a P-Cygni feature dominated by O-stars, and broad emission associated with very massive and WR stars, respectively; Section~\ref{sec:ultraviolet}).
These stellar wind features underlying the strong nebular emission lines enable a direct test of stellar population synthesis prescriptions (Section~\ref{sec:uvoptsynthesis}).
The degree to which these models are able to match both the wind features and the nebular emission powered by the massive stars provides crucial insight into the uncertain nature of the very young stellar populations that dominate both these star-forming regions and many galaxies at $z>6$, which we discuss in Section~\ref{sec:discuss}.

For comparison with solar metallicity, we assume a solar gas-phase oxygen abundance of $12+\log_{10}\left([\text{O/H}]_\odot\right)  = 8.69$ \citep{asplundChemicalCompositionSun2009} unless otherwise noted.\footnote{In the context of the Charlot \& Bruzual (in-prep.) stellar population synthesis models, this is very near the value of the gas-phase oxygen abundance $12+\log_{10}(\mathrm{O/H})=8.71$ for $Z=Z_\odot=0.01524$ and near-solar dust-to-metal mass ratio $\xi_d=0.3$ \citep[see Table~2 in][]{gutkinModellingNebularEmission2016}.}
For distance calculations and related quantities, we adopt a flat cosmology with $H_0 = 70$ \unit{km \, s^{-1} \, Mpc^{-1}}.
All equivalent widths are measured in the rest-frame and we choose to associate emission with positive values ($W_0>0$).

\section{Sample Selection}
\label{sec:sampleselect}

While rare in absolute terms, star-forming regions dominated by massive stars at metallicities and ages approaching the conditions found at very high-redshift can be found in appreciable numbers in modern large surveys.
Relatively-shallow optical spectroscopic observations such as those provided by the Sloan Digital Sky Survey (SDSS) can detect emission from highly-ionized gas excited by hard stellar radiation or entrained in dense stellar winds for sufficiently bright nearby systems.
In particular, as recombination lines of \heplus{} with an ionization potential of 54.4 eV, the \ion{He}{ii} $\lambda4686$~\AA{} optical line and its sibling at $1640$~\AA{} in the UV are a powerful signature of galaxies dominated by very hot stars, and broad emission near this line and in the red probe the highly-ionized winds of WR stars \citep[e.g.][]{crowtherPhysicalPropertiesWolfRayet2007}.
These features can be leveraged to efficiently locate the optimum laboratories for the study of prominent populations of young, massive stars.

The Sloan Digital Sky Survey (SDSS) spectroscopic database has been mined for star-forming galaxies displaying both \heii{} $\lambda 4686$ \citep[][hereafter \citetalias{shiraziStronglyStarForming2012}]{shiraziStronglyStarForming2012} and broad WR emission \citep[][hereafter \citetalias{brinchmannGalaxiesWolfRayetSignatures2008}]{brinchmannGalaxiesWolfRayetSignatures2008} separately, yielding hundreds of examples of both including some overlap.
High-resolution ultraviolet and optical spectra of ten \heii{}-emitters from \citetalias{shiraziStronglyStarForming2012} spanning $7.7<12+\log\mathrm{O/H}<8.5$ were presented in \citetalias{senchynaUltravioletSpectraExtreme2017}.
Among the higher-metallicity subset at $Z/Z_\odot \simeq 0.2$--$0.5$, these spectra revealed clear stellar wind diagnostics, providing direct insight into the massive stars in these systems.
The \heii{} profiles of these systems were dominated by broad emission (FWHM$\sim 1600$ km/s) characteristic of populations of envelope-stripped WR and helium stars and very massive stars near the Eddington limit which drive strong, optically-thick stellar winds.
The youngest of these stellar \heii{}-emitters were also found to power nebular \ion{C}{iii}] at equivalent widths well exceeding that of other systems with less prominent wind emission at these metallicities, suggestive of further potential commonalities with \ion{C}{iii}] emitters in the reionization era.

The clear detection of optical and UV massive stellar wind signatures alongside nebular gas emission provides a unique opportunity to test stellar population synthesis predictions at the very young ages relevant for modeling galaxies dominated by bursts at high-redshift.
To establish a statistical sample of UV spectra for galaxies with substantial massive stellar populations to conduct such a test, we selected an additional seven systems from the \citetalias{shiraziStronglyStarForming2012} \ion{He}{ii}-emitting sample which also showed prominent WR stellar wind signatures in the optical.
In particular, \citetalias{shiraziStronglyStarForming2012} identified 116 systems with both \ion{He}{ii} emission and evidence for broad emission at the blue or red WR bumps.
These two bumps, located at 4600--4700 and 5650--5800 \AA{}, are composed of blended emission lines of \ion{He}{ii}, \ion{N}{iii}, \ion{N}{v}, and \ion{C}{iii,iv} in the highly-ionized winds of WR stars \citep[e.g.][]{crowtherPhysicalPropertiesWolfRayet2007,brinchmannGalaxiesWolfRayetSignatures2008}.
Specifically, we selected galaxies from the star-forming \citetalias{shiraziStronglyStarForming2012} sample with:
\begin{itemize}
    \item Clearly-detected WR features in the SDSS spectra: \texttt{WRcl}$\geq 2$.
    \item Sufficiently bright magnitude in the blue for study in a single orbit per grating with \hstcos{}: $u<18.5$ (aperture magnitude in the $3''$-diameter SDSS fiber).
    \item Very high specific star formation rate (sSFR), as selected by [\ion{O}{iii}] $\lambda 5007$ rest-frame equivalent width in-excess of $600$ \AA{}.
\end{itemize}

\begin{figure*}
    \includegraphics[width=1.0\textwidth]{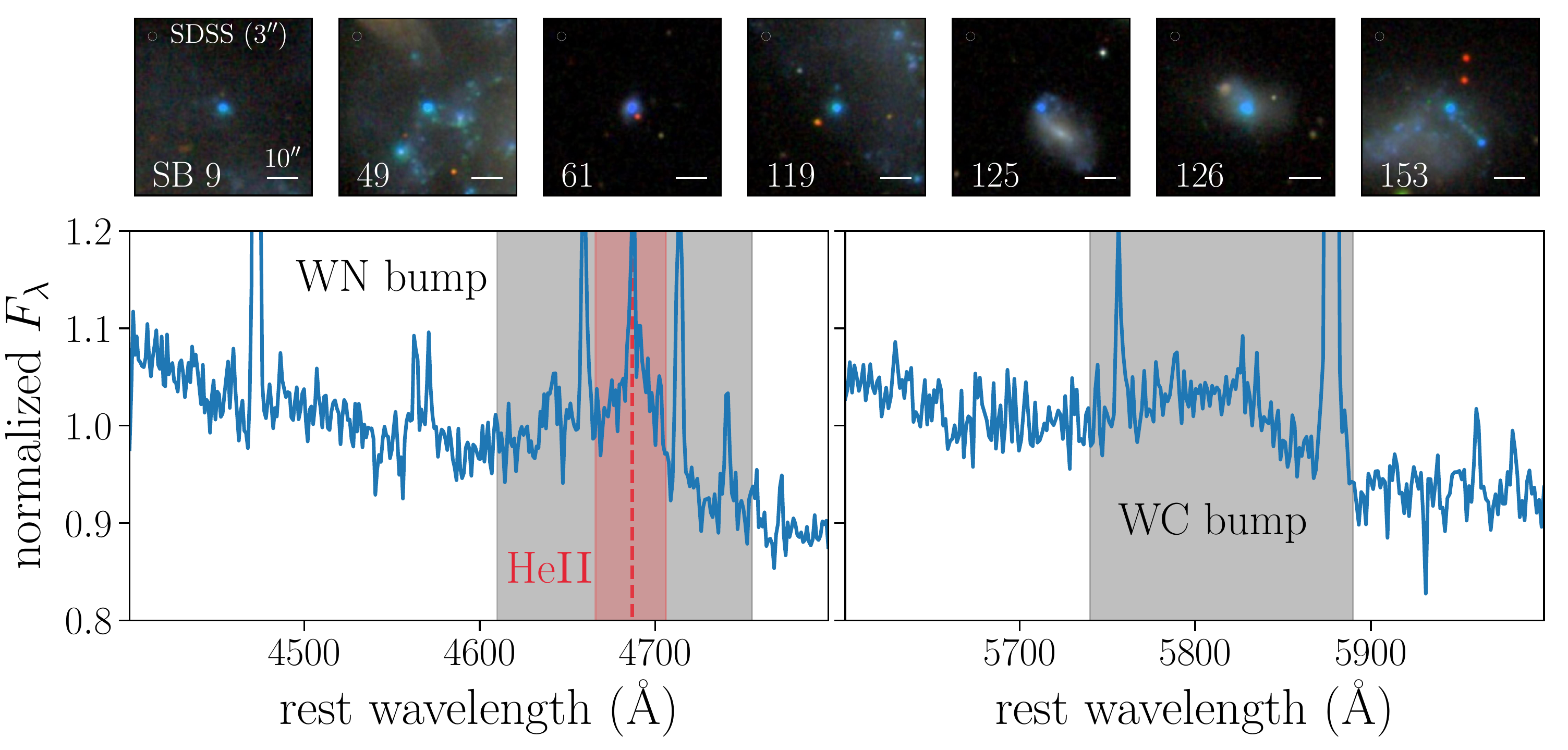}
    \caption{
        Optical $g,r,i$ cutouts from SDSS centered on our targets, ordered by the identification number from the \citet{shiraziStronglyStarForming2012} sample (top row).
        A 10$''$ scalebar and a 3$''$-diameter circle representing the size of the SDSS spectroscopic aperture are displayed for reference in white over each cutout.
        On the bottom row, we highlight the broad blue and red WR bumps in the composite SDSS spectrum of these systems after normalization and median-stacking.
        The presence of this emission from highly-ionized gas entrained in dense stellar winds (in particular, broad \ion{He}{ii} $\lambda 4686$ clearly visible under the narrow nebular component) is clear evidence that these regions are dominated by massive stars formed in a recent burst.
    }
    \label{fig:montage_wspec}
\end{figure*}

A total of 11 galaxies satisfied these requirements, three of which were included in the \citetalias{senchynaUltravioletSpectraExtreme2017} UV sample (SB 80, 179, and 191).
We selected the seven highest-equivalent width systems of the remaining eight for follow-up with \hstcos{} to complete a UV survey of ten galaxies with extreme WR emission.
The new targets are summarized in Table~\ref{tab:basicprop}, and a summary plot showing SDSS cutouts and their stacked SDSS spectra are displayed in Figure~\ref{fig:montage_wspec}.
We will refer to these systems by their identification number from \citetalias{shiraziStronglyStarForming2012} (prefaced SB) throughout this paper.
Note that by construction, our targets are also all present in the earlier sample of SDSS spectra with prominent WR features assembled in \citetalias{brinchmannGalaxiesWolfRayetSignatures2008}.
We present cross-matched identification numbers from that sample and discuss the targets and their host galaxies in the context of other studies in Appendix~\ref{app:targetcontext}.
Together with the targets from \citetalias{senchynaUltravioletSpectraExtreme2017}, our sample comprises the highest-sSFR systems with evidence for prominent stellar \ion{He}{ii} emission in the SDSS spectroscopic survey, ideal for confronting models for very young stellar populations applied at high redshift.

\begin{table*}
\centering
\caption{Basic properties of the target star-forming regions, including RA and Dec; redshift; distance and comoving size of the COS aperture (from redshift, corrected for the Local Flow); aperture magnitudes in the SDSS $u$ and $i$ bands; and exposure time in each of the two \hstcos{} gratings.
In addition to the seven systems for which new spectra are presented in this paper, we also include the other three systems with identical selection (\citetalias{shiraziStronglyStarForming2012} \heii{}-emitters with optical WR emission) that were analyzed in \citetalias{senchynaUltravioletSpectraExtreme2017} and which we incorporate in our analysis presented here.
}
\label{tab:basicprop}

\begin{tabular}{lrrccccc}
\hline
SBID  & RA & Dec & z & Distance & 2.5\arcsec{} & $u$, $i$ & NUV/G185M, FUV/G160M\\ 
 & (J2000) & (J2000) &  & (Mpc) & comoving kpc & AB mag & exposure/s (dataset ID)\\ 
\hline
9   & 13:04:32.27 & -3:33:22.1 & 0.0045 & 20 & 0.24 & 18.1, 18.7 & 2260 (LDI204010), 2597 (LDI204020) \\ 
49  & 1:15:33.82 & -0:51:31.2 & 0.0056 & 31 & 0.38 & 17.1, 17.6 & 2160 (LDI201010), 2597 (LDI201020) \\ 
61  & 14:48:05.38 & -1:10:57.7 & 0.0274 & 117 & 1.41 & 17.9, 17.6 & 2260 (LDI209010), 2597 (LDI209020) \\ 
119 & 14:32:48.36 & 9:52:57.1 & 0.0047 & 22 & 0.27 & 18.1, 18.5 & 2260 (LDI206010), 2601 (LDI206020) \\ 
125 & 11:32:35.35 & 14:11:29.8 & 0.0176 & 78 & 0.94 & 18.2, 18.3 & 2264 (LDI202010), 2604 (LDI202020) \\ 
126 & 11:50:02.73 & 15:01:23.5 & 0.0025 & 18 & 0.21 & 16.7, 17.1 & 2200 (LDI203010), 2120 (LDI208010) \\ 
153 & 13:14:47.37 & 34:52:59.8 & 0.0029 & 11 & 0.13 & 17.6, 18.1 & 2084 (LDI205010), 2497 (LDI205020) \\ 

\hline
\multicolumn{8}{c}{From \citetalias{senchynaUltravioletSpectraExtreme2017}:}\\
\hline

80  & 9:42:56.74 & 9:28:16.2 &  0.0108  & 46 & 0.56 & 17.9, 18.1 & 2132 (LCY401010), 2608 (LCY401020) \\
179 & 11:29:14.15 & 20:34:52.0 & 0.0047 & 25 & 0.30 & 18.3, 18.5 & 2160 (LCY402010), 2588 (LCY402020)  \\
191 & 12:15:18.60 & 20:38:26.7 & 0.0028 & 10 & 0.12 & 17.7, 18.2 & 2156 (LCY404010), 2616 (LCY404020) \\

\hline
\end{tabular}

\end{table*}

\section{Observations and Analysis}
\label{sec:observations}

\subsection{Optical Spectra and Imaging}
\label{sec:optneb}

By virtue of their selection from the SDSS spectroscopic survey, high-quality imaging and optical spectra are already available for all seven new targets.
The broadband photometric magnitudes of the target regions provide important constraints on the total stellar mass and SFR enclosed in the spectroscopic apertures.
We measure aperture photometry centered on each target region from SDSS sky-subtracted and calibrated $ugriz$ frames covering each target, adopting a 3$''$-diameter aperture corresponding to the size of the SDSS spectroscopic aperture.

The public SDSS spectra for these systems provides $R\simeq 1500$--$2500$ coverage over 3800--9200 \AA{}.
As our targets reside at very low-redshifts $z<0.03$, the [\ion{O}{ii}] $\lambda\lambda 3727,3729$ doublet (hereafter [\ion{O}{ii}] $\lambda 3727$ as we observe it as an unresolved single line) remains outside the SDSS spectroscopic range for all but the highest-redshift, SB 61.\footnote{While [\ion{O}{ii}] $\lambda 3727$ is indeed barely shifted into the wavelength range of the SDSS spectrum for the second highest-redshift system SB 125 at $z=0.018$, the bluest part of the line profile is cut-off at the edge of the SDSS spectrum.}
The 3727 doublet is an important component of gas-phase oxygen abundance and ionization parameter determination.
Therefore, in addition to the SDSS spectra, we obtained supplementary blue spectra with the Blue Channel spectrograph mounted on the 6.5m MMT telescope on Mount Hopkins.
Spectra for four of the targets (SB 119, 125, 126, and 153) were obtained on the night of April 9, 2019 with the 800gpm grating and $1.5''\times 180''$ slit, yielding spectra spanning 3400--5200 \AA{}.
Seeing as-measured from the wavefront sensor ranged from 1.1$''$--1.5$''$ over the course of the night.
In addition, SB 49 was observed on the night of October 8, 2019 with the 300gpm grating and $1.0''\times 180''$ slit, resulting in coverage of 3500--8500 \AA{}.
All observations were reduced using standard longslit data reduction techniques, including wavelength calibration using HeAr/Ne lamp exposures and flux calibration with standard star observations conducted at similar airmass (Feige 34 and BD+33 2642 on April 9, and HZ 14 on October 8).
To correct for potential differences in aperture and flux calibration between the SDSS and MMT spectra, we correct the flux of [\ion{O}{ii}] $\lambda 3727$ measured in the MMT data by the median ratio between other strong lines measured in both spectra.
We find a median scatter in this ratio measured between different lines of 8\%, which we add in-quadrature to the corrected flux uncertainty.

\begin{table}
    \centering
    \caption{Summary of supplementary MMT observations used to measure [\ion{O}{ii}] $\lambda 3727$.}
    \label{tab:mmt_obs}
    \begin{tabular}{lcc}
        \hline
        Name & Airmass & Integration time (s)\\
        \hline
        \multicolumn{3}{c}{April 9, 2019} \\
        \hline
        SB 125 & 1.2 & 3600 \\
        SB 126 & 1.1 & 3600 \\
        SB 119 & 1.2 & 1800 \\
        SB 153 & 1.3 & 1800 \\
        \hline
        \multicolumn{3}{c}{October 8, 2019} \\
        \hline
        SB 49 & 1.1 & 900 \\
        \hline
    \end{tabular}
\end{table}

Nebular line flux and equivalent width measurements were performed using the technique described in more detail in \citetalias{senchynaUltravioletSpectraExtreme2017}.
In brief, a base model consisting of a Gaussian emission line (or two in the case of doublets or fitting for multiple components to a single line, as specified) and a linear continuum are fitted to the spectroscopic data in the region of each emission line of interest using the Markov chain Monte Carlo sampling framework \texttt{emcee} \citep{foreman-mackeyEmceeMCMCHammer2013}.
The fits are examined by-eye to ensure that the fit wavelength range is sufficient to accurately capture the continuum level and ensure that the correct features are identified.
The resulting posterior distributions after removing a nominal burn-in period straightforwardly provide confidence intervals for the parameters of interest, including fluxes, equivalent widths, and line widths.

With this photometry and nebular line information in-hand, important constraints can be placed on the bulk star formation history and ionized gas properties of these star forming regions.
As a first step towards constraining the timescale of recent star formation in these systems, we fit the broadband photometric measurements and H$\beta$ equivalent widths measured from the SDSS data.
In particular, we use the BayEsian Analysis of GaLaxy sEds \citep[\beagle{}, v0.23.0;][]{chevallardModellingInterpretingSpectral2016} tool to model each system.
The underlying stellar models and population synthesis methodology will be discussed in more detail in Section~\ref{sec:uvoptsynthesis} and have been previously described by \citet{gutkinModellingNebularEmission2016}.
We assume a constant star formation history (CSFH) with an SMC extinction curve \citep{gordonQuantitativeComparisonSmall2003} and allow the age of this star formation period ($0<\log_{10}[t/\mathrm{yr}]<9$),  metallicity ($-2.2<\log_{10} [Z/Z_\odot] < 0$), stellar mass ($0<\log_{10}[M/M_\odot]<10$), nebular ionization parameter ($-4<\log U < -1$) and $V$-band attenuation ($0<\hat{\tau}_V<2$) to vary over the indicated uniform prior ranges.
We inflate the photometric flux uncertainties by a conservative 5\% added in-quadrature to both more heavily-weight the spectroscopic H$\beta$ equivalent width and account for possible differences in spectroscopic versus photometric aperture, and find good agreement with all fit fluxes in the posterior PDFs.

The results of these SED fits are displayed in Table~\ref{tab:sedres}.
By selection, these systems display extremely high equivalent width optical nebular line emission.
Our measurements of their SDSS H$\beta$ equivalent widths range from 125 to 285 \AA{}, yielding correspondingly strong broadband nebular line contamination.
In particular, the target $g$-band photometric measurements (contaminated by H$\beta$ and [\ion{O}{iii}] $\lambda \lambda 4959,5007$) are brighter than measurements in the adjacent $u$-band (close to the continuum, contaminated only by weaker [\ion{O}{ii}] $\lambda 3727$) by 0.3--1.0 (median 0.5) magnitudes.
The results of our SED fitting reflect this, indicating that the photometry of all seven objects is consistent with entirely young stellar populations.
The median inferred age of our assumed constant star formation history model range from a very young $3.7$ Myr in the most extreme case (SB 153) to $\sim 20$Myr for systems with lower equivalent-width optical emission (SB 125, 126), consistent with the presence of WR wind signatures in the optical spectra (Figure~\ref{fig:montage_wspec}).
These ages represent an upper limit for the young stellar component of the target systems, as adopting a single-age simple stellar population would require ages uniformly $<10$ Myr to reproduce the observed optical line equivalent widths.
The stellar masses inferred from this fitting are uniformly low, spanning $10^{4.9}$--$10^{7.0}$ $M_\odot$, as expected for very young stellar populations with low mass-to-light ratios.
This fitting strongly supports the idea that the spectroscopic apertures centered on these systems are entirely dominated by continuum and nebular gas emission from massive stars formed in a very recent star formation episode.

\begin{table}
\caption{Results of fits to the broadband photometric $ugriz$ SED and H$\beta$ equivalent widths with \beagle{}.
    These results indicate that the optical photometry of the target systems is in all cases consistent with very young stellar populations formed within the last 3--20 Myr.
}
\label{tab:sedres}

\begin{tabular}{lccc}
\hline
Target & $\log_{10}(M/M_{\odot})$ & $\log_{10}(\mathrm{SFR}/(M_{\odot}/\mathrm{yr}))$ & age/Myr\\ 
\hline
SB 9 & $5.05 \pm 0.22$ & $-1.95 \pm 0.12$ & $5.5^{+10.8}_{-0.3}$\\ 
SB 49 & $5.80 \pm 0.10$ & $-1.20 \pm 0.10$ & $5.5^{+3.5}_{-0.3}$\\ 
SB 61 & $6.95 \pm 0.05$ & $-0.05 \pm 0.05$ & $6.2^{+1.4}_{-0.9}$\\ 
SB 119 & $5.12 \pm 0.11$ & $-1.88 \pm 0.10$ & $5.6^{+4.0}_{-0.5}$\\ 
SB 125 & $6.66 \pm 0.16$ & $-0.65 \pm 0.05$ & $20.2^{+11.4}_{-5.7}$\\ 
SB 126 & $5.81 \pm 0.16$ & $-1.42 \pm 0.03$ & $17.1^{+9.2}_{-4.2}$\\ 
SB 153 & $4.93 \pm 0.15$ & $-2.07 \pm 0.15$ & $3.7^{+0.3}_{-0.8}$\\ 
\hline
\end{tabular}

\end{table}

Measurements of the prominent collisionally-excited nebular lines provides direct constraints on the ionization state and chemical abundances of this ionized gas.
As in \citetalias{senchynaUltravioletSpectraExtreme2017}, we use the software \pyneb{} \citep{luridianaPyNebNewTool2015} to derive gas-phase oxygen abundances.
We use the temperature-sensitive ratios of [\ion{O}{ii}] $\lambda 7325$ / $\lambda 3727$ and [\ion{O}{iii}] $\lambda 4363$ / $\lambda 5007$ alongside the density-sensitive [\ion{S}{ii}] $\lambda 6731$ / $\lambda 6716$ ratio to derive electron temperatures and densities appropriate for $\mathrm{O^{+}}$ and $\mathrm{O^{2+}}$, respectively.
\footnote{We assume atomic transition probabilities for \ion{O}{ii} from \citet{wieseAtomicTransitionProbabilities1996,froesefischerBreitPauliEnergyLevels2004} and effective collision strengths from \citet{tayalOscillatorStrengthsElectron2007}. For \ion{O}{iii} we adopt the atomic data of \citet{storeyTheoreticalValuesOIII2000,froesefischerBreitPauliEnergyLevels2004} and collision strengths of \citet{storeyCollisionStrengthsNebular2014}. And for \ion{S}{ii}, we utilize the transition probabilities presented by \citet{podobedovaCriticallyEvaluatedAtomic2009,tayalBreitPauliTransitionProbabilities2010} and collision strengths from \citet{tayalBreitPauliTransitionProbabilities2010}.}
In the one case for which we lack a measurement of [\ion{O}{ii}] $\lambda 3727$ (SB 9), we instead estimate $\mathrm{T_e}$(\ion{O}{ii}) from that measured for \ion{O}{iii} using the empirical relationship derived by \citet[][equation 14, for the $12+\log\mathrm{O/H}>8$ subsample]{izotovChemicalCompositionMetalpoor2006}.
Attenuation corrections are derived from the Balmer decrement assuming an SMC internal extinction curve \citep{gordonQuantitativeComparisonSmall2003} and an intrinsic value of H$\alpha$/H$\beta$ computed with \pyneb{} using our derived $\mathrm{T_e, n_e}$ after correcting first for Galactic extinction towards each object using the \citet{schlaflyMeasuringReddeningSloan2011} maps and the $R_V=3.1$ extinction curve of \citet{fitzpatrickCorrectingEffectsInterstellar1999}.

The resulting metallicities and other optical nebular line properties are presented in Table~\ref{tab:optneb}.
In addition to high equivalent-width emission indicative of an extremely young effective age, these systems exhibit signs of very highly-ionized gas, with dust-corrected \ott{}$=$ [\ion{O}{iii}] $\lambda 4959$ + $\lambda 5007$ / [\ion{O}{ii}] $\lambda 3727$ values of 5.3--10.7.
Their gas-phase metallicities span the regime $8.0<12+\log_{10}\mathrm{O/H}<8.3$, which assuming a solar abundance of $12+\log(\mathrm{O/H})_\odot = 8.69$ corresponds to a relative metallicity range of $0.2<Z/Z_\odot<0.4$.

\begin{table*}
\caption{
    Optical nebular line properties measured from SDSS and MMT spectra where available.
    Note that since we do not have a measurement of [\ion{O}{ii}] $\lambda 3727$ for SB 9, the metallicity for this system (\textdagger) was determined using only [\ion{O}{ii}] $\lambda\lambda 7320, 7330$.
}
\label{tab:optneb} 

\begin{tabular}{lccccccc}
\hline
Name & \ott{} & \rtt{} & [\ion{O}{iii}] $\lambda 5007$ & H$\beta$ & E(B-V) & $\mathrm{T_e}$(\ion{O}{iii}) & $12+\log_{10}(\mathrm{O/H})$\\ 
  &  &  & $W_0$ (\AA{}) & $W_0$ (\AA{}) &  & $10^4$ K & \\ 
\hline
SB 9& -- & --& $920\pm52$& $209\pm9$& 0.04 & $1.04 \pm 0.02$ & $8.30 \pm 0.05$\textdagger\\ 
SB 49& $6.9\pm0.9$& $6.3\pm0.2$& $1001\pm39$& $223\pm5$& 0.21 & $0.98 \pm 0.02$ & $8.20 \pm 0.04$\\ 
SB 61& $10.7\pm0.6$& $10.4\pm0.5$& $1127\pm82$& $146\pm8$& 0.11 & $1.29 \pm 0.02$ & $8.11 \pm 0.04$\\ 
SB 119& $5.5\pm0.5$& $7.4\pm0.4$& $947\pm54$& $208\pm15$& 0.13 & $1.10 \pm 0.02$ & $8.11 \pm 0.04$\\ 
SB 125& $8.6\pm0.7$& $9.2\pm0.3$& $824\pm34$& $125\pm4$& 0.25 & $1.13 \pm 0.02$ & $8.19 \pm 0.03$\\ 
SB 126& $8.6\pm0.6$& $7.7\pm0.3$& $723\pm36$& $134\pm4$& 0.18 & $1.21 \pm 0.02$ & $8.02 \pm 0.04$\\ 
SB 153& $5.3\pm0.7$& $9.1\pm0.4$& $1719\pm133$& $286\pm12$& 0.00 & $1.12 \pm 0.02$ & $8.20 \pm 0.04$\\ 

\hline
\multicolumn{8}{c}{From \citetalias{senchynaUltravioletSpectraExtreme2017}:}\\
\hline

SB 80   & $3.7_{-0.3}^{+0.3}$ & $9.6_{-1.1}^{+1.2}$ & $1195 \pm 132$  & $243 \pm 17$ & 0.13 &  $1.15 \pm 0.02$ & $8.24 \pm 0.06$ \\
SB 179  & $2.7_{-0.1}^{+0.1}$ & $8.5_{-1.4}^{+1.9}$ & $770 \pm 81$ & $196 \pm 8$  & 0.17 &  $1.07 \pm 0.02$ & $8.35 \pm 0.07$ \\
SB 191  & $9.5_{-0.9}^{+0.7}$ & $9.0_{-1.5}^{+1.8}$ & $1649 \pm 239$ & $393 \pm 23$ & 0.02 &  $1.05 \pm 0.03$ & $8.30 \pm 0.07$ \\

\hline
\end{tabular}

\end{table*}

\subsection{Ultraviolet Spectra}
\label{sec:uvspec}

\begin{figure*}
    \includegraphics[width=1.0\textwidth]{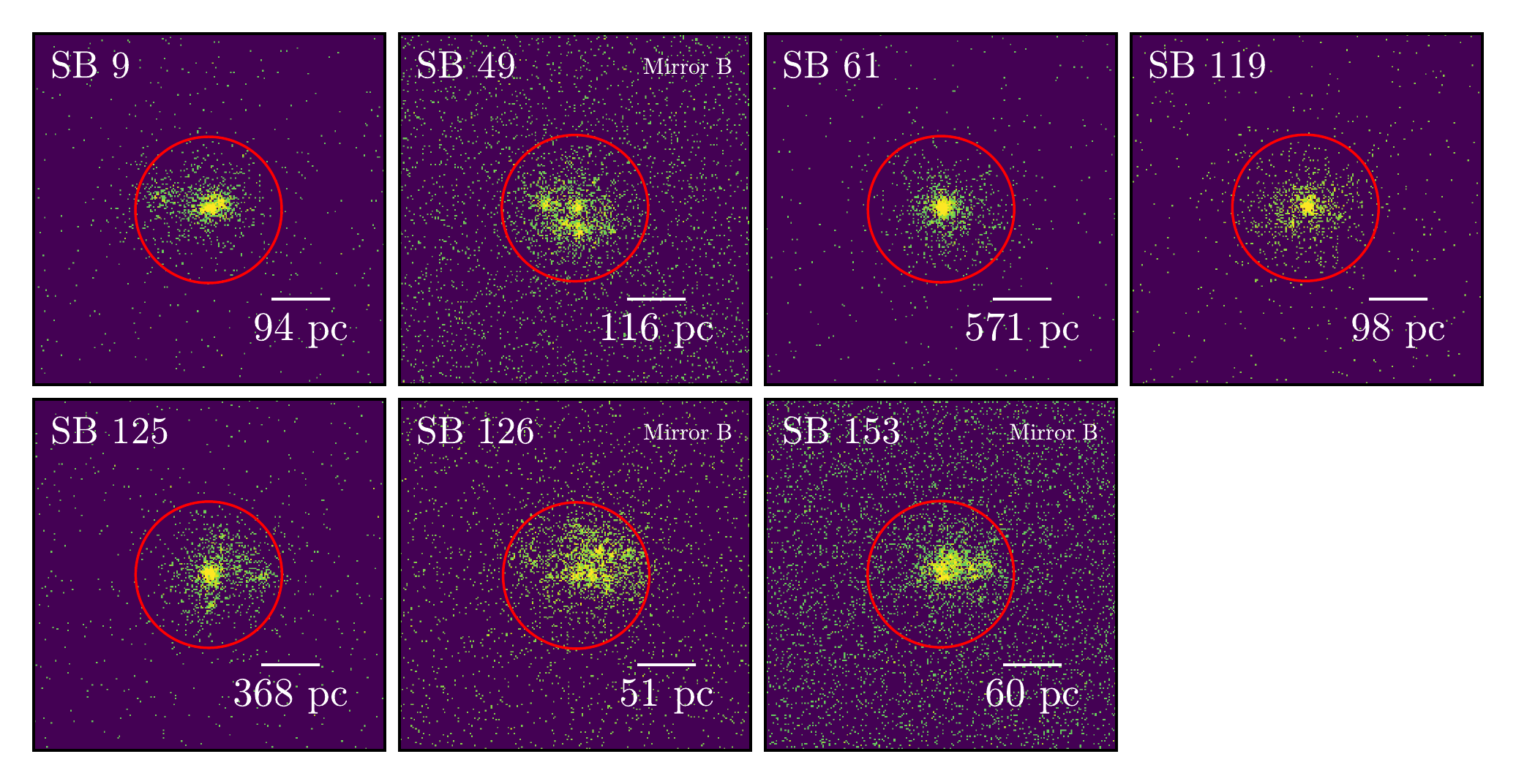}
    \caption{
        Target acquisition images taken in the NUV with \hstcos{}.
        A 1.0$''$ scalebar is drawn in each case, labeled with the comoving distance this angle corresponds to at the estimated distance of each object.
        Note that the brightest three sources were observed with Mirror B to ensure instrument safety, which produces a fainter second image at about half-intensity displaced by 20 pixels from the primary image.
        Thus, some of the structure visible in these images (noted) is artificial.
        While each object is clearly dominated by a bright central object, the additional diffuse flux and fainter compact sources visible around each suggest that we are viewing light from a complex of star-forming regions.
    }
    \label{fig:montage_tacq}
\end{figure*}

We secured moderate-resolution UV spectra with \hstcos{} for all seven target regions in a program approved in HST Cycle 25 (GO:15185, PI:Stark).
Our observations were conducted in the NUV with the G185M grating and in the FUV with G160M, with a 2-orbit visit devoted to each object.
Each target was first centered with a 7--81 second \texttt{ACQ/IMAGE} exposure with either Mirror A or B, with the optical element and exposure time chosen with the COS ETC based upon an assumed compact source with flat SED in $F_\nu$ normalized to the SDSS $u$-band fiber magnitude.
The resulting target acquisition images are shown in Fig.~\ref{fig:montage_tacq}, and reveal a range of morphologies at 10--100 pc scales (though note that the marked Mirror B images are complicated by the fainter, displaced secondary image this optical element generates).
The remainder of the first orbit of each visit was devoted to G185M observation, with the second orbit occupied by G160M.
The central wavelength of each observation was selected with the SDSS redshift in mind to ensure that the full stellar \ion{C}{iv} $\lambda \lambda 1548,1550$ profile, \ion{He}{ii} $\lambda 1640$ and \ion{O}{iii}] $\lambda\lambda 1661,1666$ were located in G160M segments A and B, and that the [\ion{C}{iii}] $\lambda 1907$ + \ion{C}{iii}] $\lambda 1909$ doublet (hereafter \ion{C}{iii}] $\lambda 1908$) was centered in the middle segment NUVB.

Two visits of our program required repeat observation.
On 29 April 2018, guide star tracking was lost during reacquisition at the beginning of the second orbit for SB 126.
While the G185M integration was successfully completed before fine guidance loss occurred, this prevented the G160M integration (\texttt{LDI203020}) from proceeding.
A single-orbit repeat was scheduled as-per HOPR 90495, and a G160M spectrum was successfully recorded for this object on 27 May 2018 (\texttt{LDI208010}).
During the original scheduled visit targeting SB 61 on 20 June 2018, the FGS acquired the guide stars after target acquisition had been attempted.
As a result, the aperture was not correctly centered on-target for the spectroscopic exposures, resulting in an unexpectedly low continuum level and signal-to-noise in both the G160M and G185M spectra (\texttt{LDI207020} and \texttt{LDI207010}, respectively).
These observations were repeated following HOPR 90852 with a successful target acquisition on 9 August 2018, yielding spectra that reach the expected S/N which we use in this paper (\texttt{LDI209020} and \texttt{LDI209010}).

All spectroscopic observations were taken in time-tagged photon-address mode (\texttt{TIME-TAG}) with wavelength calibration lamp exposures (\texttt{FLASH=YES}) to allow for optimal data correction post-observation.
We cycle through all four focal plane offset positions for each grating (\texttt{FP-POS=ALL}) to mitigate the effect of fixed pattern noise.
The data were reduced with the default settings for calibration and extraction using \textsc{CALCOS} v3.3.5 and \textsc{CRDS} v7.3.0 (\textsc{HSTDP} v2019.2).
The final one-dimensional spectra, with resolution elements (resels) of 73.4 m\AA{} in the FUV/G160M and 102 m\AA{} in the NUV/G185M, achieve a typical effective spectral resolution of 0.25 \AA{} as-inferred from the width of Milky Way absorption lines \citepalias[identical to that found in][]{senchynaUltravioletSpectraExtreme2017} and median signal-to-noise of 11 (4) per resel at 1450 \AA{} (1700 \AA{}) in G160M and 1.7 per resel at 1900 \AA{} in G185M.

We measure emission lines in the \hstcos{} spectra in the same manner as for the optical emission lines.
The high spectral resolution afforded by the G160M and G185M gratings allows us to easily identify and separate Milky Way (MW) and ISM absorption lines, alleviating blending concerns except in the most extreme cases.
In addition to our measurement of \ion{O}{iii}] and \ion{C}{iii}] semi-forbidden nebular line emission, we define two integration regions designed to measure the strength of the two key stellar wind features accessed by the G160M spectra.
In order to ensure that ISM and MW absorption lines do not contaminate these measurements, we first mask-out the spectra $\pm 1.5$ \AA{} around all identified strong absorption lines and interpolate the spectrum over these windows.
We estimate the equivalent width of the \ion{C}{iv} P-Cygni wind absorption trough by integrating this cleaned spectrum between 1530 and 1550 \AA{} relative to a linear continuum determined by evaluating the median flux in two windows chosen to avoid wind and nebular line features: [1500:1525] and [1565:1575].
Likewise, we measure the strength of emission in the \ion{He}{ii} $\lambda 1640$ wind and nebular emission line in the window 1630--1650 \AA{} relative to a continuum estimated from flux at [1620:1630] and [1650:1660] \footnote{We use only the blueward continuum window to integrated \ion{He}{ii} $\lambda 1640$ for SB~80 from \citetalias{senchynaUltravioletSpectraExtreme2017} as strong \ion{Al}{ii} $\lambda 1670$ absorption contaminates a significant portion of the redward window.}.
We present the resulting measurements in Table~\ref{tab:uvmeas}.

\begin{table*}
\caption{Line measurements from the new ultraviolet \hstcos{} spectra presented in this paper along with those for the three other extreme WR galaxies included in \citetalias{senchynaUltravioletSpectraExtreme2017}.
In particular, we present equivalent widths integrated over two broad regions capturing the \ion{C}{iv} resonant stellar P-Cygni absorption trough and broad \ion{He}{ii} stellar wind emission; and flux and equivalent width measurements for nebular \ion{O}{iii}] and \ion{C}{iii}] emission.
Where these lines are undetected, we present 2.5$\sigma$ upper-limits thereon.
Nebular lines marked by a \textdagger{} were contaminated by MW \ion{Al}{ii} $\lambda 1670.79$ absorption and either entirely suppressed or likely significantly affected, and should be interpreted with caution.
We also present C/O estimates from population synthesis fits to the UV and optical nebular lines as described in Section~\ref{sec:synth_meth} and adopted in the remainder of the fits presented in this paper.
}
\label{tab:uvmeas}
\scriptsize

\begin{tabular}{lcccccc}
\hline
Name & \ion{C}{iv} $\lambda 1549$ P-Cygni & \ion{He}{ii} $\lambda 1640$ & \ion{O}{iii}] 1661 & \ion{O}{iii}] 1666 & \ion{C}{iii}] 1908 & $\log_{10}(\mathrm{C/O})$\\ 
  & Absorption, $W_0/$\AA{} & Emission, $W_0/$\AA{} & $10^{-15}$ ergs/s/cm$^2$ ($W_0/$\AA{}) & $10^{-15}$ ergs/s/cm$^2$ ($W_0/$\AA{}) & $10^{-15}$ ergs/s/cm$^2$ ($W_0/$\AA{}) & (Sec.~\ref{sec:synth_meth})\\ 
\hline
\hline
SB 9 & $-4.40 \pm 0.07$ & $2.16 \pm 0.16$ & $0.59 \pm 0.06$ ($0.32 \pm 0.03$) & $1.89 \pm 0.11$ ($1.05 \pm 0.07$) & $6.61 \pm 0.44$ ($4.86 \pm 0.35$) & $-0.36\pm0.02$ \\ 
SB 49 & $-6.51 \pm 0.04$ & $3.20 \pm 0.10$ & $<0.97$ ($<0.28$) \textdagger & $4.50 \pm 0.23$ ($1.46 \pm 0.08$) & $6.86 \pm 0.41$ ($3.12 \pm 0.20$) & $-0.67\pm 0.06$\\ 
SB 61 & $-2.28 \pm 0.09$ & $1.99 \pm 0.18$ & $1.19 \pm 0.09$ ($0.60 \pm 0.05$) & $4.12 \pm 0.11$ ($2.19 \pm 0.07$) & $15.16 \pm 0.80$ ($9.43 \pm 0.52$) & $-0.29\pm0.02$ \\ 
SB 119 & $-4.26 \pm 0.10$ & $4.27 \pm 0.24$ & $0.71 \pm 0.05$ ($0.57 \pm 0.04$) & $0.79 \pm 0.04$ ($0.65 \pm 0.03$) & $4.24 \pm 0.47$ ($3.79 \pm 0.45$) & $-0.36\pm0.03$ \\ 
SB 125 & $-4.62 \pm 0.09$ & $2.67 \pm 0.21$ & $<0.47$ ($<0.30$) & $1.50 \pm 0.09$ ($1.05 \pm 0.07$) & $6.11 \pm 0.39$ ($4.80 \pm 0.33$) & $-0.29\pm 0.04$ \\ 
SB 126 & $-3.96 \pm 0.03$ & $1.93 \pm 0.07$ & $<1.24$ ($<0.19$) & $<1.24$ ($<0.20$) \textdagger & $11.68 \pm 0.62$ ($2.44 \pm 0.13$) & $-0.29 \pm 0.06$ \\ 
SB 153 & $-3.75 \pm 0.06$ & $3.94 \pm 0.13$ & $1.11 \pm 0.09$ ($0.43 \pm 0.04$) & $1.20 \pm 0.06$ ($0.50 \pm 0.03$) \textdagger & $17.69 \pm 0.41$ ($8.17 \pm 0.23$) & $-0.30\pm 0.03$\\ 

\hline
\multicolumn{6}{c}{From \citetalias{senchynaUltravioletSpectraExtreme2017}:}\\
\hline

SB 80 & $-3.12\pm0.13$ & $3.29\pm0.27$ & $0.60\pm0.04$ ($0.55\pm0.04$) & $1.69\pm0.07$ ($1.68\pm0.08$) & $<4.1$ ($<4.0$) & $-0.48\pm 0.27$\\
SB 179 & $-5.22\pm0.11$ & $4.67\pm0.23$ & $<1.1$ ($<0.9$) & $1.36\pm0.06$ ($1.17\pm0.06$) & $6.80\pm0.23$ ($8.71\pm0.42$) & $-0.24\pm0.04$ \\
SB 191 & $-5.20\pm0.08$ & $4.02\pm0.16$ & $1.19\pm0.07$ ($0.64\pm0.04$) & -- (--) \textdagger & $15.91\pm0.39$ ($11.33\pm0.34$) & $-0.31\pm 0.01$ \\

\hline
\end{tabular}

\end{table*}

\begin{figure}
    \includegraphics[width=0.5\textwidth]{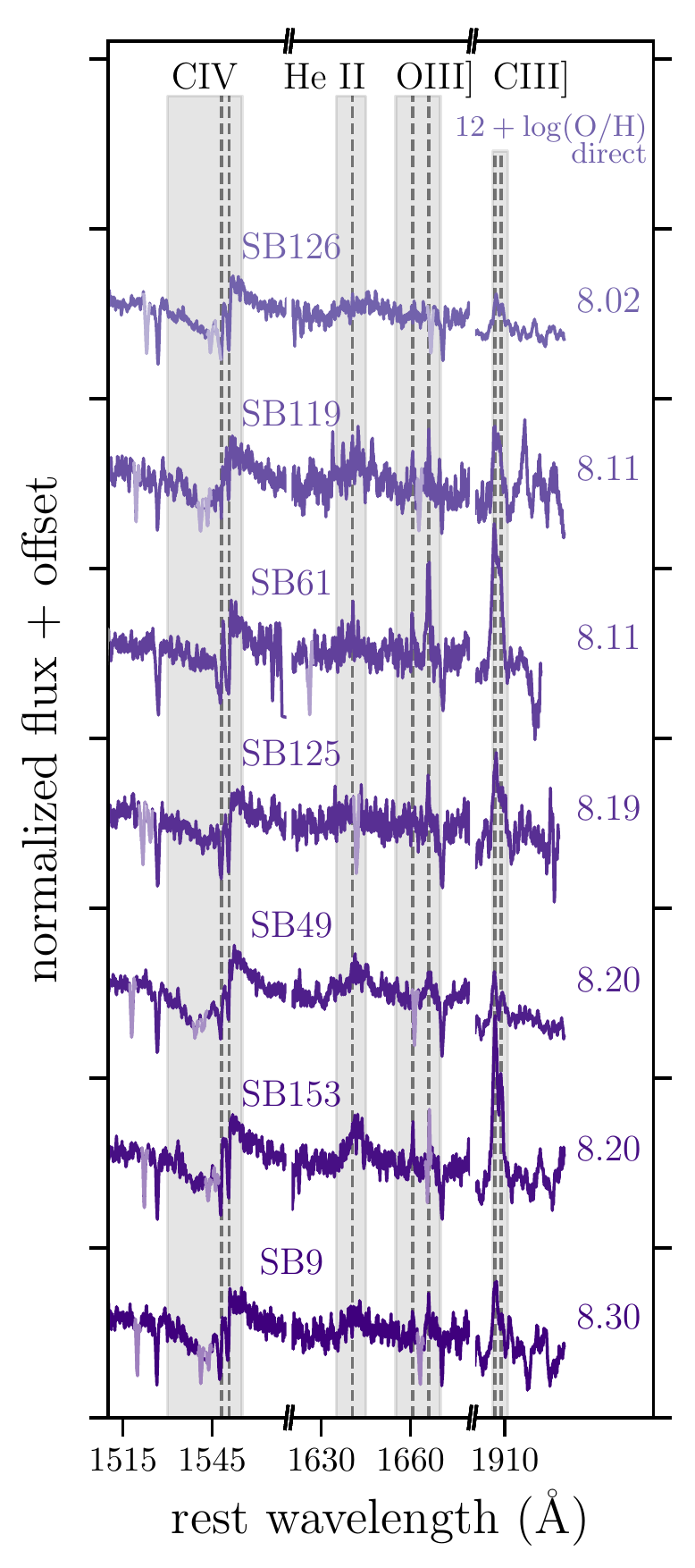}
    \caption{
    Ultraviolet \hstcos{} spectra for the seven new star-forming regions targeted in this paper.
    Sections of the spectra impacted by intervening Milky Way absorption lines are plotted in a lighter shade.
    The G160M+G185M data reveal \ion{O}{iii}] and \ion{C}{iii}] nebular emission in addition to very prominent stellar wind signatures both in resonant \ion{C}{iv} P-Cygni profiles and broad wind emission in \ion{He}{ii}.
    }
    \label{fig:uvspec}
\end{figure}

The measured properties of the UV spectra of our sample (Figure~\ref{fig:uvspec}) further evince a dominant population of massive stars in these galaxies.
We detect \ion{C}{iii}] emission in all seven systems, at equivalent widths ranging from 2--9 \AA{} and \ion{O}{iii}] $\lambda 1666$ in all but one (SB 126, where it is fully absorbed by the MW \ion{Al}{ii} $\lambda 1671$ line).
Including the three additional \citetalias{senchynaUltravioletSpectraExtreme2017} objects reveals a similar detection rate, extending the \ion{C}{iii}] equivalent width range up to just over 11 \AA{} (in SB~191) with only one system undetected in this doublet (SB~80).
The essentially unanimous detection of this doubly-ionized carbon and oxygen emission in the UV confirms the presence of a strong ionizing continuum at $\sim$20--40 eV.

As mentioned, the other high-ionization line complexes in our G160M spectra provide a direct window onto the massive stars powering this nebular gas emission.
The \ion{C}{iv} $\lambda\lambda 1548,1550$ doublet complexes show no clear sign of narrow nebular emission, instead revealing a characteristic P-Cygni profile consisting of broad redshifted emission and blueshifted absorption extending down to $\sim 1530$--1535 \AA{} ($\sim 2500$--3500 km/s), reaching equivalent widths in absorption of 2.3--6.5 \AA{}.
This feature is formed in the dense winds of massive O stars, and is observed to extend to comparable terminal velocities in LMC stars at $Z/Z_\odot \simeq 0.5$ \citep[e.g.][]{masseyPhysicalPropertiesEffective2004,crowtherR136StarCluster2016}.
Likewise, rather than the nebular gas emission typically observed in \ion{He}{ii} $\lambda 1640$ at metallicities below $\sim 0.2 Z_\odot$, the target systems all show broad emission with FWHM $\sim 1500$--2000 km/s.
This emission is characteristic of massive WN and O If stars, and is formed in extremely dense and optically thick stellar winds driven by stars approaching the Eddington limit.
The prominence of this emission is particularly extraordinary, with three systems exceeding 4 \AA{}.
We will discuss the context and implications of these detections in more detail in the following section.

\section{The ultraviolet stellar wind features}
\label{sec:ultraviolet}

Imprints of extremely hot stellar atmospheres dominate the UV spectra of these $10^5$--$10^7$ $M_\odot$ star-forming complexes.
These features represent a powerful probe of the massive stars populating these galaxies and an opportunity to test and inform stellar population synthesis models at very young effective ages.
Before proceeding to describe the experiment we will conduct leveraging these stellar wind profiles in more detail, in this section we will first explore their strength in our sample relative to other star-forming galaxies and determine whether modern population synthesis models approach their equivalent width distribution at all.

While obtaining profiles at this signal-to-noise for individual systems at cosmological redshifts is extremely challenging, prominent stellar wind signatures are routinely encountered in the integrated light spectra of galaxies dominated by recent star formation.
Indeed, \ion{C}{iv} P-Cygni and broad \ion{He}{ii} emission have been detected in gravitationally-lensed and stacked spectra of galaxies out to $z\sim 4$ \citep[e.g.][]{shapleyRestFrameUltravioletSpectra2003,erbPhysicalConditionsYoung2010,dessauges-zavadskyRestframeUltravioletSpectrum2010,jonesKeckSpectroscopyFaint2012,steidelReconcilingStellarNebular2016}, at strengths that have challenged the accuracy of canonical stellar population synthesis models.
These $z\simeq 2$--3 galaxies reach stellar \heii{} $\lambda 1640$ equivalent widths of 1--3 \AA{}, often significantly in-excess of canonical stellar population synthesis predictions without invoking solar or supersolar stellar metallicities \citep[e.g.][]{leithererStarburst99SynthesisModels1999,brinchmannNewInsightsStellar2008}.

\begin{figure}
    \includegraphics[width=0.5\textwidth]{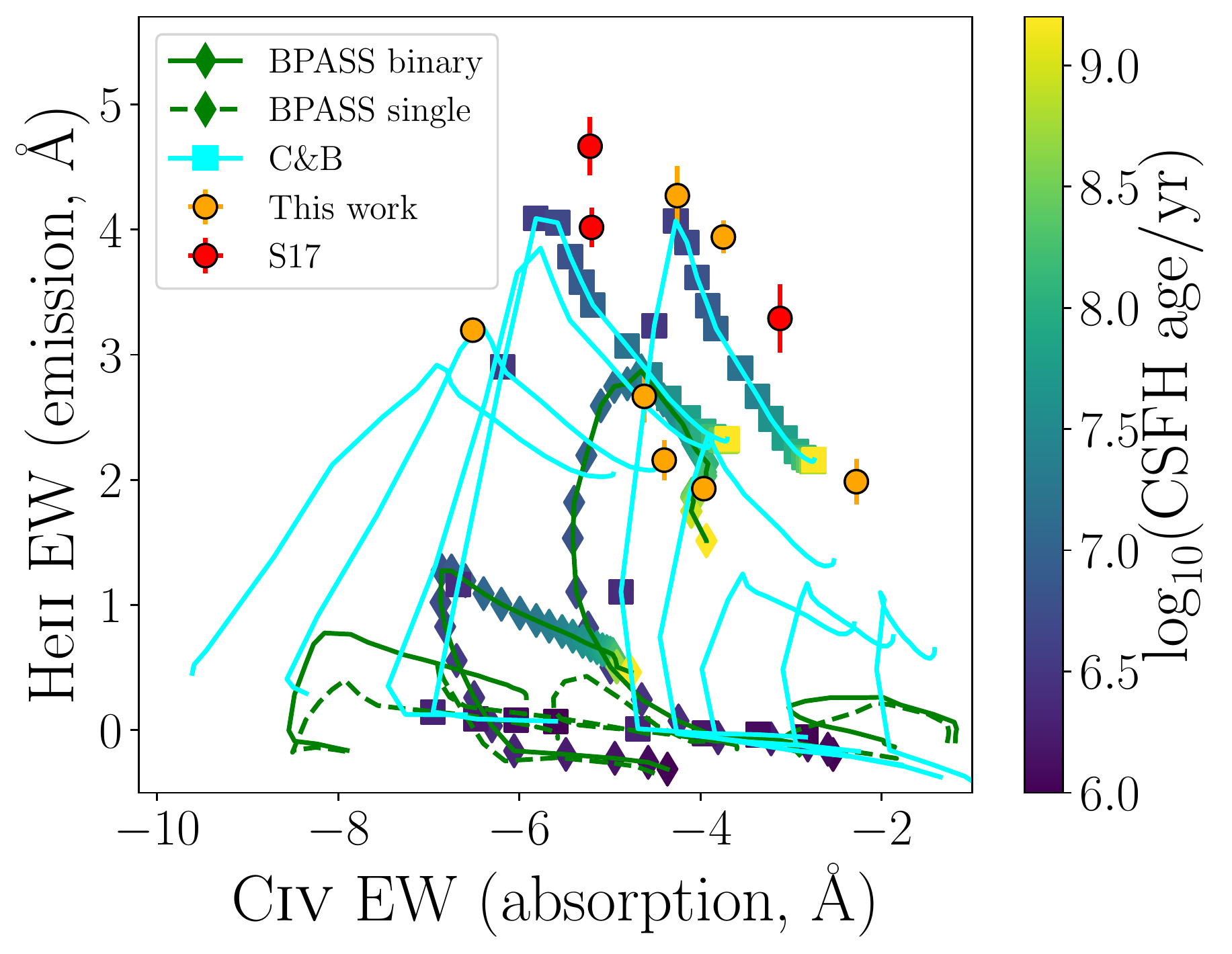}
    \caption{
    The strength of stellar \heii{} emission versus \civ{} stellar absorption for our extreme WR galaxies and selected population synthesis models.
    For each set of models, we plot tracks of age assuming constant star formation for different stellar metallicities.
    Each track rises to peak at a maximal \heii{} equivalent width at an age corresponding to the dominant source of this optically-thick wind emission (canonical massive WR/Of stars or QHE binary products; ages are displayed in color-coded points for a subset), and lower-metallicity model tracks tend to reside farther to the right in median value (weaker \civ{} P-Cygni absorption due to weaker stellar winds overall).
    The observed galaxies (orange and red points) reach broad stellar \heii{} emission equivalent widths well in-excess of canonical model prediction (e.g.\ the BPASS single-star models, green dashed).
    However, comparison with BPASS and \cb{} models suggests that accounting for rapidly-rotating stars produced by binary mass transfer or adopting new predictions for the evolution and atmospheres of very massive stars can both boost these predictions close to the equivalent width regime observed.
    }
    \label{fig:stellarwindstrengths}
\end{figure}

The star-forming regions assembled here reveal even more prominent stellar wind emission, and are among the most intense stellar \heii{}-emitters known.
In Figure~\ref{fig:stellarwindstrengths}, we plot the equivalent width of the stellar \heii{} emission against that of the P-Cygni \civ{} absorption trough, measured as described in Section~\ref{sec:uvspec}, for the full sample of extreme WR galaxies in Table~\ref{tab:uvmeas}.
All of the galaxies in this sample power stellar \heii{} emission at equivalent widths $\geq 2$ \AA{}, reaching 4.7 \AA{} in the most extreme case (SB~179).
The only comparable or more prominent \heii{} emission in integrated light spectra has been found in similar star-forming complexes at low redshift; in particular, NGC~3125-1 \citep[$7.1$ \AA{};][]{chandarNGC31251Most2004,woffordRareEncounterVery2014}, SSC-N in II~Zw~40 \citep[7.1 \AA{};][]{leithererPhysicalPropertiesII2018}, NGC~5253-\#5 \citep[5.1 \AA{};][]{smithVeryMassiveStar2016}, and the coadded spectrum of stars in the center of R136 in the LMC \citep[$4.5$ \AA{};][]{crowtherR136StarCluster2016}.
This powerful though not unprecedented emission places our targets uniformly in-excess of the predictions of canonical stellar population synthesis models \citep[e.g.][]{brinchmannNewInsightsStellar2008}, which even including bursts struggle to exceed 2 \AA{} in the 1640 line.

However, recent developments in stellar population modeling dramatically alter this picture.
Canonically, broad \heii{} emission in integrated galaxy spectra is produced primarily by very massive Of stars and classical WR stars; initially very massive ($>25 M_\odot$) stars which remove their own envelopes through intense stellar wind mass loss exposing their He-burning core after evolving off the main sequence \citep[e.g.][]{crowtherPhysicalPropertiesWolfRayet2007}.
In this case, prominent \heii{} emission appears only for a short period $\sim 5$ Myr following an instantaneous burst of star formation.
However, wind emission in \heii{} is a common trait of very hot stellar cores without a thick envelope of hydrogen, and such stars can also be produced in abundance by binary interaction.
Both mass donor stars stripped of their outer envelopes and mass gainers rejuvenated and driven to high rotation rates by conservative mass transfer can evolve into helium stars, potentially driving strong wind emission in \heii{} \citep[e.g.][]{cantielloBinaryStarProgenitors2007,gotbergSpectralModelsBinary2018}.
In particular, \citet{eldridgeEffectStellarEvolution2012} demonstrate that including stars spun-up to very high rotation rates by mass accretion which become fully-mixed and consequently undergo quasi-homogeneous evolution \citep[QHE; e.g.][]{yoonEvolutionRapidlyRotating2005} can substantially impact on the strength of this wind emission.
They find that incorporating a prescription for generating such stars into their Binary Population And Spectral Synthesis (BPASS) code brings model predictions assuming a constant star formation history into much better agreement with observations at $z\sim 2$--3.

Changes to the modeling of very massive single stars can also significantly alter predictions for stellar \heii{} and other high-ionization wind lines.
In addition to massive stars which evolve through a WR phase during core helium-burning, very massive stars on the MS but so luminous as to reside close to the Eddington limit can also drive dense winds with \heii{} in prominent emission, dominating \heii{} emission in very young ($<5$ Myr) star clusters \citep[Of/WNh stars: e.g.][]{crowtherR136StarCluster2016}.
The latest version of the \citet{bruzualStellarPopulationSynthesis2003} stellar population synthesis models (Charlot \& Bruzual, in-preparation; hereafter \cb{}) adopt new prescriptions for the evolution, mass loss rates, and atmospheres of massive stars \citep[as described in Section~\ref{sec:synth_meth} and][]{vidal-garciaModellingUltravioletlineDiagnostics2017} which together significantly affect the strength of optically thick wind lines like \heii{}.
In particular, the underlying PARSEC stellar evolutionary models for massive stars adopt a new Eddington-factor formalism to predict mass loss rates for the most massive stars, which allows strong stellar winds to be driven at lower metallicities and very young ages for sufficiently luminous stars \citep{chenPARSECEvolutionaryTracks2015}.

It is instructive to compare the predictions of these modern stellar population synthesis codes for the UV stellar wind features in the context of our new measurements.
In Figure~\ref{fig:stellarwindstrengths}, we plot both BPASS (v1.1, with and without binary interaction included\footnote{In particular, we plot the results for BPASS v1.1 for consistency with \citet{eldridgeEffectStellarEvolution2012}, but note that we measured \heii{} $\lambda 1640$ in the latest v2.2 results and did not find significantly higher equivalent widths than reported in v1.1.}) and \cb{} model tracks for comparison with our \hstcos{} stellar wind measurements.
For each set of models, we plot tracks in age assuming a constant star formation history at the relevant stellar metallicities provided by the models ($0.05<Z/Z_\odot<1.0$ for BPASS; and for \cb{}, $0.16\leq Z/Z_\odot\leq1$).
Each fixed-metallicity track (line) spans a limited range of \civ{} equivalent widths, with higher-metallicity models with stronger O-star winds presenting deeper \civ{} absorption on average.
Each track generally begins at low-EW in \heii{} and first moves nearly horizontally to increasingly strong \civ{} absorption as the O-star population is built-up.
The tracks then progress to higher \heii{} equivalent widths which peak at a certain age before settling-down to a more moderate equilibrium value in \heii{} and \civ{} as the continuum contribution of somewhat older stellar populations with weaker winds is built-up.

As can be seen for the selected tracks for which we plot individual model points with color-coded ages, this peak value in \heii{} occurs at different ages for different model assumptions.
In particular, the BPASS models reach this value at $\sim$10--50 Myr as the contribution from initially lower-mass stars $\sim 10$--$20 \; M_\odot$ that are spun-up into QHE by mass transfer reach a maximum.
These models with QHE boost \heii{} substantially over the BPASS single-star predictions, and this effect is most prominent at $Z/Z_\odot=0.2$ due to the competing effects of increasing numbers of stars undergoing QHE at lower metallicities and the diminishing strength of their stellar winds \citep{eldridgeEffectStellarEvolution2012}.
While this does push the models closer to the observed population, intriguingly at a similar range of \civ{} equivalent widths as observed for our strongest \heii{} emitters, the maximum \heii{} equivalent widths reached by these models still fall short of the most extreme emitters by at least 1 \AA{}.

In contrast, the \cb{} models reach a maximum in \heii{} at much younger ages.
Because the dominant source of the stellar \heii{} emission in these single star models is initially-massive Of and WR stars rather than a broader variety of stars spun-up by binary interaction, the tracks boost to their respective maxima at $\sim 4$--5 Myr rather than $>10$ Myr.
Additionally, both the peak and equilibrium values of the \heii{} equivalent width for a given track are uniformly greater than predicted by the other models, extending to 3--4 \AA{} and 2--3 \AA{}, respectively.
As in the BPASS predictions, the maximum in \heii{} wind emission is reached at an intermediate metallicity rather than at solar; in the case of the \cb{} models, this occurs at $Z=0.006$--$0.008$, or $Z/Z_\odot \simeq 0.5$ (approximately the metallicity of the LMC).
While the tracks still fall slightly short of the most intense emission observed, these models have now entered a  similar part of this stellar wind parameter space as that occupied by our observed star-forming regions.

The \cb{} models come close to spanning the stellar \civ{}-\heii{} equivalent width regime that our target galaxies occupy.
This suggests that the improved treatment of the evolution of single massive stars and their winds has alleviated at least some of the tension previously noted between predicted and observed stellar \heii{} strengths in very young star-forming systems.
However, these models do not include binary mass transfer, mergers, or initially-high rotation rates, which the BPASS models discussed above demonstrate could substantially boost wind emission in \heii{} as well.
The impact of these processes on the evolution of the most massive stars remains highly uncertain, yet increasingly relevant to our ability to interpret high-redshift galaxies dominated by very young stellar populations.
In the following section, we describe an experiment testing whether the \cb{} models can self-consistently reproduce the strong \civ{} and \heii{} wind lines alongside the nebular emission lines powered by the same massive stars.
By searching for discrepancies with the latest single-star population synthesis predictions, we will determine whether the ionizing spectrum and stellar wind lines are impacted significantly by binary evolution or other uncertain physics relevant at these young inferred stellar ages.
The results of this experiment will illustrate the impact that ignoring such processes may have on the validity of spectral inference for galaxies undergoing bursts of star formation in the distant Universe.

\section{UV--optical spectral synthesis}
\label{sec:uvoptsynthesis}

As outlined in the sections above, our high-quality UV--optical spectroscopy access powerful diagnostics of both the winds driven by massive stars and the ionizing radiation fields they produce by way of reprocessed nebular emission.
The very high equivalent width of the optical lines these systems power implies a dominant young $\lesssim 10$ Myr stellar population, comparable to the extremely high sSFR systems glimpsed in the reionization era \citep[e.g.][]{roberts-borsaniGalaxiesRedSpitzer2016,endsleyOIIIBetaEquivalent2020} and probing conditions where the uncertain treatment of very massive stars in population synthesis will have an outsized impact on our ability to correctly interpret and model observations.

As discussed in Section~\ref{sec:ultraviolet}, in this paper we will focus on comparisons to the latest predictions of the \cb{} stellar population synthesis framework.
While these state-of-the-art single-star evolution models come far closer to reproducing the observed distribution of stellar wind line strengths than previous generations, they still do not incorporate the impact of mass transfer, mergers, or rapid rotation.
Systematic issues in reproducing the spectra of very young star-forming regions such as these can provide valuable insight into the physics at work in shaping the evolution of the most massive stars.

In particular, when the strong nebular emission line fluxes are fitted with the \cb{} models and used to constrain the properties of the underlying stellar population, we are interested in whether the strength of the UV stellar wind features of the responsible massive stars can be successfully reproduced.
If the \cb{} models are unable to match both sets of features, the discrepancies may reveal the impact of stellar interactions commonly neglected in population synthesis at these metallicities and ages.
We describe our methodology and the results of these fits in the following two subsections.

\subsection{Methodology}
\label{sec:synth_meth}

We will conduct this comparison experiment against the updated \cb{} single-star models using the \beagle{} inference framework.
As implied by our findings in Section~\ref{sec:ultraviolet}, the improvements in the treatment of very massive star evolution adopted by these models has alleviated some of the tension previously found with the stellar \heii{} wind line; but these models still neglect the physics of binary evolution and rotation, which may have a significant impact even at very young stellar population ages.
The version of the \cb{} models used in this work is described in more detail in \citet[][their Section~2.1 and Appendix~A, updating \citealt{gutkinModellingNebularEmission2016}]{vidal-garciaModellingUltravioletlineDiagnostics2017}.
The most important features of this code for the purposes of this work concern the revised predictions for the evolution and atmospheres of the most massive stars (extended now to 300 $M_\odot$).
The latest predictions from the PAdova and TRieste Stellar Evolution Code (\parsec{}) are now adopted to predict the evolution of individual massive stars \citep{bressanPARSECStellarTracks2012,chenPARSECEvolutionaryTracks2015}.
Notably, these latest \parsec{} results include a revised formalism for the prediction of stellar wind driven mass loss rates based upon the stellar Eddington factor \citep{vinkWindModellingVery2011}.
This is in contrast to most other stellar evolutionary codes which predict O and WR mass loss rates using older calibrations \citep[][]{vinkMasslossPredictionsStars2001} derived under assumptions about wind driving which likely break down for stars at the highest luminosities and sub-solar metallicities \citep[see e.g.][]{grafenerMassLossLatetype2008,mullerConsistentSolutionVelocity2008,sanderDrivingClassicalWolfRayet2020}.
In addition, the \cb{} models leverage the large library of non-LTE, line-blanketed, spherically-expanding WR model atmospheres produced by the Potsdam Wolf Rayet group (PoWR) spanning metallicities from $Z=0.001$--$0.014$, effective temperatures from $10^{4.4}$--$10^{5.3}$ K, and a range of CNO mass fractions\footnote{see \url{http://www.astro.physik.uni-potsdam.de/~wrh/PoWR/powrgrid1.php}.} \citep[][]{grafenerLineblanketedModelAtmospheres2002,hamannTemperatureCorrectionMethod2003,hainichWolfRayetStarsLarge2014,hainichWolfRayetStarsSmall2015,sanderHydrodynamicModelingMassive2015,todtPotsdamWolfRayetModel2015}.
These updated mass loss and atmosphere prescriptions for massive stars near the Eddington limit have significantly improved bulk agreement with observations for stellar lines like \heii{} $\lambda 1640$ produced in the optically-thick winds such Of and WR stars drive, as we discussed in Section~\ref{sec:ultraviolet} above.

The spectral inference framework \beagle{} \citep[][introduced in Section~\ref{sec:optneb}]{chevallardModellingInterpretingSpectral2016} is readily able to fit spectrophotometric observables using the \cb{} stellar population models and associated \cloudy{} nebular line predictions, and provides the inference framework we require for this experiment.
For this analysis, we choose to adopt the fiducial \citet{chabrierGalacticStellarSubstellar2003} initial mass function (IMF) with upper mass cutoff of 300 $M_\odot$ \citep[as we will discuss below, the contribution of very massive stars $>100 M_\odot$ to high-ionization stellar wind lines can be dominant; e.g.][]{crowtherR136StarCluster2016}.
For all fundamental variables describing the stellar population, we adopt a uniform prior over the adopted range for simplicity.
Since the initial photometric fits including nebular emission were well-fit by a CSFH model (Section~\ref{sec:optneb}) and this likewise populates the region of stellar wind parameter space occupied by our observations (Figure~\ref{fig:stellarwindstrengths}), we parameterize our assumed SFH as such.
However, in addition to the variable start time (which we allow to vary from $10^{6.5}$--$10^{9.0}$ years) and stellar mass ($M\leq 10^{10} M_\odot$) for this recent period of star formation, we allow an early truncation of up to 10 Myr before the present time to allow additional flexibility in the mix of very massive stars present.
We allow the stellar metallicity to freely vary with $-2.2\leq\log_{10}(Z/Z_\odot)\leq 0$.
While this framework explicitly couples the interstellar metallicity to this stellar metallicity, depletion of refractory elements onto dust grains can change the gas-phase oxygen abundance at fixed $Z$. 
We allow the dust-to-metal mass ratio $\xi_d$ to vary from 0.1--0.5, which effectively corresponds to a factor of 1.6 range in the ratio of gas-phase oxygen to stellar iron \citep{gutkinModellingNebularEmission2016}.
This ratio plays an important role in modulating the joint nebular and stellar spectra of star-forming systems \citep[e.g.][]{steidelReconcilingStellarNebular2016,stromMeasuringPhysicalConditions2018,sandersMOSDEFSurveyDirectmethod2020}, since oxygen is the primary species probed in the nebular spectrum while the winds and ionizing spectra of massive stars are shaped primarily by iron.
Thus a model preference for low $\xi_d$ can be interpreted in part as a preference for enhanced $\alpha$/Fe in this context.
Since we are studying relatively small star-forming regions, we assume an SMC extinction curve with variable $0\leq\tauV{}\leq 2$.
Finally, we allow the nebular ionization parameter to vary over $-4\leq \log U \leq -1$.

\begin{figure}
    \includegraphics[width=0.5\textwidth]{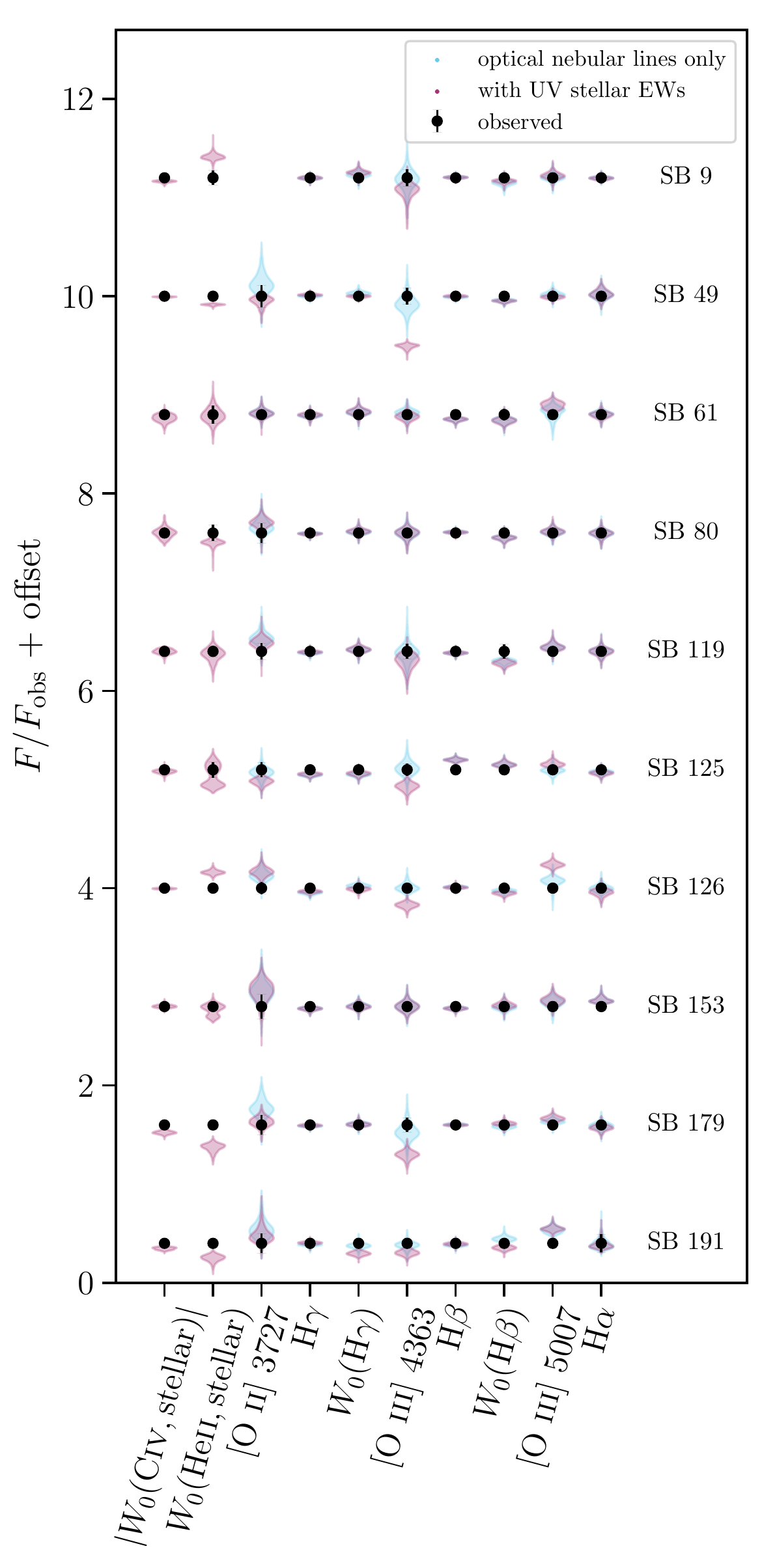}
    \caption{
        Results of our fits to nebular line measurements for each target system with \beagle{}.
        In each row, we plot the fitted measurement as a black point with errorbar (often small enough to be hidden by the point); and behind it, we plot the fit model distribution for each normalized quantity drawn from the model posterior.
        We present results from two fits; one including only the optical nebular line information (presented in blue) and an otherwise identical run but incorporating the equivalent width of the UV stellar wind features presented on the right side of this plot (red).
        Agreement is very good with fits to the nebular line information only, though some offsets with the data are apparent when jointly fitting these measurements with the stellar wind equivalent widths.
    }
    \label{fig:beagle_fitlinecomp}
\end{figure}

\begin{figure}
    \includegraphics[width=0.5\textwidth]{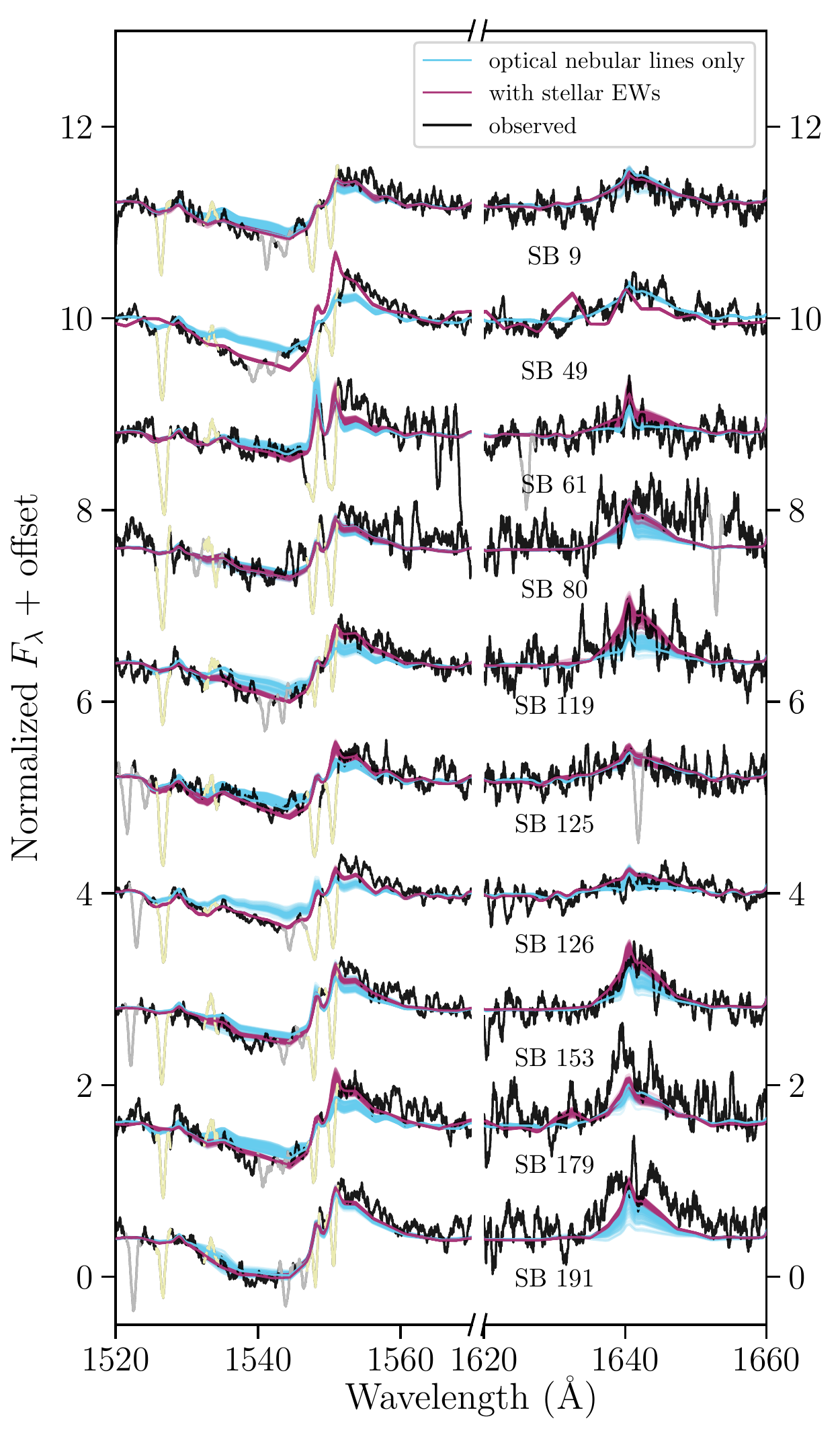}
    \caption{
        Comparison of our \beagle{} model results to the UV \civ{} and \heii{} complexes, which in these galaxies are both dominated by broad massive stellar wind features.
        We plot both the data and models after first normalizing by the continuum as described in the text.
        The fits in blue incorporate no information about these wind features, whereas those in red include the equivalent width measured for both.
        Interstellar absorption lines in the data from the galaxy CGM and the MW are masked in yellow and white, respectively.
        Given only the nebular line constraints, the \cb{} models systematically underestimate the equivalent width of both stellar wind tracers.
        While incorporating these strengths into the fits generally improves the agreement, as discussed in the text this is in most cases requires metallicities and effective ages in significant tension with the optical data.
    }
    \label{fig:beagle_uvcomp}
\end{figure}

In this experiment, we are interested first in fitting the nebular line fluxes which constrain the stellar ionizing spectrum and gas-phase metallicity.
In particular, we focus on the set of strong optical nebular lines necessary to measure the gas-phase oxygen abundance via the direct method: [\ion{O}{ii}] $\lambda 3727$, [\ion{O}{iii}] $\lambda 4363$, H$\beta$, [\ion{O}{iii}] $\lambda\lambda 5007$, and H$\alpha$
Since the gas densities inferred from the [\ion{S}{ii}] $\lambda \lambda 6716,6731$ doublet are all consistent with 100--200 $\unit{cm}^{-3}$, for simplicity we exclude these lines from the fits and instead fix the gas density to the nearest grid value of 100 $\unit{cm}^{-3}$.
For additional leverage on the dust attenuation, we add H$\gamma$; and to better anchor the recent star formation history of the models, we also fit equivalent widths measured in SDSS for H$\gamma$, H$\beta$, [\ion{O}{iii}] $\lambda 5007$, and H$\alpha$.
In each case, the observed line flux or equivalent width is compared to the flux measured via simple integration using a linear continuum in the model spectra.
We choose the integration region and continuum windows for each line manually to ensure that they capture the entire line flux and that the continuum is not biased by other features.
Since our objective is to determine whether the UV stellar wind lines can be reproduced, we also perform a second fit for each object including the equivalent widths of these features measured as for the data (as in Section~\ref{sec:uvspec}, though obviously without the need for ISM or MW absorption line masking in the model spectra).
For each object, we extract constraints on the underlying model parameters from the posterior (displayed in Table~\ref{tab:beagle_nebfit}), as well as the flux distributions for the fitted lines to examine fit quality (Figure~\ref{fig:beagle_fitlinecomp}).
Finally, we also sample the posterior model distribution and extract predicted UV spectra at \civ{} $\lambda\lambda 1548,1550$ and \heii{} $\lambda 1640$.
Analysis of the UV spectra in a continuum-normalized space minimizes issues with potential flux calibration mismatch between the optical and UV.
We fit a cubic spline to the UV spectrum with the stellar wind lines masked both for the data and the model spectra, and normalize the spectra by this continuum before comparison in Figure~\ref{fig:beagle_uvcomp}.

Thus far, we have focused on fits to only the optical nebular lines and the equivalent widths of the UV stellar wind lines.
This relatively straightforward approach allows us to ignore the complications that reddening and aperture matching uncertainties introduce when fitting both UV and optical line fluxes from \hstcos{} and SDSS.
However, we are also interested in whether the UV \ion{O}{iii}] and \ion{C}{iii}] lines can be reproduced, and the ratio of these lines provides valuable constraints on the gas-phase C/O ratio.
Thus, we perform an additional set of fits with the above-described optical line information along with the flux and equivalent width of both \ion{O}{iii}] $\lambda 1661,1666$ and the total flux of the \ion{C}{iii}] $\lambda 1908$ doublet (with upper limits employed where lines are undetected or impacted by MW absorption), and allow the C/O abundance to vary over the full range of the models (0.1--1.4).
We verify that the strength of the \ion{O}{iii} and \ion{C}{iii} lines are reasonably well-reproduced by this fit within the measurement uncertainties, confirming that the prominent \ion{C}{iii}] detections here can be reproduced by our population synthesis models and that the resulting C/O abundances are reasonable.
We present these C/O constraints in Table~\ref{tab:uvmeas}.
We also verify that the results of the experiment we present in the following section, comparing nebular line fits with or without including constraints on the UV stellar equivalent widths, produce qualitatively-similar results for the stellar wind lines when the UV nebular lines are included.
For simplicity of interpretation, we focus on fits only to the optical nebular line and UV equivalent width fits for the remainder of this paper.
However, we utilize the results of the joint optical and UV nebular line fits to first fix the C/O abundance assumed in the population synthesis fits presented and discussed here, adopting the median values presented in Table~\ref{tab:uvmeas} for each.

The parameter estimates from the optical nebular line fits (Table~\ref{tab:beagle_nebfit}) for the new systems in this paper are generally in reasonable agreement with the fits to the photometry and $W_0(\mathrm{H\beta})$ described in Section~\ref{sec:optneb} (Table~\ref{tab:sedres}).
The star formation rates are generally well-matched between the two methods within their uncertainty.
The nebular line fits without UV information generally prefer a slightly earlier initiation of the current period of star formation than the photometry fits (median age of 20 Myr compared to 4 Myr), and the inferred stellar masses tend to be correspondingly higher by $\sim 0.3$--0.5 dex. \footnote{With the exception of SB 153, where the nebular line fit prefers a star formation rate 0.4 dex lower and a lower mass by 0.3 dex.}
The nebular-only fits uniformly prefer a truncation of the current period of star formation within the last Myr, and suggest gas-phase oxygen abundances $12+\log\mathrm{O/H}\simeq 8.3$--$8.5$ modestly larger than those inferred from the direct-$T_e$ method (Table~\ref{tab:optneb}).
With $\xi_d\simeq 0.1$--0.4, this places the estimates of the stellar metallicity $Z_\star$ at $25$--50\% solar.
Adding the UV stellar equivalent widths changes this picture, as will be discussed in detail in the following section and Appendix~\ref{sec:casebycase}.

\begin{table*}
\caption{Constraints on model parameters of interest from our \beagle{} fits to nebular line emission, with or without including the UV stellar equivalent width measurements in the fit.
In particular, we highlight the stellar mass, SFR, and beginning and end of the CSFH period; as well as the gas phase oxygen abundance, stellar metallicity, dust-to-metal mass ratio, and ionization parameter
}
\label{tab:beagle_nebfit}

\begin{tabular}{lcccccccc}
\hline

Target & $\log_{10}(\mathrm{M/M_\odot})$ & $\log_{10}(\mathrm{SFR}/\mathrm{M_\odot yr^{-1}})$ & $\mathrm{t}_1$ &  $\mathrm{t}_0$  & $12+\log_{10}(\mathrm{O/H})$ & $\log_{10}(Z_\star/Z_\odot)$ & $\xi_d$ & $\log_{10} U$ \\
\hline
\multicolumn{9}{c}{Optical nebular lines only} \\
\hline 

SB 9& $5.29^{+0.05}_{-0.05}$& $-1.95^{+0.06}_{-0.03}$& $7.2^{+0.1}_{-0.1}$& $<6.0$& $8.39^{+0.04}_{-0.04}$& $-0.37^{+0.07}_{-0.06}$& $0.29^{+0.13}_{-0.13}$& $-2.95^{+0.08}_{-0.06}$\\ 
SB 49& $6.26^{+0.04}_{-0.04}$& $-0.93^{+0.04}_{-0.04}$& $7.2^{+0.0}_{-0.0}$& $<6.0$& $8.53^{+0.03}_{-0.03}$& $-0.30^{+0.04}_{-0.03}$& $0.16^{+0.06}_{-0.04}$& $-2.74^{+0.06}_{-0.06}$\\ 
SB 61& $7.56^{+0.06}_{-0.08}$& $-0.05^{+0.05}_{-0.03}$& $7.6^{+0.1}_{-0.1}$& $<6.0$& $8.25^{+0.04}_{-0.07}$& $-0.57^{+0.05}_{-0.07}$& $0.13^{+0.04}_{-0.02}$& $-2.06^{+0.06}_{-0.07}$\\ 
SB 80& $6.10^{+0.04}_{-0.04}$& $-0.91^{+0.04}_{-0.04}$& $6.9^{+0.0}_{-0.1}$& $<6.0$& $8.33^{+0.03}_{-0.03}$& $-0.47^{+0.08}_{-0.04}$& $0.21^{+0.15}_{-0.08}$& $-2.59^{+0.05}_{-0.05}$\\ 
SB 119& $5.60^{+0.06}_{-0.07}$& $-1.71^{+0.07}_{-0.04}$& $7.3^{+0.1}_{-0.1}$& $<6.0$& $8.37^{+0.05}_{-0.04}$& $-0.43^{+0.09}_{-0.05}$& $0.17^{+0.18}_{-0.05}$& $-2.76^{+0.07}_{-0.05}$\\ 
SB 125& $7.06^{+0.08}_{-0.07}$& $-0.63^{+0.07}_{-0.08}$& $7.7^{+0.1}_{-0.1}$& $<6.0$& $8.38^{+0.03}_{-0.03}$& $-0.34^{+0.07}_{-0.07}$& $0.35^{+0.10}_{-0.14}$& $-2.58^{+0.06}_{-0.06}$\\ 
SB 126& $6.44^{+0.05}_{-0.11}$& $-1.42^{+0.05}_{-0.04}$& $7.9^{+0.0}_{-0.2}$& $<6.0$& $8.28^{+0.03}_{-0.03}$& $-0.52^{+0.03}_{-0.03}$& $0.15^{+0.06}_{-0.04}$& $-2.65^{+0.05}_{-0.06}$\\ 
SB 153& $4.58^{+0.04}_{-0.03}$& $-2.42^{+0.04}_{-0.03}$& $6.8^{+0.0}_{-0.0}$& $<6.0$& $8.36^{+0.03}_{-0.03}$& $-0.45^{+0.05}_{-0.04}$& $0.16^{+0.09}_{-0.04}$& $-2.51^{+0.07}_{-0.07}$\\ 
SB 179& $5.70^{+0.06}_{-0.06}$& $-1.62^{+0.05}_{-0.03}$& $7.3^{+0.1}_{-0.1}$& $<6.0$& $8.44^{+0.03}_{-0.03}$& $-0.35^{+0.04}_{-0.04}$& $0.17^{+0.08}_{-0.05}$& $-2.75^{+0.06}_{-0.06}$\\ 
SB 191& $4.51^{+0.05}_{-0.04}$& $-2.49^{+0.05}_{-0.04}$& $6.5^{+0.0}_{-0.0}$& $<6.0$& $8.51^{+0.02}_{-0.02}$& $-0.32^{+0.02}_{-0.02}$& $0.12^{+0.02}_{-0.01}$& $-2.09^{+0.05}_{-0.06}$\\ 

\hline
\multicolumn{9}{c}{With UV stellar equivalent widths} \\
\hline

SB 9& $5.25^{+0.03}_{-0.03}$& $-1.87^{+0.02}_{-0.02}$& $7.1^{+0.0}_{-0.0}$& $<6.0$& $8.45^{+0.03}_{-0.03}$& $-0.22^{+0.02}_{-0.02}$& $0.46^{+0.03}_{-0.04}$& $-2.83^{+0.03}_{-0.03}$\\ 
SB 49& $6.17^{+0.04}_{-0.04}$& $-0.83^{+0.04}_{-0.04}$& $6.3^{+0.0}_{-0.0}$& $6.3^{+0.0}_{-0.0}$& $8.67^{+0.01}_{-0.00}$& $-0.00^{+0.00}_{-0.00}$& $0.49^{+0.01}_{-0.01}$& $-2.29^{+0.02}_{-0.01}$\\ 
SB 61& $7.33^{+0.08}_{-0.12}$& $0.08^{+0.04}_{-0.05}$& $7.3^{+0.1}_{-0.2}$& $<6.0$& $8.32^{+0.02}_{-0.04}$& $-0.46^{+0.02}_{-0.04}$& $0.22^{+0.03}_{-0.03}$& $-1.96^{+0.04}_{-0.04}$\\ 
SB 80& $6.09^{+0.04}_{-0.04}$& $-0.91^{+0.04}_{-0.04}$& $6.9^{+0.0}_{-0.0}$& $<6.0$& $8.33^{+0.03}_{-0.03}$& $-0.39^{+0.01}_{-0.01}$& $0.38^{+0.05}_{-0.06}$& $-2.59^{+0.04}_{-0.04}$\\ 
SB 119& $5.31^{+0.04}_{-0.04}$& $-1.69^{+0.04}_{-0.04}$& $6.8^{+0.0}_{-0.0}$& $6.1^{+0.0}_{-0.0}$& $8.37^{+0.04}_{-0.03}$& $-0.30^{+0.02}_{-0.01}$& $0.46^{+0.03}_{-0.06}$& $-2.56^{+0.04}_{-0.04}$\\ 
SB 125& $6.72^{+0.06}_{-0.04}$& $-0.33^{+0.03}_{-0.03}$& $7.0^{+0.1}_{-0.1}$& $<6.0$& $8.51^{+0.02}_{-0.01}$& $-0.13^{+0.02}_{-0.01}$& $0.49^{+0.01}_{-0.01}$& $-2.35^{+0.02}_{-0.02}$\\ 
SB 126& $6.17^{+0.04}_{-0.04}$& $-1.15^{+0.03}_{-0.03}$& $7.3^{+0.0}_{-0.0}$& $<6.0$& $8.48^{+0.01}_{-0.01}$& $-0.16^{+0.00}_{-0.00}$& $0.49^{+0.01}_{-0.01}$& $-2.40^{+0.02}_{-0.02}$\\ 
SB 153& $4.60^{+0.03}_{-0.02}$& $-2.40^{+0.03}_{-0.02}$& $6.7^{+0.0}_{-0.0}$& $<6.0$& $8.36^{+0.04}_{-0.04}$& $-0.37^{+0.01}_{-0.01}$& $0.35^{+0.07}_{-0.06}$& $-2.47^{+0.06}_{-0.06}$\\ 
SB 179& $5.41^{+0.03}_{-0.03}$& $-1.59^{+0.03}_{-0.03}$& $6.6^{+0.0}_{-0.0}$& $6.1^{+0.0}_{-0.0}$& $8.55^{+0.02}_{-0.01}$& $-0.09^{+0.01}_{-0.01}$& $0.49^{+0.01}_{-0.02}$& $-2.46^{+0.03}_{-0.03}$\\ 
SB 191& $4.54^{+0.03}_{-0.02}$& $-2.46^{+0.03}_{-0.02}$& $6.5^{+0.0}_{-0.0}$& $<6.0$& $8.55^{+0.00}_{-0.01}$& $-0.28^{+0.00}_{-0.00}$& $0.11^{+0.01}_{-0.01}$& $-2.01^{+0.03}_{-0.04}$\\

\hline

\end{tabular}

\end{table*}

\begin{figure}
    \includegraphics[width=0.5\textwidth]{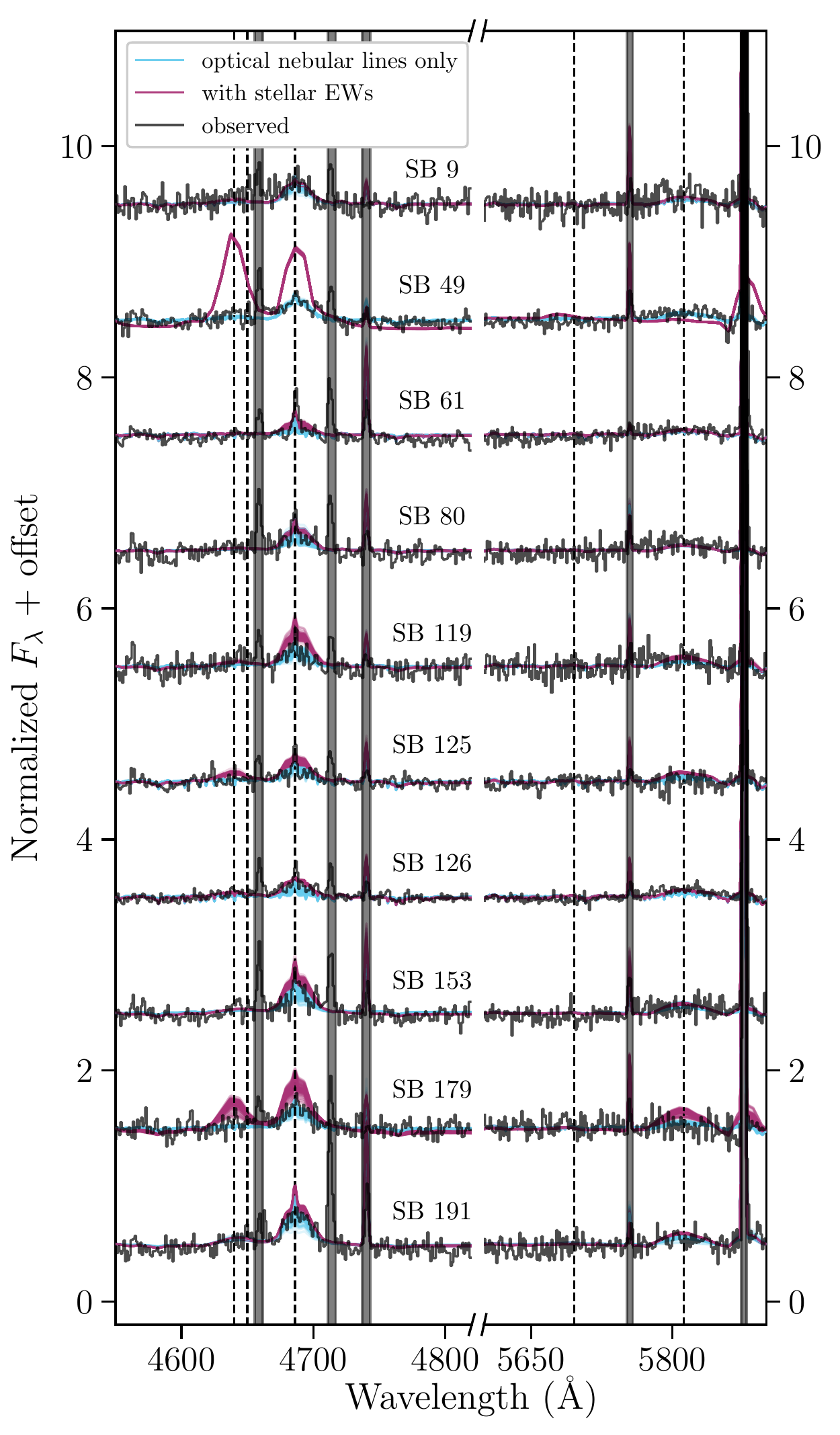}
    \caption{
        Comparison of our \beagle{} model results to the optical WR stellar wind features at $~4600$--$4700$ (the blue bump, left) and $\sim 5650$--$5800$ \AA{} (red bump, right).
        As in Figure~\ref{fig:beagle_uvcomp}, we plot both the data and models respectively after first normalizing by the continuum.
        The fits in blue incorporate only nebular line information, whereas those in red also include the equivalent width of the \civ{} and \heii{} stellar features measured in the UV.
        We highlight with dashed lines the approximate centers of the broad stellar wind features we aim to reproduce (\ion{N}{iii} 4640, \ion{C}{iii/iv} 4650, \ion{He}{ii} 4686 in the blue; and \ion{C}{iv} 5808 in the red).
        The fits with the nebular line information-only provide a generally reasonable fit to the observed optical stellar wind features, in which \ion{He}{ii} is by far the most prominent; while the fits with the UV stellar wind strengths incorporated in some cases dramatically over-predict the strength of several of the WR stellar winds lines.
    }
    \label{fig:beagle_optwrcomp}
\end{figure}

\subsection{Results}
\label{sec:synth_res}

In this subsection, we will analyze the results of the population synthesis fits described above and explore possible explanations for any discrepancies with the observations this uncovers.
We will focus first on the ability of the \cb{} models to reproduce the nebular lines and massive stellar wind signatures in \ref{sec:synth_res_stellar}.
In \ref{sec:metallicitycomp} we will compare the metallicities derived using the \cb{} photoionization models and those computed with the direct-$T_e$ method (Section~\ref{sec:optneb}).
Finally, we will discuss the peculiar profile of broad \heii{} emission detected in two of the strongest emitters in our sample in \ref{sec:peculiarheii}.

\subsubsection{Matching the strong UV stellar wind lines}
\label{sec:synth_res_stellar}

In this section we will summarize the joint analysis of the nebular emission lines and stellar wind signatures fit as-described in the previous section to determine whether the \cb{} models can self-consistently reproduce both.
Matching both in local systems is a benchmark that stellar population synthesis prescriptions must meet if the constraints extracted from their application to nebular emission lines alone at high redshift are to be interpreted physically.
While the \cb{} models we focus on do approximately reproduce the range of stellar wind strengths observed in our systems (Section~\ref{sec:ultraviolet}), they do not include the impact of binary mass transfer or rotation.
If these or other neglected physics play a significant role in shaping the evolution of the most massive stars, this may manifest in discrepancies with the spectroscopic properties of very young star-forming regions such as those presented in this paper.
We defer a detailed case-by-case discussion of the objects to Appendix~\ref{sec:casebycase}, and summarize the results for the sample as a whole here.

The strong nebular emission lines powered by the massive stars in these systems provide leverage on both the gas-phase metallicity and ionization state as well on the stellar ionizing spectrum.
In the first part of our experiment, we fit only the fluxes and equivalent widths of the optical nebular lines, and highlight the resulting model flux and equivalent width distributions relative to those observed in Figure~\ref{fig:beagle_fitlinecomp} (in blue).
In every case, the models are able to match this gas emission well within the measured uncertainties.
All of the fits prefer a continuous star formation history extending to the present (with the flexible star formation interval end uniformly constrained within $t_0<1$ Myr), but with very young maximum stellar ages ranging from 3--80 Myr (median 18 Myr).
This is broadly congruent with the results of our broadband SED fits including H$\beta$ equivalent widths (Table~\ref{tab:sedres}), though with a preference for somewhat older maximum ages and consequently slightly higher total stellar masses.
The gas-phase metallicities inferred range from $12+\log\mathrm{O/H}=8.25$--$8.53$, systematically higher than those measured with the same lines using the direct-$T_e$ method by $\sim 0.2$ dex on average (Table~\ref{tab:optneb}).
We discuss this offset in more detail in Section~\ref{sec:metallicitycomp}, but conclude in summary that this is an expected consequence of comparing to estimates from photoionization models.
The stellar metallicity $Z$ is tied to this gas-phase metallicity in our model, with the caveat that increasing depletion onto dust grains with increasing dust-to-metal mass ratios ($\xi_d$) can allow for higher stellar $Z$ at fixed gas-phase oxygen abundance (Section~\ref{sec:synth_meth}).
But the models find good agreement with the nebular lines at $\xi_d$ values uniformly below solar \citep[$\xi_{d,\odot}=0.36$;][]{gutkinModellingNebularEmission2016}, as expected for metal-poor star-forming galaxies; yielding stellar metallicities $Z/Z_\odot = 0.3$--$0.5$.
In all cases, the gas ionization parameter indicates highly-ionized gas, with $\log U$ ranging from $-3.0$ up to $-2.1$.
In summary, the fits to the nebular lines alone paint a well-constrained picture of systems dominated by continuous star formation initiated recently with metallicities of 30--50\% solar.

As expected, our fits reveal that the properties of the stellar populations dominating these systems are specified fairly precisely by the strength and equivalent widths of the strong nebular emission lines that their ionizing radiation powers, in the context of our stellar models.
But the strong stellar wind lines accessed by our spectra provide an independent view of these massive stars against which these models can also be compared.
In Figure~\ref{fig:beagle_uvcomp} we highlight the crucial \civ{} and \heii{} complexes in the \hstcos{} data, and compare them to continuum-normalized model spectra sampled from the posterior of our fits to the nebular lines only (in blue).
If the models have derived the correct mix of massive stars and treat their evolution and atmospheres correctly, we should expect to find good agreement with these wind lines.

Instead, the UV stellar wind lines reveal significant tension with the \cb{} model fits.
In nearly every case, the strength of these features is underestimated by the models constrained by the nebular line properties; and there are no cases in which the model wind features are stronger than observed.
The observations generally reveal more prominent broad \heii{} emission as well as deeper absorption and stronger emission in the \civ{} P-Cygni profile than predicted.
Both stellar wind diagnostics are systematically more prominent relative to the continuum than the \cb{} models predict given the nebular emission properties.

Before interpreting this result, we must determine whether there is any part of the model parameter space which would reconcile these differences with the observed stellar wind features.
As described above, we repeat the same nebular line fits with the equivalent widths of the \civ{} absorption trough and \heii{} emission line included alongside the optical line fluxes and equivalent widths.
The resulting fits are displayed in red in Figure~\ref{fig:beagle_uvcomp}, and as-expected, generally bring the UV stellar wind lines closer to agreement with the data.
However, the only way that the models are able to approach the observed UV stellar wind equivalent widths is through changes in metallicity and age which bring the data into new tension with the optical data, as described in detail in Appendix~\ref{sec:casebycase} and summarized here.

First, in order to match the strong stellar wind profiles observed, the models in most cases attempt to move to higher stellar metallicities.
Since the strength of line-driven stellar winds increases significantly with higher iron-peak element abundances, this allows the models to reach higher equivalent widths in both \civ{} and \heii{}.
This increase in the stellar iron abundance is accomplished for most systems by a combination of both an increase in the total metallicity $Z$ (by up to $0.3$ dex in the most extreme cases), as well as an increase in dust-to-metal mass ratio $\xi_d$ (Table~\ref{tab:beagle_nebfit}).
As described above, since the gas-phase and stellar metallicity are tied, such an increase in depletion onto dust grains will lead to an increase in the stellar iron abundance if the gas-phase O/H is held constant.
However, the preferred $\xi_d$ is increased to above solar value in a total of seven of the ten systems, six of which reach the upper end of our prior distribution and the available models at $\xi_d=0.5$.
Such super-solar dust content is highly unlikely in these moderately low-metallicity star-forming regions.
In addition, despite this shift in $\xi_d$, the increase in stellar Fe/H still induces a significant shift upwards in gas-phase O/H as well.
In six of the ten cases, this shift upwards in gas-phase metallicity is large enough to bring the model confidence interval for the temperature-sensitive [\ion{O}{iii}] $\lambda 4363$ line flux out of agreement with the data due to enhanced cooling (Figure~\ref{fig:beagle_fitlinecomp}).
If we were to adopt a more flexible model with variable $\alpha/\mathrm{Fe}$ which would effectively allow for decoupled stellar Fe/H and gas-phase O/H \citep[e.g.][]{steidelReconcilingStellarNebular2016}, some of this tension could be resolved.
However, achieving these results would require much higher iron content in the stars than suggested by the gas-phase O/H, implying a highly unlikely sub-solar $\alpha$/Fe abundance (opposite that inferred for star-forming systems and theoretically expected from yields dominated by Type II supernovae).

In addition, the prominent UV stellar wind features force the models to different effective stellar age distributions.
The \civ{} and \heii{} profiles are dominated by very massive Of stars and short-lived wind-stripped massive WR stars in the \cb{} models.
As a result, the models for galaxies with initially underestimated UV wind features almost uniformly prefer a younger effective stellar ages, with the inferred start of star formation shifting down to a median of 7 Myr across all systems.
In addition, the best fits for 3 systems with particularly discrepant UV wind features (SB~49, 119, 179) shifted from star formation proceeding continuously until $t_0<1$ Myr ago to preferring an early truncation 1--2 Myr ago; and in the case of SB~49, the preferred star formation history collapses to an essentially instantaneous burst occurring 2 Myr ago.
However, these star formation history shifts result in very different populations of WR stars, and consequently different predictions for the optical WR wind features.
For all three of these systems with the largest shift in star formation history, individual wind lines in the blue or red WR bumps are substantially over-estimated relative to those observed (Figure~\ref{fig:beagle_optwrcomp}).
This suggests that the improvements to the UV stellar wind lines for these systems are only accomplished in the models with populations of Of/WR stars that are either present in the wrong numbers or at too high a metallicity, leading to significant disagreement in the optical.

Several other explanations also fall short of providing an adequate explanation for this difficulty in simultaneously matching the nebular lines and stellar wind features.
It is unlikely these discrepancies are due to issues with our treatment of dust extinction or inter-aperture flux calibration, as we have focused only on the equivalent width of the stellar features in the UV.
Our model included a flexible star-formation history allowed to vary from essentially instantaneous bursts to a continuous model; and adding an additional independent burst component allowed to vary in age on top of this model did not change the results qualitatively.
While stochastic IMF sampling in very low-mass $\lesssim 10^5 \unit{M_\odot}$ clusters could lead to peculiar populations of massive stars and spectroscopic properties disjoint from those of a continuous star-formation model \citep[e.g.][]{vidal-garciaModellingUltravioletlineDiagnostics2017,paalvastMetallicityCalibrationsGalaxies2017}, the most discrepant objects are not biased towards particularly small numbers of massive stars.
These systems extend to relatively large inferred masses $\sim 10^6 M_\odot$ (in particular, SB~49 and 179 at $10^{6.3}$ and $10^{5.7} M_\odot$, respectively) where stochastic sampling should be relatively unimportant.
In particular, using a dust correction derived from the optical with an SMC extinction curve (Section~\ref{sec:optneb}) and assuming a single WNL star has a luminosity in \ion{He}{ii} $\lambda 1640$ of $1.2\e{37}$ erg/s \citep{schaererNewModelsWolfRayet1998,chandarNGC31251Most2004}, the integrated luminosity of \ion{He}{ii} in these systems suggests that the number of WR stars in the discrepant systems spans the entire range displayed by our sample; from as few as $\mathrm{N(WNL)}\simeq$13 in the lowest-mass system SB~191 to as many as 300 and 1650 in SB~179 and 49, respectively.
Thus, stochasticity is also unlikely to explain these offsets from the model predictions.

Rather than a product of analysis uncertainties, this systematic mismatch may be the signature of stellar physics that the \cb{} models do not presently include.
The excess observed power in the UV stellar wind lines that the models cannot account for would be naturally produced by the massive, rejuvenated, and rapidly-rotating products of binary mass transfer and mergers.
Including such interactions especially at the high mass end would produce additional very massive stars with Of or WN spectral signatures powered in their dense winds, as demonstrated by BPASS in the steady mass-transfer case (with a different set of underlying single-star evolutionary models with far weaker baseline stellar wind emission in \heii{}).
We discuss this possibility and its implications for galaxy modeling in more detail in Section~\ref{sec:stellardiscuss}.

\subsubsection{Comparing gas-phase metallicity estimates}
\label{sec:metallicitycomp}

We note that fits to the optical nebular line information only consistently produce higher gas-phase metallicity estimates than found with the direct-$T_e$ method above (Section~\ref{sec:optneb}).
This offset ranges from 0.09--0.33 dex, with a median offset of 0.18 for our ten systems.
This is a significant difference typically well in-excess of the joint uncertainties on both metallicity determinations.
However, comparable or larger offsets are common in the literature \citep[see also][]{chevallardPhysicalPropertiesHionizingphoton2018}.
The tendency of metallicity estimates based upon photoionization model grids to provide systematically larger gas-phase oxygen abundance estimates than the canonical two-zone direct-$T_e$ method has been previously attributed to complications in interpreting [\ion{O}{iii}] $\lambda 4363$ with the oversimplified ionization structure assumed by the latter \citep[e.g.][]{blancIZIInferringGas2015,valeasariBONDBayesianOxygen2016}.
Because the optical forbidden line ratios are sensitive to temperature and density fluctuations and to the velocity distribution of free electrons, the implicit assumption in the direct-$T_e$ approach that [\ion{O}{iii}] $4363/5007$ provides an accurate bulk measure of the average $T_e$ in a perfect two-zone \Hii{} region structure is likely not always valid \citep[e.g.][]{tsamisHeavyElementsGalactic2003,nichollsResolvingElectronTemperature2012}.
Indeed, this offset may be also related to the longstanding offset between metallicities determined with recombination lines and collisionally-excited lines in nearby \Hii{} regions and planetary nebulae \citep[the abundance discrepancy factor problem; e.g.][]{peimbertAbundanceRatioGaseous1993,garcia-rojasAbundanceDiscrepancyProblem2007,gomez-llanosBiabundancePhotoionizationModels2020}.

Regardless, since the lines including [\ion{O}{iii}] $\lambda 4363$ are well-fitted by \beagle{} when considered alone, we interpret this offset found when fitting the nebular lines only as the presently unmitigable systematic uncertainty in gas-phase oxygen abundance determinations using only collisionally-excited nebular lines.
In contrast, as discussed above, the far larger metallicity offsets found when the nebular lines and stellar wind features are fitted together are not well explained by these issues, and the inability to match the observed [\ion{O}{iii}] $\lambda 4363$ line fluxes is indicative of real tension with the data.

\subsubsection{Peculiar \ion{He}{ii} profiles in SB~179 and 191}
\label{sec:peculiarheii}

The objects which present the greatest disagreement with the stellar model predictions in the UV stellar wind lines are SB~179 and 191 (Section~\ref{sec:synth_res_stellar} and Appendix~\ref{sec:casebycase}).
In particular, both power significantly more prominent \ion{He}{ii} $\lambda 1640$ emission than the models can reproduce alongside the optical nebular lines.
Closer inspection of these very high-S/N stellar wind profiles in the UV could provide additional insight into the origin of the systematic disagreement of this line with the stellar models.

\begin{figure}
    \includegraphics[width=0.5\textwidth]{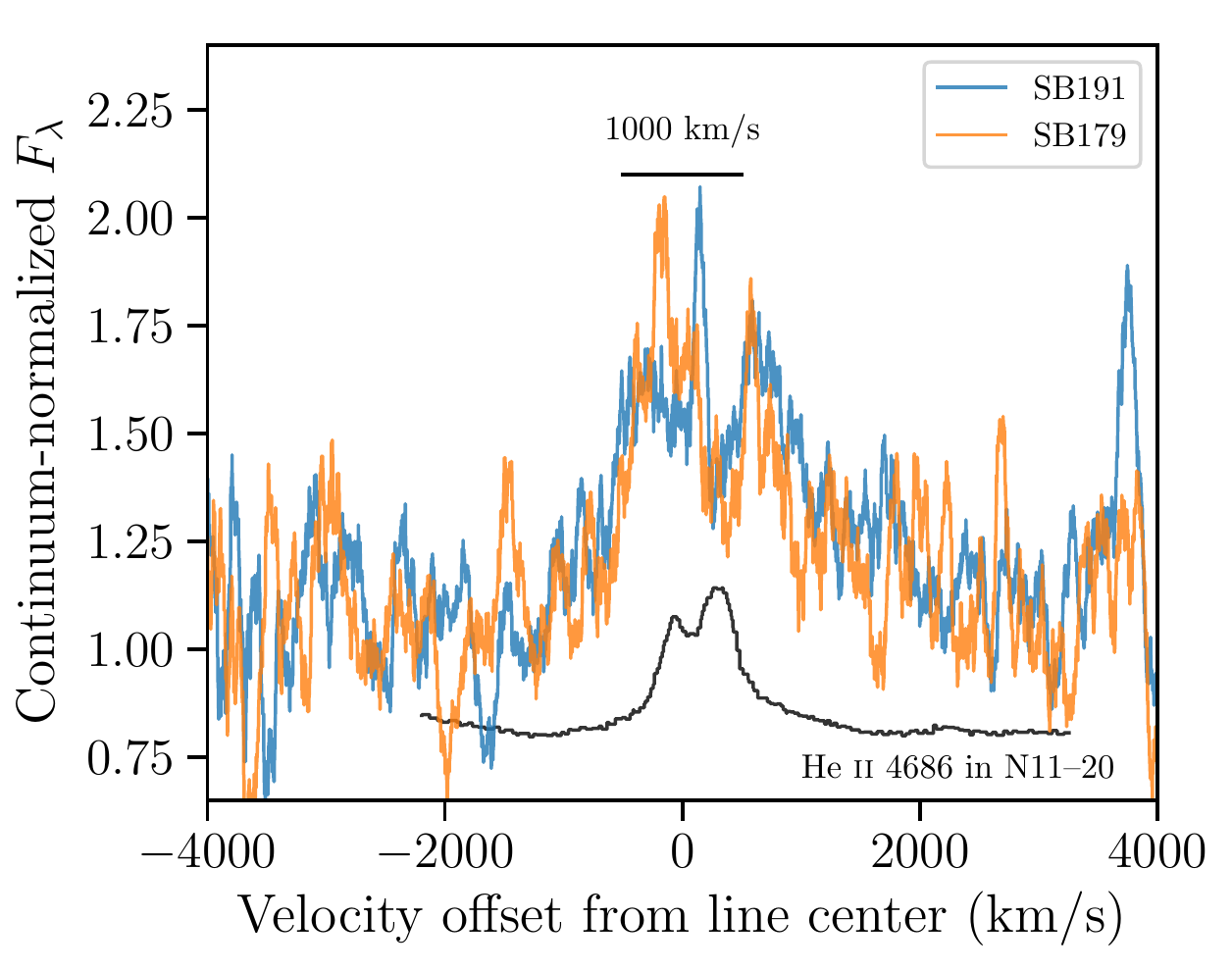}
    \caption{
        The peculiar, apparently double-peaked \heii{} $\lambda 1640$ profiles for SB~179 and 191 (note SB~191 also shows a narrow nebular component in addition).
        For comparison, we plot the \heii{} $\lambda 4686$ profile of the star N11--20, an rapidly-rotating ($v\sin i=260 \unit{km/s}$) Onfp giant in the LMC with the centrally-absorbed profile characteristic of this spectral class \citep[adapted from][]{walbornOnfpClassMagellanic2010}.
    }
    \label{fig:peculiar_heii}
\end{figure}

Intriguingly, both of these systems display a clear double-peak in the broad emission at \heii{} $\lambda 1640$, with peak separation of-order 1000 km/s (Figure~\ref{fig:peculiar_heii}).
No intervening MW absorption is expected at this wavelength for these systems (which reside at different redshifts, $z=0.0047$ and $0.0028$).
If interpreted first as two kinematic emission components, this profile is inconsistent with a stellar origin.
The two peaks are individually far more narrow than the other 1640 profiles, and 1000 km/s is well beyond the rotational velocity a massive star can maintain.
Roughly symmetric double-peaked \ion{He}{ii} profiles are observed in some accretion disks surrounding neutron stars in low-mass X-ray binaries \citep[e.g.][]{soriaOpticalSpectroscopyGRO2000,valbakerMassNeutronStar2005} and in other nebular lines surrounding high-mass X-ray binaries \citep[e.g.][]{filippenkoPossibleEvidenceDisk1988}.
However, these systems typically display lower peak velocity separation ($\lesssim 500$ km/s) than observed here.
In addition, SB~191 has archival \chandra{} observations which constrain its X-ray emission to $L_{\mathrm{X,0.5--8 keV}}<1.3\e{38}$ erg/s, making the presence of a massive accretion disk in this system unlikely.
Aspherical supernovae explosions are also observed to produce double-peaked emission line profiles with even larger velocity separations at late times in cases where the explosion is viewed close to down the jet axis \citep[e.g.][]{maedaAsphericitySupernovaExplosions2008}, but we see no other strong evidence for a recent supernovae in the nebular spectra.

Rather than a double-peak in emission, this profile may instead be characterized by central absorption.
In fact, such a \ion{He}{ii} profile is observed in a rare subset of hot O star spectra, defining the spectral designation Onfp with \ion{He}{ii} $\lambda 4686$ in emission but with a peculiar central reversal \citep[``f'' and ``p''; see Figure~\ref{fig:peculiar_heii}:][]{walbornSpaceDistributionStars1973,contiSpectroscopicObservationsOtype1974,walbornOnfpClassMagellanic2010}.
As indicated by the ``n'', this profile is commonly found alongside significant broadening in the absorption and emission lines indicative of rapid rotation.
Detailed atmosphere modeling has demonstrated that such a profile is naturally produced for rotating stars driving an isotropic wind \citep{hillierInfluenceRotationOptical2012}, supporting the notion that such stars are rapid rotators driving dense stellar winds.
Such strong winds should efficiently brake rotation, but many of the stars of this class in both the Milky Way and Magellanic Clouds show evidence of binarity or runaway velocities \citep{walbornOnfpClassMagellanic2010}.
They may then be the product of binary mass transfer or stellar mergers, rejuvenated in luminosity and spun-up to very high velocity by this interaction.

While a promising potential clue as to the nature of the dominant massive stars in these systems, several aspects of the observed integrated \ion{He}{ii} profiles remain puzzling in this context.
The \ion{He}{ii} profiles for individual Onfp stars typically show slightly blueshifted absorption, and thus an asymmetric emission profile with the red component brighter.
However, this effect may be washed-out when integrated over a population of stars at a spread in velocities powering various \ion{He}{ii} profiles.
In addition, the optical \ion{He}{ii} 4686 profiles for SB~179 and 191 do not show the same morphology, even at Keck/ESI resolution \citepalias{senchynaUltravioletSpectraExtreme2017}.
Individual Onfp stars show significant time variability in their wind profiles on timescales of days however, sometimes transitioning out of this peculiar profile entirely \citep{walbornOnfpClassMagellanic2010}.
While the circular apertures of these instruments are similar in size (2.5\arcsec{} and 3.0\arcsec{}, respectively) and the objects appear compact on this scale in the \hstcos{} target acquisition imaging (Figure~\ref{fig:montage_tacq}), the \hstcos{} primary science aperture is subject to vignetting beyond $0.4$\arcsec{} from the aperture center \citep{dashtamirovaCosmicOriginsSpectrograph2020}.
This may be the part of the cause of the puzzling lack of a nebular \heii{} $\lambda 1640$ detection in the \hstcos{} spectrum of SB~125 and 126 when this emission is present at \heii{} $\lambda 4686$ in the SDSS spectrum, though dust extinction likely plays a significant role as well.
In the case of SB~179 and 191, it is possible that the COS spectrum is dominated by a subcluster within the SDSS aperture.
And if Onfp stars are indeed likely runaways expelled from their natal clusters by binary interaction, they may see a substantially smaller column of dust than other massive stars and thus are more likely to dominate the \ion{He}{ii} profile in the ultraviolet.
High spatial and spectral resolution optical IFU observations probing the \ion{He}{ii} $\lambda 4686$ line across these systems are necessary to confidently confirm and narrow the origins of this peculiar emission profile.
However, this detection may lend unexpected support to the notion that the excess emission in \heii{} is indeed due at least in part to rapidly-rotating products of mass transfer, as discussed further in Section~\ref{sec:stellardiscuss}.

\section{Discussion}
\label{sec:discuss}

We have presented new ultraviolet spectra for a sample of nearby star-forming regions hosting very young stellar populations at moderately sub-solar ($\sim 20$--50\% $Z_\odot$) metallicity.
Such integrated spectra provide a key opportunity to study the properties of the massive star populations that dominate these clusters and similarly high specific star formation rate systems at cosmological redshifts.
Joint fits to both the strong stellar wind emission signatures and the nebular emission the massive stars power reveal significant tension with the data, suggestive of potentially missing physics in the stellar population synthesis models at these very young ages.
We also uniformly detect nebular \ion{C}{iii}] emission in this sample, enabling an investigation of the strength of this doublet for galaxies dominated by very recent star formation.
We first discuss the implications for models of very young stellar populations and massive stars in Section~\ref{sec:stellardiscuss}, before turning our attention to interpreting strong \ion{C}{iii}] emission at high-redshift in Section~\ref{sec:ciiiemission}.

\subsection{Evidence for an overabundance of very massive stars}
\label{sec:stellardiscuss}

Nebular spectroscopy at cosmological distances, soon to be extended to large samples at $z>6$ with \jwst{}, represents an opportunity to study early star-forming systems in much greater physical detail than photometry alone can afford.
However, the accuracy of physical inferences drawn from fitting this emission will be fundamentally limited by the uncertainties in stellar population synthesis models, which still vary substantially with different prescriptions for stellar evolution especially at ionizing energies \citep[e.g.][]{stanwayInitialMassFunction2019}.
Evidence from both photometric bands contaminated by rest-optical line emission and the first rest-UV spectra in the reionization era suggests that a significant fraction of $z>6$ galaxies are dominated by populations of massive stars formed in bursts initiated within $<20$ Myr of observation \citep[e.g.][]{starkSpectroscopicDetectionIV2015,endsleyOIIIBetaEquivalent2020}.
Thus, our understanding of the most distant galaxies JWST will observe will depend sensitively on models for the evolution of populations of massive stars on much shorter timescales than typical star-forming galaxies even at $z\sim 2$ \citep[e.g.][]{steidelReconcilingStellarNebular2016}.

Study of stars and clusters in the Local Group has provided important clues as to the physics shaping the evolution of very young stellar populations.
In particular, the 30 Doradus star-forming region in the LMC and its central young cluster R136 ($\sim 1$--3 Myr, $Z/Z_\odot \sim 0.5$) has long represented a benchmark analogue for intense starbursts observed at high redshift.
Intriguingly, the resolved stellar contents of this regions reveal evidence for a significant overabundance of massive stars $>30 M_\odot$ relative to a Salpeter IMF \citep{schneiderExcessMassiveStars2018}, including a sizable population of very massive stars with initial masses 100--320 $M_\odot$ in R136 \citep{crowtherR136StarCluster2016}.
Simulations suggest that such an overpopulation can be readily produced by mass transfer and mergers in binary systems, yielding numerous very massive stars even exceeding the upper limit of the initial mass function on timescales of Myrs \citep[e.g.][]{schneiderAgesYoungStar2014,deminkIncidenceStellarMergers2014}.
These very massive Of stars drive dense winds yielding prominent \heii{} and \civ{} \citep{crowtherR136StarCluster2016} and, if they acquire sufficient angular momentum during mass transfer, may undergo quasi-chemically homogeneous evolution and effectively become an extremely massive Wolf-Rayet star.
The stars produced by this mechanism are very good candidates for long gamma-ray burst progenitors \citep[e.g.][]{cantielloBinaryStarProgenitors2007}, and may have an outsized impact on the properties of recently-formed star clusters.
However, without additional resolved star-forming region approaching the age and luminosity of 30 Doradus \citep[e.g.][]{kennicuttStructuralPropertiesGiant1984}, the prevalence of these extreme mass transfer products and their impact on the integrated spectra of young star-forming regions as a function of metallicity remains unclear.

While unresolved, young star-forming regions such as those presented in this work provide the next best laboratories in which the evolution of populations of massive stars can be constrained.
The systems presented in this paper are by-selection dominated by very young stellar populations, with [\ion{O}{iii}]+H$\beta$ equivalent widths and photometric sSFRs at or exceeding the highest values inferred at $z>6$ and suggesting these systems provide a clear window onto very recently-formed $\lesssim 10$ Myr stellar populations (Tables~\ref{tab:sedres} and \ref{tab:optneb}).
And also by selection, all reside at sufficiently high metallicities ($Z/Z_\odot\sim$ 20--50\%) for radiatively-driven stellar wind signatures to be detectable in the integrated spectra alongside strong ionized gas emission.
The \ion{He}{ii} $\lambda 1640$ stellar wind line is a particularly useful probe of both initially very massive and luminous massive stars rejuvenated by mass transfer \citep[see also][]{stanwayEvaluatingImpactBinary2020}.
Reproducing this stellar wind line along with the other available constraints is a crucial litmus test for stellar population synthesis, but matching it at high equivalent width in both local and $z\sim 2$ galaxies has long stymied stellar models (Section~\ref{sec:ultraviolet}).

In Section~\ref{sec:uvoptsynthesis}, we described an experiment designed to leverage both the nebular and stellar wind lines to search for evidence of stars missing from current stellar population synthesis models.
We focus on the \cb{} stellar population synthesis models, which adopt the latest prescriptions for single-star evolution but do not include binary mass transfer or rotation.
While BPASS incorporates a model for the impact of mass transfer on very massive stars, the predicted impact on \heii{} on $<10$ Myr timescales in this framework is minimal, and insufficient to reach the equivalent widths observed in our local young star-forming regions \citep[Section~\ref{sec:ultraviolet} and][]{eldridgeEffectStellarEvolution2012}.
Significant recent improvements in wind-driven mass loss predictions have allowed the \cb{} models to resolve much of the tension with the strength of \heii{} $\lambda 1640$ relative to older stellar population models (Section~\ref{sec:ultraviolet}), so residual issues in matching this line may be a relatively clean probe of additional physics these models are missing at very young ages.

Our comparisons with the \cb{} models revealed significant tension with the observed star-forming regions.
When the nebular lines alone were fit, we found that the models systematically underestimated the strength of both the UV stellar \civ{} P-Cygni and \heii{} emission lines (Section~\ref{sec:synth_res_stellar}).
Explicitly including the UV stellar equivalent widths in the fits failed to identify a consistent model able to match both the nebular line measurements and the UV--optical stellar wind strengths, suggesting that the models would require unreasonably-high metallicity stars to reach the observed wind strengths.
We discussed other possible explanations for this offset, but find no satisfactory alternative explanation, suggesting that even with their improved treatment of massive stellar winds, the \cb{} models simply cannot self-consistently reproduce the nebular emission and observed wind line equivalent widths.

The mismatch between the \cb{} predictions and our data in the UV stellar wind lines is consistent with a significant overabundance of very massive stars relative to the models.
In R136, stars in excess of $100 M_\odot$ completely dominate \heii{} and contribute strongly to \civ{} due to the particularly strong winds these luminous stars drive \citep{crowtherR136StarCluster2016}.
A top-heavy initial mass function could explain this, though the existing evidence for significant IMF variations at subsolar metallicities is not clear-cut \citep[e.g.][]{gunawardhanaGalaxyMassAssembly2011,marksEvidenceTopheavyStellar2012,hopkinsDawesReviewMeasuring2018}; but as outlined above, a high incidence of mass transfer and mergers among initially very massive stars would also naturally yield such a distribution.
If binary evolution has substantially affected the massive stars present, we would expect many to be rapidly rotating due to angular momentum transfer.
The detection of peculiar double-peaked profiles which may be produced by rapidly-rotating stars in two of the most discrepant systems presented lends additional support to the notion that mass transfer occurring on very short timescales may be at work (Section~\ref{sec:peculiarheii}).

Our results suggest that the stellar populations in very young systems with sSFR comparable to bursts observed at $z\sim 6$ may host populations of very massive stars significantly enhanced by mass transfer and mergers.
Much of the work on the observational consequences of binary evolution for the SEDs of star-forming galaxies has focused on the impact of these processes on $\sim 10$--100 Myr timescales when numerous initially lower-mass stars ($\lesssim 20 M_\odot$) become significantly hotter than canonically expected \citep[e.g.][]{eldridgeEffectStellarEvolution2012,maBinaryStarsCan2016,xiaoEmissionlineDiagnosticsNearby2018}.
However, our difficulty in matching the ultraviolet spectra of these young star-forming regions suggests that binary evolution may profoundly alter the properties and relative number of very massive stars on significantly shorter timescales as well.
The fact that this results in detectable differences in the integrated spectra of young star-forming regions has dramatic consequences for attempts to model systems dominated by similar populations at high-redshift.

These data indicate that neither \cb{} nor BPASS can fully describe the properties of young stellar populations dominated by very massive stars.
This suggests that the mapping from metallicity and age to stellar population composition and the emergent ionizing spectrum may be significantly in-error at ages $<10$ Myr.
At yet lower metallicities ($<20\%\; Z_\odot$) where stellar winds weaken and become increasingly transparent to hard ionizing radiation, the same stellar products of binary evolution will likely have an even more dramatic impact on the integrated ionizing spectrum and thus observed nebular emission.
It is crucial that these models be carefully tested at these young ages to mitigate the impact of these systematic uncertainties on our ability to understand the spectra of high-sSFR systems in the early Universe.

The nebular emission and rich continuum signatures accessible simultaneously in extreme local objects such as these likely represent one of our best opportunities to directly constrain models for the evolution of populations of very massive stars.
Combined with observations of additional UV wind lines which can provide more detailed leverage on the detailed makeup and age of the massive star population \citep[e.g.][]{smithVeryMassiveStar2016,chisholmConstrainingMetallicitiesAges2019}, self-consistently reproducing UV--optical spectra for young star-forming regions such as those presented here will be an important benchmark for next-generation stellar population synthesis models incorporating the substantial improvements made both to prescriptions for single and binary star evolution over the past decade.
Successfully matching the stellar and nebular features in these systems would be strong evidence that the mapping between stellar metallicity, age, and ionizing spectrum predicted by the models is accurate at very young ages, directly addressing one of the chief sources of systematic uncertainty in the modeling of high-redshift galaxies.

\subsection{Nebular \ion{C}{iii}] emission powered by moderately metal-poor massive stars}
\label{sec:ciiiemission}

The growing body of rest-frame UV line detections at $z\gtrsim 6$ has ignited substantial interest in locating and understanding nearby galaxies that power similar emission.
Particularly puzzling among the four \ion{C}{iii}] detections at these redshifts are those in EGS-zs8-1 and z7\_GND\_42912, two very bright $L^\star_{UV}\simeq 3$, 2 reionization-era galaxies with \ion{C}{iii}] measured at 22, 16 \AA{} \citep[respectively;][]{starkSpectroscopicDetectionsIII2015,hutchisonNearinfraredSpectroscopyGalaxies2019}.
This is an order of magnitude higher than typical galaxies at similar luminosities at lower redshift \citep[e.g.][]{shapleyRestFrameUltravioletSpectra2003,duIIIEmissionStarforming2017}.
The emerging picture from extensive work at both $z\sim 2-4$ \citep[e.g.][]{nakajimaVIMOSUltraDeep2018,mainaliRELICSSpectroscopyGravitationallylensed2019,duSearchingAnalogsPeak2020} and in the local $z\sim 0$ Universe \citep[e.g.][]{rigbyIIIEmissionStarforming2015,senchynaUltravioletSpectraExtreme2017,senchynaExtremelyMetalpoorGalaxies2019} suggests that such prominent \ion{C}{iii}] emission requires a combination of both very high specific star formation rates $\gtrsim 10-100 \mathrm{Gyr}^{-1}$ and very metal-poor ($Z/Z_\odot\lesssim 0.2$, $12+\log\mathrm{O/H}\lesssim8.0$) stars and gas.
If such low metallicities are indeed necessary, these prominent \ion{C}{iii}] detections at $z>6$ imply that these $\gtrsim L^\star$ systems reside at sub-SMC metallicities, implying a large shift in the luminosity-metallicity relationship from $z\sim 2$--6.

However, the most extreme of the local galaxies with UV spectroscopy hint at a possible alternate solution.
The detection of very strong $\sim 9$-- $11$ \AA{} \ion{C}{iii}] emission alongside prominent stellar \ion{He}{ii} in SB~179 and 191, very young local star-forming regions at $12+\log\mathrm{O/H}=8.3$ ($Z/Z_\odot \simeq 0.4$), suggested that very massive stars could potentially power stronger emission than expected at well above SMC metallicities  \citepalias{senchynaUltravioletSpectraExtreme2017}.
The spectra presented in this work for seven additional systems selected in the same manner as SB~179 and 191 allow us to test this possibility.
In addition to their WR wind signatures in the optical, our targets extend to very high sSFRs, and reach equivalent widths in [\ion{O}{iii}] $\lambda 5007$ up to $1719\pm 133$ \AA{} (in SB~153) higher than in any of the targets studied in \citetalias{senchynaUltravioletSpectraExtreme2017}.
According to the trend towards more prominent \ion{C}{iii}] emission at high sSFR or $W_0$([\ion{O}{iii}]) displayed by that sample and observed by others \citep[e.g.][]{mainaliRELICSSpectroscopyGravitationallylensed2019}, such young effective stellar populations should be capable of powering equivalent widths in \ion{C}{iii}] approaching that observed in the reionization era.

\begin{figure}
    \includegraphics[width=0.5\textwidth]{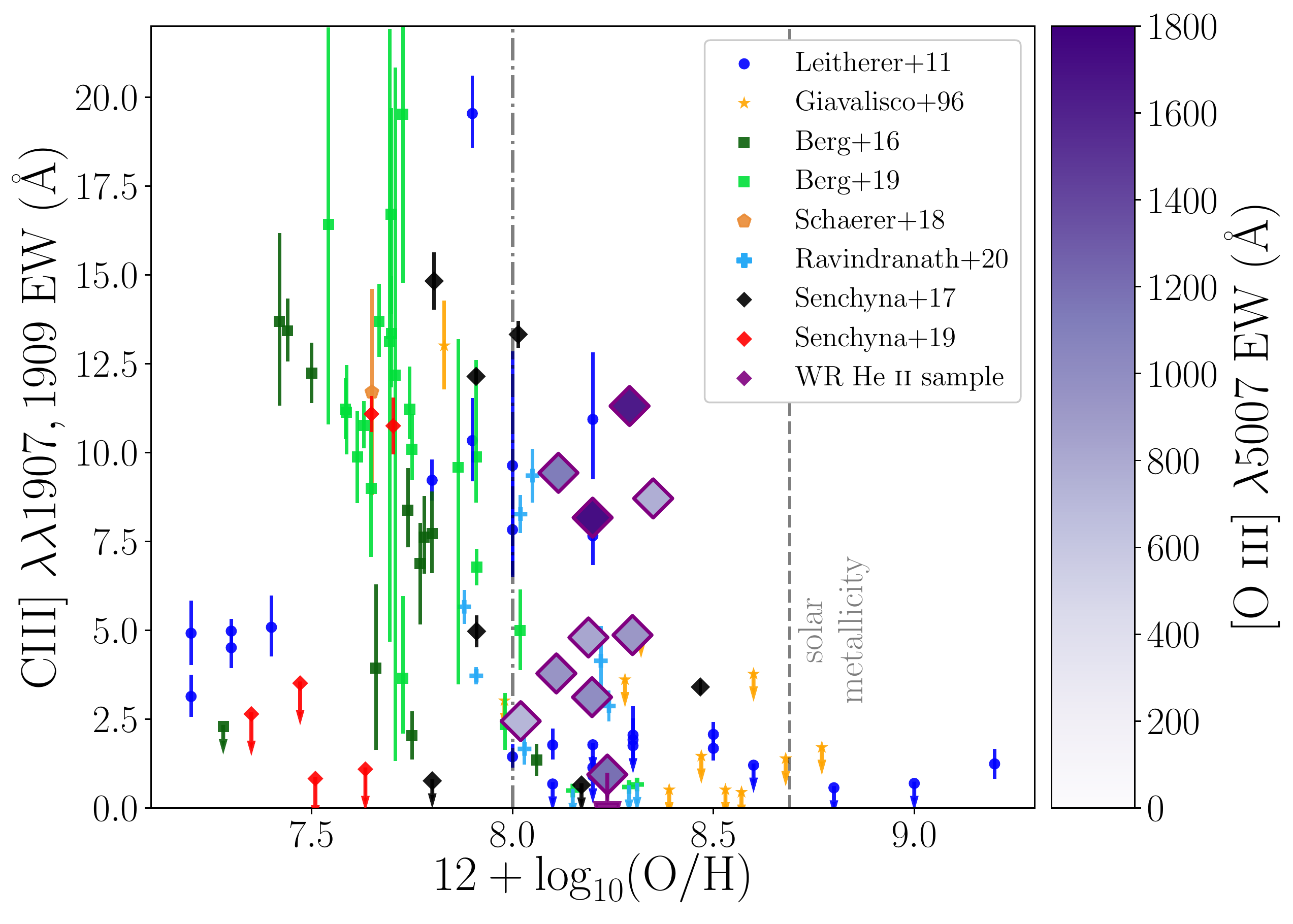}
    \caption{
        The equivalent width of nebular \ion{C}{iii}] emission as a function of metallicity for the ten extreme WR galaxies presented in this work.
        We also show measurements from UV spectra of $z\lesssim 0.3$ galaxies assembled by \citet[][]{leithererUltravioletSpectroscopicAtlas2011}, \citet{giavaliscoObscurationLYAlpha1996}, \citet{bergCarbonOxygenAbundances2016}, \citet{senchynaUltravioletSpectraExtreme2017}, \citet{schaererIntenseIIIEnsuremath2018}, \citet{bergChemicalEvolutionCarbon2019}, \citet{senchynaExtremelyMetalpoorGalaxies2019}, and \citet{ravindranathSemiforbiddenIIIL19092020}.
        Oxygen abundances corresponding to solar and 20\% solar (SMC metallicity) are indicated by vertical dashed lines.
        While the galaxies presented here extend to extremely high sSFR (indicated by their [\ion{O}{iii}] $\lambda 5007$ equivalent widths) and all but one are detected in \ion{C}{iii}], they all remain below 12 \AA{} in equivalent width in this doublet.
        This suggests that even galaxies dominated by very massive stars cannot power emission in \ion{C}{iii}] at the level observed at $z>6$ at the moderately-low metallicities of our sample ($Z/Z_\odot \sim 0.2$--$0.5$)
    }
    \label{fig:ciii_metal}
\end{figure}

In Figure~\ref{fig:ciii_metal} we plot the observed equivalent width of \ion{C}{iii}] emission in these systems and other UV observations of $z\lesssim 0.3$ star-forming galaxies.
Despite their extraordinarily high equivalent width optical emission lines and correspondingly young effective stellar ages, none of these objects power \ion{C}{iii}] emission above 12 \AA{}.
Even SB~153, with extremely prominent wind emission at $3.9\pm0.1$ \AA{} in \ion{He}{ii} $\lambda 1640$ and an effective young stellar population age of $\lesssim 5$ Myr (assuming a recent CSFH), only reaches $8.7\pm 0.4$ \AA{} in \ion{C}{iii}].
This suggests that we are observing the upper envelope of the expected strength of \ion{C}{iii}] at metallicities $12+\log\mathrm{O/H}>8.0$.

Besides metallicity and sSFR, several other factors play a role in regulating \ion{C}{iii}] emission.
The ISM abundance of carbon has a direct effect on the strength of \ion{C}{iii}], and can suppress this emission significantly at very low C/O.
However, our galaxies reside at moderate C/O abundances ($-0.7<\log_{10}(\mathrm{C/O})<-0.2$, with median $-0.31$; Table~\ref{tab:uvmeas}), consistent with other local systems at comparable oxygen abundances but significantly higher than the bulk of more metal-poor galaxies at $12+\log\mathrm{O/H}<8$ which extend down to $\log_{10}\mathrm{C/O}=-1.0$ \citep[e.g.][]{estebanCarbonOxygenAbundances2014,bergChemicalEvolutionCarbon2019}.
Finding star-forming galaxies at high redshift with substantially higher C/O than these systems is unlikely given this observed pseudo-secondary behavior and low-metallicity plateau in C/O in local systems.
We do expect to encounter $\alpha$/Fe enhancement in high-redshift galaxies dominated by a rising star formation history, which can lead to a lower-metallicity stellar population and thus harder ionizing radiation field than expected from the gas-phase oxygen abundance \citep[e.g.][]{steidelReconcilingStellarNebular2016,stromMeasuringPhysicalConditions2018,sandersMOSDEFSurveyDirectmethod2020}.
The presence of lower-metallicity stars may act to further enhance \ion{C}{iii}] even at moderate gas-phase oxygen abundance, though the magnitude of the effect will depend on the interplay between both gas ionization shaped by these stars and cooling efficiency determined largely by O/H.

Even for systems dominated by very young stellar populations with prominent optical [\ion{O}{iii}] emission, it appears unlikely that \ion{C}{iii}] significantly in-excess of 10 \AA{} can be powered for stars and gas at metallicities above $12+\log\mathrm{O/H}>8.0$ ($Z/Z_\odot \gtrsim 0.2$).
The number of systems at very high sSFR studied at metallicities $12+\log\mathrm{O/H}>8.4$ remains relatively small (Fig.~\ref{fig:ciii_metal}), but all are consistent with even weaker \ion{C}{iii}] $\lesssim 5$ \AA{}.
This is broadly consistent with the picture painted by photoionization models \citep[e.g.][]{jaskotPhotoionizationModelsSemiforbidden2016,nakajimaVIMOSUltraDeep2018,platConstraintsProductionEscape2019}, which indicate that both softer ionization radiation fields and lower electron temperatures encountered at higher metallicities conspire to suppress collisionally-excited emission from the \ion{C}{iii}] doublet.
While \ion{C}{iii} and \ion{O}{iii} have similar ionization potentials and our systems power extremely prominent [\ion{O}{iii}] emission, the \ion{C}{iii}] doublet is far more sensitive to electron temperature than the optical [\ion{O}{iii}] lines \citep[e.g.][]{jaskotPhotoionizationModelsSemiforbidden2016,tangRestframeUVSpectroscopy2020}.
As a result, \ion{C}{iii}] is suppressed more strongly by the drop in temperatures associated with the increased efficiency of metal-line cooling at these high gas-phase metal abundances, producing the relatively sharp evolution with metallicity we observe.
The possibility remains that some of the strongest emission $\gtrsim 20$ \AA{} encountered in this line at high-redshift may require additional photoionization and heating from AGN or fast radiative shocks \citep[][]{nakajimaVIMOSUltraDeep2018,lefevreVIMOSUltraDeepSurvey2019,platConstraintsProductionEscape2019}.
However, in galaxies with nebular line ratios consistent with star formation, detection of \ion{C}{iii}] emission at equivalent widths in-excess of 15 \AA{} can be considered strong positive evidence for metallicities below that of the SMC ($12+\log\mathrm{O/H}\lesssim 8.0$).
Along with \ion{C}{iv} and \ion{He}{ii} emission which become prominent at yet lower metallicities \citep[$12+\log\mathrm{O/H}<7.7$; e.g.][]{senchynaExtremelyMetalpoorGalaxies2019,bergIntenseIVHe2019}, these strong UV nebular lines may be a powerful probe of early chemical evolution in the JWST era when detection of [\ion{O}{iii}] $\lambda 4363$ is intractable due to faintness or when the optical lines are entirely inaccessible at the highest redshifts.

\section{Summary}
\label{sec:summary}

Nearby star-forming regions at very young ages represent a crucial testbed for models of massive stellar populations and a benchmark for understanding high-redshift galaxies.
In this paper, we presented new ultraviolet spectra probing the highly-ionized gas and stellar winds in 7 systems \citep[combined with an additional 3 systems originally presented in][]{senchynaUltravioletSpectraExtreme2017} which are all dominated by very young stellar populations identified via high equivalent width nebular emission alongside WR wind signatures in the optical (Section~\ref{sec:sampleselect}).
The optical and ultraviolet spectra both reveal clear signatures of massive stars at gas-phase metallicities corresponding to 25--40\% solar (Section~\ref{sec:observations}), and effective stellar population ages $\lesssim 10$ Myr comparable to that inferred for systems undergoing intense bursts of star formation at $z>6$.
Our spectra access both \ion{C}{iii}] and \ion{O}{iii}] nebular emission as well as stellar \heii{} $\lambda 1640$ emission and \civ{} P-Cygni features produced in dense stellar winds (Section~\ref{sec:ultraviolet}).
These high-quality spectra enable a unique test of stellar population synthesis models, leveraging constraints on both the ionizing spectrum and metallicity through nebular gas emission and on the strong winds driven by massive stars through the wind diagnostic lines.
Our main conclusions are summarized as follows:

\begin{itemize}
    \item We detect nebular \ion{C}{iii}] in all of the targeted systems, confirming its ubiquity at high sSFR in galaxies below 50\% solar metallicity.
    However, the equivalent widths in this doublet only reach a maximum of 12 \AA{} for the entire sample of local galaxies in this metallicity range ($8.1<12+\log\mathrm{O/H}<8.4$), likely due to efficient gas cooling at these metallicities.
    This suggests that the most prominent emission powered by stars in this line ($>15$ \AA{}) requires gas and/or stars at or below SMC metallicity ($Z/Z_\odot \lesssim 20\%$), suggesting that the massive reionization-era systems detected at these equivalent widths may harbor significantly subsolar gas and stars.
    \item We detect \heii{} $\lambda 1640$ stellar wind emission at equivalent widths ranging up to $4.7$ \AA{}, among the most prominent detections of this line known and well in-excess of previous generations of stellar population synthesis predictions.
    We find that the latest Charlot \& Bruzual (\cb{}, in-prep) models are able to largely reproduce this observed range, while the current predictions from BPASS are not.
    This suggests that improvements in the treatment of massive star evolution and spectra in the \cb{} models may play an important role in reproducing the spectra of the most extreme star-forming regions both locally and at $z>6$.
    \item However, we are unable to simultaneously match both the strong UV stellar wind signatures and the nebular line measurements with the \cb{} stellar population models.
    These models, which neglect binary mass transfer, systematically under-predict the strength of both \civ{} and \heii{} in the UV.
    Any \cb{} solution able to match these wind strengths resides at too large a metallicity, bringing the optical nebular lines and WR features out of alignment with the data.
    We conclude that the most plausible explanation for this discrepancy is an overabundance of the very massive stars that dominate these wind lines at young ages, which would be naturally expected from a high incidence of binary mass transfer and mergers.
    This suggests that binary evolution may substantially alter the properties and appearance of populations of massive stars on very short timescales $<10$ Myr to which observations of high-sSFR galaxies are sensitive.
\end{itemize}

As stellar population synthesis models become increasingly sophisticated and their applications to observables at high-redshift become increasingly varied, it is ever more important to anchor their predictions empirically where possible.
Our results suggest that current prescriptions for massive stellar populations fall short of reproducing the properties of the highest-sSFR systems nearby at $\sim 30\%$ solar metallicity.
Resolving this discrepancy will likely involve synthesizing the progress made by different modeling groups in treating the evolution and properties of binary mass transfer products and very massive stars.
The high-quality spectra attainable for nearby star-forming systems such as these represent one of our only opportunities to test this next generation of stellar population models as a function of age and metallicity, with important consequences both for our understanding of massive star evolution and for correctly modeling the very young galaxies in the early Universe that \jwst{} is poised to uncover.

\section*{Acknowledgments}

PS thanks Selma de Mink and Benjamin Johnson for insightful conversations during the drafting of this manuscript.
This research is based on observations made with the NASA/ESA Hubble Space Telescope and supported by a grant from the Space Telescope Science Institute, which is operated by the Association of Universities for Research in Astronomy, Inc., under NASA contract NAS 5-26555.
These observations are associated with GO programs 14168 and 15185.
Observations reported here were obtained at the MMT Observatory, a joint facility of the University of Arizona and the Smithsonian Institution.
We thank Jennifer Andrews for obtaining the MMT spectrum of SB~49 presented in this paper, and Erin Martin for her excellent work as telescope operator during the April 2019 observing run published here.

D.\ P.\ S.\ acknowledges support from the National Science Foundation through the grant AST-1410155.
JC, SC and AVG acknowledge support from the ERC via an Advanced Grant under grant agreement no.\ 321323-NEOGAL.

This research made use of \textsc{Astropy}, a community-developed core \textsc{python} package for Astronomy \citep{astropycollaborationAstropyCommunityPython2013}; Matplotlib \citep{hunterMatplotlib2DGraphics2007}; \textsc{Numpy} and \textsc{SciPy} \citep{jonesSciPyOpenSource2001}; the SIMBAD database, operated at CDS, Strasbourg, France; and NASA's Astrophysics Data System.

\section*{Data Availability}

The \hst{} and SDSS data underlying this article are available through their respective data repositories (see \url{https://archive.stsci.edu/} and \url{https://www.sdss.org/dr16/}).
Data will also be shared upon reasonable request to the corresponding author.

\bibliographystyle{mnras}
\bibliography{zoterolib}

\begin{thebibliography}{}
\makeatletter
\relax
\def\mn@urlcharsother{\let\do\@makeother \do\$\do\&\do\#\do\^\do\_\do\%\do\~}
\def\mn@doi{\begingroup\mn@urlcharsother \@ifnextchar [ {\mn@doi@}
  {\mn@doi@[]}}
\def\mn@doi@[#1]#2{\def\@tempa{#1}\ifx\@tempa\@empty \href
  {http://dx.doi.org/#2} {doi:#2}\else \href {http://dx.doi.org/#2} {#1}\fi
  \endgroup}
\def\mn@eprint#1#2{\mn@eprint@#1:#2::\@nil}
\def\mn@eprint@arXiv#1{\href {http://arxiv.org/abs/#1} {{\tt arXiv:#1}}}
\def\mn@eprint@dblp#1{\href {http://dblp.uni-trier.de/rec/bibtex/#1.xml}
  {dblp:#1}}
\def\mn@eprint@#1:#2:#3:#4\@nil{\def\@tempa {#1}\def\@tempb {#2}\def\@tempc
  {#3}\ifx \@tempc \@empty \let \@tempc \@tempb \let \@tempb \@tempa \fi \ifx
  \@tempb \@empty \def\@tempb {arXiv}\fi \@ifundefined
  {mn@eprint@\@tempb}{\@tempb:\@tempc}{\expandafter \expandafter \csname
  mn@eprint@\@tempb\endcsname \expandafter{\@tempc}}}

\bibitem[\protect\citeauthoryear{Allen, Wright  \& Goss}{Allen
  et~al.}{1976}]{allenDwarfEmissionGalaxy1976}
Allen D.~A.,  Wright A.~E.,   Goss W.~M.,  1976, \mn@doi [\mnras]
  {10.1093/mnras/177.1.91}, 177, 91

\bibitem[\protect\citeauthoryear{Asplund, Grevesse, Sauval  \& Scott}{Asplund
  et~al.}{2009}]{asplundChemicalCompositionSun2009}
Asplund M.,  Grevesse N.,  Sauval A.~J.,   Scott P.,  2009, \mn@doi [\araa]
  {10.1146/annurev.astro.46.060407.145222}, 47, 481

\bibitem[\protect\citeauthoryear{{Astropy Collaboration} et~al.,}{{Astropy
  Collaboration} et~al.}{2013}]{astropycollaborationAstropyCommunityPython2013}
{Astropy Collaboration} et~al., 2013, \mn@doi [\aap]
  {10.1051/0004-6361/201322068}, 558, A33

\bibitem[\protect\citeauthoryear{Berg, Skillman, Henry, Erb  \& Carigi}{Berg
  et~al.}{2016}]{bergCarbonOxygenAbundances2016}
Berg D.~A.,  Skillman E.~D.,  Henry R. B.~C.,  Erb D.~K.,   Carigi L.,  2016,
  \mn@doi [\apj] {10.3847/0004-637X/827/2/126}, 827, 126

\bibitem[\protect\citeauthoryear{Berg, Erb, Henry, Skillman  \& McQuinn}{Berg
  et~al.}{2019a}]{bergChemicalEvolutionCarbon2019}
Berg D.~A.,  Erb D.~K.,  Henry R. B.~C.,  Skillman E.~D.,   McQuinn K. B.~W.,
  2019a, \mn@doi [\apj] {10.3847/1538-4357/ab020a}, 874, 93

\bibitem[\protect\citeauthoryear{Berg, Chisholm, Erb, Pogge, Henry  \&
  Olivier}{Berg et~al.}{2019b}]{bergIntenseIVHe2019}
Berg D.~A.,  Chisholm J.,  Erb D.~K.,  Pogge R.,  Henry A.,   Olivier G.~M.,
  2019b, \mn@doi [\apj] {10.3847/2041-8213/ab21dc}, 878, L3

\bibitem[\protect\citeauthoryear{Blanc, Kewley, Vogt  \& Dopita}{Blanc
  et~al.}{2015}]{blancIZIInferringGas2015}
Blanc G.~A.,  Kewley L.,  Vogt F. P.~A.,   Dopita M.~A.,  2015, \mn@doi [\apj]
  {10.1088/0004-637X/798/2/99}, 798, 99

\bibitem[\protect\citeauthoryear{Bowyer, Drake  \& Vennes}{Bowyer
  et~al.}{2000}]{bowyerExtremeUltravioletAstronomy2000}
Bowyer S.,  Drake J.~J.,   Vennes S.,  2000, \mn@doi [\araa]
  {10.1146/annurev.astro.38.1.231}, 38, 231

\bibitem[\protect\citeauthoryear{Bresolin, Kennicutt  \& Garnett}{Bresolin
  et~al.}{1999}]{bresolinIonizingStarsExtragalactic1999}
Bresolin F.,  Kennicutt Jr. R.~C.,   Garnett D.~R.,  1999, \mn@doi [\apj]
  {10.1086/306576}, 510, 104

\bibitem[\protect\citeauthoryear{Bressan, Marigo, Girardi, Salasnich, Dal~Cero,
  Rubele  \& Nanni}{Bressan et~al.}{2012}]{bressanPARSECStellarTracks2012}
Bressan A.,  Marigo P.,  Girardi L.,  Salasnich B.,  Dal~Cero C.,  Rubele S.,
  Nanni A.,  2012, \mn@doi [\mnras] {10.1111/j.1365-2966.2012.21948.x}, 427,
  127

\bibitem[\protect\citeauthoryear{Brinchmann, Pettini  \& Charlot}{Brinchmann
  et~al.}{2008a}]{brinchmannNewInsightsStellar2008}
Brinchmann J.,  Pettini M.,   Charlot S.,  2008a, \mn@doi [\mnras]
  {10.1111/j.1365-2966.2008.12914.x}, 385, 769

\bibitem[\protect\citeauthoryear{Brinchmann, Kunth  \& Durret}{Brinchmann
  et~al.}{2008b}]{brinchmannGalaxiesWolfRayetSignatures2008}
Brinchmann J.,  Kunth D.,   Durret F.,  2008b, \mn@doi [\aap]
  {10.1051/0004-6361:200809783}, 485, 657

\bibitem[\protect\citeauthoryear{Bruzual \& Charlot}{Bruzual \&
  Charlot}{2003}]{bruzualStellarPopulationSynthesis2003}
Bruzual G.,  Charlot S.,  2003, \mn@doi [\mnras]
  {10.1046/j.1365-8711.2003.06897.x}, 344, 1000

\bibitem[\protect\citeauthoryear{Byler, Dalcanton, Conroy  \& Johnson}{Byler
  et~al.}{2017}]{bylerNebularContinuumLine2017}
Byler N.,  Dalcanton J.~J.,  Conroy C.,   Johnson B.~D.,  2017, \mn@doi [\apj]
  {10.3847/1538-4357/aa6c66}, 840, 44

\bibitem[\protect\citeauthoryear{Cantiello, Yoon, Langer  \& Livio}{Cantiello
  et~al.}{2007}]{cantielloBinaryStarProgenitors2007}
Cantiello M.,  Yoon S.-C.,  Langer N.,   Livio M.,  2007, \mn@doi [\aap]
  {10.1051/0004-6361:20077115}, 465, L29

\bibitem[\protect\citeauthoryear{Chabrier}{Chabrier}{2003}]{chabrierGalacticStellarSubstellar2003}
Chabrier G.,  2003, \mn@doi [\pasp] {10.1086/376392}, 115, 763

\bibitem[\protect\citeauthoryear{Chandar, Leitherer  \& Tremonti}{Chandar
  et~al.}{2004}]{chandarNGC31251Most2004}
Chandar R.,  Leitherer C.,   Tremonti C.~A.,  2004, \mn@doi [\apj]
  {10.1086/381723}, 604, 153

\bibitem[\protect\citeauthoryear{Charlot \& Longhetti}{Charlot \&
  Longhetti}{2001}]{charlotNebularEmissionStarforming2001}
Charlot S.,  Longhetti M.,  2001, \mn@doi [\mnras]
  {10.1046/j.1365-8711.2001.04260.x}, 323, 887

\bibitem[\protect\citeauthoryear{Chen, Bressan, Girardi, Marigo, Kong  \&
  Lanza}{Chen et~al.}{2015}]{chenPARSECEvolutionaryTracks2015}
Chen Y.,  Bressan A.,  Girardi L.,  Marigo P.,  Kong X.,   Lanza A.,  2015,
  \mn@doi [\mnras] {10.1093/mnras/stv1281}, 452, 1068

\bibitem[\protect\citeauthoryear{Chevallard \& Charlot}{Chevallard \&
  Charlot}{2016}]{chevallardModellingInterpretingSpectral2016}
Chevallard J.,  Charlot S.,  2016, \mn@doi [\mnras] {10.1093/mnras/stw1756},
  462, 1415

\bibitem[\protect\citeauthoryear{Chevallard et~al.,}{Chevallard
  et~al.}{2018}]{chevallardPhysicalPropertiesHionizingphoton2018}
Chevallard J.,  et~al., 2018, \mn@doi [\mnras] {10.1093/mnras/sty1461}, 479,
  3264

\bibitem[\protect\citeauthoryear{Chisholm, Rigby, Bayliss, Berg, Dahle,
  Gladders  \& Sharon}{Chisholm
  et~al.}{2019}]{chisholmConstrainingMetallicitiesAges2019}
Chisholm J.,  Rigby J.~R.,  Bayliss M.,  Berg D.~A.,  Dahle H.,  Gladders M.,
  Sharon K.,  2019, \mn@doi [\apj] {10.3847/1538-4357/ab3104}, 882, 182

\bibitem[\protect\citeauthoryear{Conti \& Leep}{Conti \&
  Leep}{1974}]{contiSpectroscopicObservationsOtype1974}
Conti P.~S.,  Leep E.~M.,  1974, \mn@doi [\apj] {10.1086/153135}, 193, 113

\bibitem[\protect\citeauthoryear{Crowther}{Crowther}{2007}]{crowtherPhysicalPropertiesWolfRayet2007}
Crowther P.~A.,  2007, \mn@doi [\araa]
  {10.1146/annurev.astro.45.051806.110615}, 45, 177

\bibitem[\protect\citeauthoryear{Crowther et~al.,}{Crowther
  et~al.}{2016}]{crowtherR136StarCluster2016}
Crowther P.~A.,  et~al., 2016, \mn@doi [\mnras] {10.1093/mnras/stw273}, 458,
  624

\bibitem[\protect\citeauthoryear{Dashtamirova, Fischer  \& {et
  al.}}{Dashtamirova et~al.}{2020}]{dashtamirovaCosmicOriginsSpectrograph2020}
Dashtamirova D.,  Fischer W.~J.,   {et al.} 2020, Cosmic {{Origins Spectrograph
  Instrument Handbook}}, {{Version}} 12.1.
{STScI}, {Baltimore}

\bibitem[\protect\citeauthoryear{Dessauges-Zavadsky, D'Odorico, Schaerer,
  Modigliani, Tapken  \& Vernet}{Dessauges-Zavadsky
  et~al.}{2010}]{dessauges-zavadskyRestframeUltravioletSpectrum2010}
Dessauges-Zavadsky M.,  D'Odorico S.,  Schaerer D.,  Modigliani A.,  Tapken C.,
    Vernet J.,  2010, \mn@doi [\aap] {10.1051/0004-6361/200913337}, 510, A26

\bibitem[\protect\citeauthoryear{Du, Shapley, Martin  \& Coil}{Du
  et~al.}{2017}]{duIIIEmissionStarforming2017}
Du X.,  Shapley A.~E.,  Martin C.~L.,   Coil A.~L.,  2017, \mn@doi [\apj]
  {10.3847/1538-4357/aa64cf}, 838, 63

\bibitem[\protect\citeauthoryear{Du, Shapley, Tang, Stark, Martin, Mobasher,
  Topping  \& Chevallard}{Du et~al.}{2020}]{duSearchingAnalogsPeak2020}
Du X.,  Shapley A.~E.,  Tang M.,  Stark D.~P.,  Martin C.~L.,  Mobasher B.,
  Topping M.~W.,   Chevallard J.,  2020, \mn@doi [\apj]
  {10.3847/1538-4357/ab67b8}, 890, 65

\bibitem[\protect\citeauthoryear{Eldridge \& Stanway}{Eldridge \&
  Stanway}{2012}]{eldridgeEffectStellarEvolution2012}
Eldridge J.~J.,  Stanway E.~R.,  2012, \mn@doi [\mnras]
  {10.1111/j.1365-2966.2011.19713.x}, 419, 479

\bibitem[\protect\citeauthoryear{Eldridge, Stanway, Xiao, McClelland, Taylor,
  Ng, Greis  \& Bray}{Eldridge
  et~al.}{2017}]{eldridgeBinaryPopulationSpectral2017}
Eldridge J.~J.,  Stanway E.~R.,  Xiao L.,  McClelland L. A.~S.,  Taylor G.,  Ng
  M.,  Greis S. M.~L.,   Bray J.~C.,  2017, \mn@doi [\pasa]
  {10.1017/pasa.2017.51}, 34, e058

\bibitem[\protect\citeauthoryear{Endsley, Stark, Chevallard  \&
  Charlot}{Endsley et~al.}{2020}]{endsleyOIIIBetaEquivalent2020}
Endsley R.,  Stark D.~P.,  Chevallard J.,   Charlot S.,  2020

\bibitem[\protect\citeauthoryear{Erb, Pettini, Shapley, Steidel, Law  \&
  Reddy}{Erb et~al.}{2010}]{erbPhysicalConditionsYoung2010}
Erb D.~K.,  Pettini M.,  Shapley A.~E.,  Steidel C.~C.,  Law D.~R.,   Reddy
  N.~A.,  2010, \mn@doi [\apj] {10.1088/0004-637X/719/2/1168}, 719, 1168

\bibitem[\protect\citeauthoryear{Esteban, {Garc{\'i}a-Rojas}, Carigi, Peimbert,
  Bresolin, {L{\'o}pez-S{\'a}nchez}  \& {Mesa-Delgado}}{Esteban
  et~al.}{2014}]{estebanCarbonOxygenAbundances2014}
Esteban C.,  {Garc{\'i}a-Rojas} J.,  Carigi L.,  Peimbert M.,  Bresolin F.,
  {L{\'o}pez-S{\'a}nchez} A.~R.,   {Mesa-Delgado} A.,  2014, \mn@doi [\mnras]
  {10.1093/mnras/stu1177}, 443, 624

\bibitem[\protect\citeauthoryear{Filippenko, Romani, Sargent  \&
  Blandford}{Filippenko et~al.}{1988}]{filippenkoPossibleEvidenceDisk1988}
Filippenko A.~V.,  Romani R.~W.,  Sargent W. L.~W.,   Blandford R.~D.,  1988,
  \mn@doi [\aj] {10.1086/114806}, 96, 242

\bibitem[\protect\citeauthoryear{Fitzpatrick}{Fitzpatrick}{1999}]{fitzpatrickCorrectingEffectsInterstellar1999}
Fitzpatrick E.~L.,  1999, \mn@doi [\pasp] {10.1086/316293}, 111, 63

\bibitem[\protect\citeauthoryear{Foreman-Mackey, Hogg, Lang  \&
  Goodman}{Foreman-Mackey et~al.}{2013}]{foreman-mackeyEmceeMCMCHammer2013}
Foreman-Mackey D.,  Hogg D.~W.,  Lang D.,   Goodman J.,  2013, \mn@doi [\pasp]
  {10.1086/670067}, 125, 306

\bibitem[\protect\citeauthoryear{Froese~Fischer \& Tachiev}{Froese~Fischer \&
  Tachiev}{2004}]{froesefischerBreitPauliEnergyLevels2004}
Froese~Fischer C.,  Tachiev G.,  2004, \mn@doi [Atomic Data and Nuclear Data
  Tables] {10.1016/j.adt.2004.02.001}, 87, 1

\bibitem[\protect\citeauthoryear{{Garc{\'i}a-Rojas} \&
  Esteban}{{Garc{\'i}a-Rojas} \&
  Esteban}{2007}]{garcia-rojasAbundanceDiscrepancyProblem2007}
{Garc{\'i}a-Rojas} J.,  Esteban C.,  2007, \mn@doi [\apj] {10.1086/521871},
  670, 457

\bibitem[\protect\citeauthoryear{Giavalisco, Koratkar  \& Calzetti}{Giavalisco
  et~al.}{1996}]{giavaliscoObscurationLYAlpha1996}
Giavalisco M.,  Koratkar A.,   Calzetti D.,  1996, \mn@doi [\apj]
  {10.1086/177557}, 466, 831

\bibitem[\protect\citeauthoryear{{G{\'o}mez-Llanos} \&
  Morisset}{{G{\'o}mez-Llanos} \&
  Morisset}{2020}]{gomez-llanosBiabundancePhotoionizationModels2020}
{G{\'o}mez-Llanos} V.,  Morisset C.,  2020, \mn@doi [\mnras]
  {10.1093/mnras/staa2157}

\bibitem[\protect\citeauthoryear{Gordon, Clayton, Misselt, Landolt  \&
  Wolff}{Gordon et~al.}{2003}]{gordonQuantitativeComparisonSmall2003}
Gordon K.~D.,  Clayton G.~C.,  Misselt K.~A.,  Landolt A.~U.,   Wolff M.~J.,
  2003, \mn@doi [\apj] {10.1086/376774}, 594, 279

\bibitem[\protect\citeauthoryear{G\"{o}tberg, de Mink, Groh, Kupfer, Crowther,
  Zapartas  \& Renzo}{G\"{o}tberg
  et~al.}{2018}]{gotbergSpectralModelsBinary2018}
G\"{o}tberg Y.,  de Mink S.~E.,  Groh J.~H.,  Kupfer T.,  Crowther P.~A.,
  Zapartas E.,   Renzo M.,  2018, \mn@doi [\aap] {10.1051/0004-6361/201732274},
  615, A78

\bibitem[\protect\citeauthoryear{G{\"o}tberg, {de Mink}, Groh, Leitherer  \&
  Norman}{G{\"o}tberg et~al.}{2019}]{gotbergImpactStarsStripped2019}
G{\"o}tberg Y.,  {de Mink} S.~E.,  Groh J.~H.,  Leitherer C.,   Norman C.,
  2019, \mn@doi [\aap] {10.1051/0004-6361/201834525}, 629, A134

\bibitem[\protect\citeauthoryear{Gr{\"a}fener \& Hamann}{Gr{\"a}fener \&
  Hamann}{2008}]{grafenerMassLossLatetype2008}
Gr{\"a}fener G.,  Hamann W.-R.,  2008, \mn@doi [\aap]
  {10.1051/0004-6361:20066176}, 482, 945

\bibitem[\protect\citeauthoryear{Gr\"{a}fener, Koesterke  \&
  Hamann}{Gr\"{a}fener
  et~al.}{2002}]{grafenerLineblanketedModelAtmospheres2002}
Gr\"{a}fener G.,  Koesterke L.,   Hamann W.~R.,  2002, \mn@doi [\aap]
  {10.1051/0004-6361:20020269}, 387, 244

\bibitem[\protect\citeauthoryear{Gunawardhana et~al.,}{Gunawardhana
  et~al.}{2011}]{gunawardhanaGalaxyMassAssembly2011}
Gunawardhana M. L.~P.,  et~al., 2011, \mn@doi [\mnras]
  {10.1111/j.1365-2966.2011.18800.x}, 415, 1647

\bibitem[\protect\citeauthoryear{Gutkin, Charlot  \& Bruzual}{Gutkin
  et~al.}{2016}]{gutkinModellingNebularEmission2016}
Gutkin J.,  Charlot S.,   Bruzual G.,  2016, \mn@doi [\mnras]
  {10.1093/mnras/stw1716}, 462, 1757

\bibitem[\protect\citeauthoryear{Hainich et~al.,}{Hainich
  et~al.}{2014}]{hainichWolfRayetStarsLarge2014}
Hainich R.,  et~al., 2014, \mn@doi [\aap] {10.1051/0004-6361/201322696}, 565,
  A27

\bibitem[\protect\citeauthoryear{Hainich, Pasemann, Todt, Shenar, Sander  \&
  Hamann}{Hainich et~al.}{2015}]{hainichWolfRayetStarsSmall2015}
Hainich R.,  Pasemann D.,  Todt H.,  Shenar T.,  Sander A.,   Hamann W.-R.,
  2015, \mn@doi [\aap] {10.1051/0004-6361/201526241}, 581, A21

\bibitem[\protect\citeauthoryear{Hamann \& Gr{\"a}fener}{Hamann \&
  Gr{\"a}fener}{2003}]{hamannTemperatureCorrectionMethod2003}
Hamann W.-R.,  Gr{\"a}fener G.,  2003, \mn@doi [\aap]
  {10.1051/0004-6361:20031308}, 410, 993

\bibitem[\protect\citeauthoryear{Hillier, Bouret, Lanz  \& Busche}{Hillier
  et~al.}{2012}]{hillierInfluenceRotationOptical2012}
Hillier D.~J.,  Bouret J.-C.,  Lanz T.,   Busche J.~R.,  2012, \mn@doi [\mnras]
  {10.1111/j.1365-2966.2012.21646.x}, 426, 1043

\bibitem[\protect\citeauthoryear{Hopkins}{Hopkins}{2018}]{hopkinsDawesReviewMeasuring2018}
Hopkins A.~M.,  2018, \mn@doi [\pasa] {10.1017/pasa.2018.29}, 35

\bibitem[\protect\citeauthoryear{Hummer \& Storey}{Hummer \&
  Storey}{1987}]{hummerRecombinationlineIntensitiesHydrogenic1987}
Hummer D.~G.,  Storey P.~J.,  1987, \mn@doi [\mnras] {10.1093/mnras/224.3.801},
  224, 801

\bibitem[\protect\citeauthoryear{Hunter}{Hunter}{2007}]{hunterMatplotlib2DGraphics2007}
Hunter J.~D.,  2007, \mn@doi [Computing In Science & Engineering]
  {10.1109/MCSE.2007.55}, 9, 90–95

\bibitem[\protect\citeauthoryear{Hutchison et~al.,}{Hutchison
  et~al.}{2019}]{hutchisonNearinfraredSpectroscopyGalaxies2019}
Hutchison T.~A.,  et~al., 2019, \mn@doi [\apj] {10.3847/1538-4357/ab22a2}, 879,
  70

\bibitem[\protect\citeauthoryear{Izotov, Stasińska, Meynet, Guseva  \&
  Thuan}{Izotov et~al.}{2006}]{izotovChemicalCompositionMetalpoor2006}
Izotov Y.~I.,  Stasińska G.,  Meynet G.,  Guseva N.~G.,   Thuan T.~X.,  2006,
  \mn@doi [\aap] {10.1051/0004-6361:20053763}, 448, 955

\bibitem[\protect\citeauthoryear{Jaskot \& Ravindranath}{Jaskot \&
  Ravindranath}{2016}]{jaskotPhotoionizationModelsSemiforbidden2016}
Jaskot A.~E.,  Ravindranath S.,  2016, \mn@doi [\apj]
  {10.3847/1538-4357/833/2/136}, 833, 136

\bibitem[\protect\citeauthoryear{Jones, Oliphant, Peterson  et~al.}{Jones
  et~al.}{2001}]{jonesSciPyOpenSource2001}
Jones E.,  Oliphant T.,  Peterson P.,   et~al., 2001, SciPy: Open source
  scientific tools for Python

\bibitem[\protect\citeauthoryear{Jones, Stark  \& Ellis}{Jones
  et~al.}{2012}]{jonesKeckSpectroscopyFaint2012}
Jones T.,  Stark D.~P.,   Ellis R.~S.,  2012, \mn@doi [\apj]
  {10.1088/0004-637X/751/1/51}, 751, 51

\bibitem[\protect\citeauthoryear{Kennicutt}{Kennicutt}{1984}]{kennicuttStructuralPropertiesGiant1984}
Kennicutt Jr. R.~C.,  1984, \mn@doi [\apj] {10.1086/162669}, 287, 116

\bibitem[\protect\citeauthoryear{Kewley, Nicholls  \& Sutherland}{Kewley
  et~al.}{2019}]{kewleyUnderstandingGalaxyEvolution2019}
Kewley L.~J.,  Nicholls D.~C.,   Sutherland R.~S.,  2019, \mn@doi [\araa]
  {10.1146/annurev-astro-081817-051832}, 57, 511

\bibitem[\protect\citeauthoryear{Labb\'{e} et~al.,}{Labb\'{e}
  et~al.}{2013}]{labbeSpectralEnergyDistributions2013}
Labb\'{e} I.,  et~al., 2013, \mn@doi [\apjl] {10.1088/2041-8205/777/2/L19},
  777, L19

\bibitem[\protect\citeauthoryear{Le~F{\`e}vre et~al.,}{Le~F{\`e}vre
  et~al.}{2019}]{lefevreVIMOSUltraDeepSurvey2019}
Le~F{\`e}vre O.,  et~al., 2019, \mn@doi [\aap] {10.1051/0004-6361/201732197},
  625, A51

\bibitem[\protect\citeauthoryear{Leitherer et~al.,}{Leitherer
  et~al.}{1999}]{leithererStarburst99SynthesisModels1999}
Leitherer C.,  et~al., 1999, \mn@doi [\apjs] {10.1086/313233}, 123, 3

\bibitem[\protect\citeauthoryear{Leitherer, Tremonti, Heckman  \&
  Calzetti}{Leitherer
  et~al.}{2011}]{leithererUltravioletSpectroscopicAtlas2011}
Leitherer C.,  Tremonti C.~A.,  Heckman T.~M.,   Calzetti D.,  2011, \mn@doi
  [\aj] {10.1088/0004-6256/141/2/37}, 141, 37

\bibitem[\protect\citeauthoryear{Leitherer, Ekstr\"{o}m, Meynet, Schaerer,
  Agienko  \& Levesque}{Leitherer
  et~al.}{2014}]{leithererEffectsStellarRotation2014}
Leitherer C.,  Ekstr\"{o}m S.,  Meynet G.,  Schaerer D.,  Agienko K.~B.,
  Levesque E.~M.,  2014, \mn@doi [\apjs] {10.1088/0067-0049/212/1/14}, 212, 14

\bibitem[\protect\citeauthoryear{Leitherer, Byler, Lee  \& Levesque}{Leitherer
  et~al.}{2018}]{leithererPhysicalPropertiesII2018}
Leitherer C.,  Byler N.,  Lee J.~C.,   Levesque E.~M.,  2018, \mn@doi [\apj]
  {10.3847/1538-4357/aada84}, 865, 55

\bibitem[\protect\citeauthoryear{Leja, Johnson, Conroy, van Dokkum  \&
  Byler}{Leja et~al.}{2017}]{lejaDerivingPhysicalProperties2017}
Leja J.,  Johnson B.~D.,  Conroy C.,  van Dokkum P.~G.,   Byler N.,  2017,
  \mn@doi [\apj] {10.3847/1538-4357/aa5ffe}, 837, 170

\bibitem[\protect\citeauthoryear{Luridiana, Morisset  \& Shaw}{Luridiana
  et~al.}{2015}]{luridianaPyNebNewTool2015}
Luridiana V.,  Morisset C.,   Shaw R.~A.,  2015, \mn@doi [\aap]
  {10.1051/0004-6361/201323152}, 573, A42

\bibitem[\protect\citeauthoryear{Ma, Hopkins, Kasen, Quataert,
  {Faucher-Gigu{\`e}re}, Kere{\v s}, Murray  \& Strom}{Ma
  et~al.}{2016}]{maBinaryStarsCan2016}
Ma X.,  Hopkins P.~F.,  Kasen D.,  Quataert E.,  {Faucher-Gigu{\`e}re} C.-A.,
  Kere{\v s} D.,  Murray N.,   Strom A.,  2016, \mn@doi [\mnras]
  {10.1093/mnras/stw941}, 459, 3614

\bibitem[\protect\citeauthoryear{Maeda et~al.,}{Maeda
  et~al.}{2008}]{maedaAsphericitySupernovaExplosions2008}
Maeda K.,  et~al., 2008, \mn@doi [Science] {10.1126/science.1149437}, 319, 1220

\bibitem[\protect\citeauthoryear{Mainali, Kollmeier, Stark, Simcoe, Walth,
  Newman  \& Miller}{Mainali et~al.}{2017}]{mainaliEvidenceHardIonizing2017}
Mainali R.,  Kollmeier J.~A.,  Stark D.~P.,  Simcoe R.~A.,  Walth G.,  Newman
  A.~B.,   Miller D.~R.,  2017, \mn@doi [\apjl] {10.3847/2041-8213/836/1/L14},
  836, L14

\bibitem[\protect\citeauthoryear{Mainali et~al.,}{Mainali
  et~al.}{2019}]{mainaliRELICSSpectroscopyGravitationallylensed2019}
Mainali R.,  et~al., 2019, arXiv e-prints, p. arXiv:1909.09212

\bibitem[\protect\citeauthoryear{Marks, Kroupa, Dabringhausen  \&
  Pawlowski}{Marks et~al.}{2012}]{marksEvidenceTopheavyStellar2012}
Marks M.,  Kroupa P.,  Dabringhausen J.,   Pawlowski M.~S.,  2012, \mn@doi
  [\mnras] {10.1111/j.1365-2966.2012.20767.x}, 422, 2246

\bibitem[\protect\citeauthoryear{Maseda et~al.,}{Maseda
  et~al.}{2017}]{masedaMUSEHubbleUltra2017}
Maseda M.~V.,  et~al., 2017, \mn@doi [\aap] {10.1051/0004-6361/201730985}, 608,
  A4

\bibitem[\protect\citeauthoryear{Massey, Bresolin, Kudritzki, Puls  \&
  Pauldrach}{Massey et~al.}{2004}]{masseyPhysicalPropertiesEffective2004}
Massey P.,  Bresolin F.,  Kudritzki R.~P.,  Puls J.,   Pauldrach A. W.~A.,
  2004, \mn@doi [\apj] {10.1086/420766}, 608, 1001

\bibitem[\protect\citeauthoryear{McGaugh}{McGaugh}{1991}]{mcgaughIIRegionAbundances1991}
McGaugh S.~S.,  1991, \mn@doi [\apj] {10.1086/170569}, 380, 140

\bibitem[\protect\citeauthoryear{M{\"u}ller \& Vink}{M{\"u}ller \&
  Vink}{2008}]{mullerConsistentSolutionVelocity2008}
M{\"u}ller P.~E.,  Vink J.~S.,  2008, \mn@doi [\aap]
  {10.1051/0004-6361:20078798}, 492, 493

\bibitem[\protect\citeauthoryear{Nakajima et~al.,}{Nakajima
  et~al.}{2018}]{nakajimaVIMOSUltraDeep2018}
Nakajima K.,  et~al., 2018, \mn@doi [\aap] {10.1051/0004-6361/201731935}, 612,
  A94

\bibitem[\protect\citeauthoryear{Nicholls, Dopita  \& Sutherland}{Nicholls
  et~al.}{2012}]{nichollsResolvingElectronTemperature2012}
Nicholls D.~C.,  Dopita M.~A.,   Sutherland R.~S.,  2012, \mn@doi [\apj]
  {10.1088/0004-637X/752/2/148}, 752, 148

\bibitem[\protect\citeauthoryear{Paalvast \& Brinchmann}{Paalvast \&
  Brinchmann}{2017}]{paalvastMetallicityCalibrationsGalaxies2017}
Paalvast M.,  Brinchmann J.,  2017, \mn@doi [\mnras] {10.1093/mnras/stx1271},
  470, 1612

\bibitem[\protect\citeauthoryear{Peimbert, Storey  \&
  {Torres-Peimbert}}{Peimbert et~al.}{1993}]{peimbertAbundanceRatioGaseous1993}
Peimbert M.,  Storey P.~J.,   {Torres-Peimbert} S.,  1993, \mn@doi [\apj]
  {10.1086/173108}, 414, 626

\bibitem[\protect\citeauthoryear{Plat, Charlot, Bruzual, Feltre,
  {Vidal-Garc{\'i}a}, Morisset, Chevallard  \& Todt}{Plat
  et~al.}{2019}]{platConstraintsProductionEscape2019}
Plat A.,  Charlot S.,  Bruzual G.,  Feltre A.,  {Vidal-Garc{\'i}a} A.,
  Morisset C.,  Chevallard J.,   Todt H.,  2019, \mn@doi [\mnras]
  {10.1093/mnras/stz2616}, 490, 978

\bibitem[\protect\citeauthoryear{Podobedova, Kelleher  \& Wiese}{Podobedova
  et~al.}{2009}]{podobedovaCriticallyEvaluatedAtomic2009}
Podobedova L.~I.,  Kelleher D.~E.,   Wiese W.~L.,  2009, \mn@doi [Journal of
  Physical and Chemical Reference Data] {10.1063/1.3032939}, 38, 171

\bibitem[\protect\citeauthoryear{Ravindranath, Monroe, Jaskot, Ferguson  \&
  Tumlinson}{Ravindranath et~al.}{2020}]{ravindranathSemiforbiddenIIIL19092020}
Ravindranath S.,  Monroe T.,  Jaskot A.,  Ferguson H.~C.,   Tumlinson J.,
  2020, \mn@doi [\apj] {10.3847/1538-4357/ab91a5}, 896, 170

\bibitem[\protect\citeauthoryear{Rigby, Bayliss, Gladders, Sharon, Wuyts,
  Dahle, Johnson  \& Peña-Guerrero}{Rigby
  et~al.}{2015}]{rigbyIIIEmissionStarforming2015}
Rigby J.~R.,  Bayliss M.~B.,  Gladders M.~D.,  Sharon K.,  Wuyts E.,  Dahle H.,
   Johnson T.,   Peña-Guerrero M.,  2015, \mn@doi [\apjl]
  {10.1088/2041-8205/814/1/L6}, 814, L6

\bibitem[\protect\citeauthoryear{{Roberts-Borsani} et~al.,}{{Roberts-Borsani}
  et~al.}{2016}]{roberts-borsaniGalaxiesRedSpitzer2016}
{Roberts-Borsani} G.~W.,  et~al., 2016, \mn@doi [\apj]
  {10.3847/0004-637X/823/2/143}, 823, 143

\bibitem[\protect\citeauthoryear{Sander, Hamann, Hainich, Shenar  \&
  Todt}{Sander et~al.}{2015}]{sanderHydrodynamicModelingMassive2015}
Sander A.,  Hamann W.-R.,  Hainich R.,  Shenar T.,   Todt H.,  2015, Wolf-Rayet
  Stars: Proceedings of an International Workshop held in Potsdam, p.~139

\bibitem[\protect\citeauthoryear{Sander, Vink  \& Hamann}{Sander
  et~al.}{2020}]{sanderDrivingClassicalWolfRayet2020}
Sander A. A.~C.,  Vink J.~S.,   Hamann W.-R.,  2020, \mn@doi [\mnras]
  {10.1093/mnras/stz3064}, 491, 4406

\bibitem[\protect\citeauthoryear{Sanders et~al.,}{Sanders
  et~al.}{2020}]{sandersMOSDEFSurveyDirectmethod2020}
Sanders R.~L.,  et~al., 2020, \mn@doi [\mnras] {10.1093/mnras/stz3032}, 491,
  1427

\bibitem[\protect\citeauthoryear{Schaerer \& Vacca}{Schaerer \&
  Vacca}{1998}]{schaererNewModelsWolfRayet1998}
Schaerer D.,  Vacca W.~D.,  1998, \mn@doi [\apj] {10.1086/305487}, 497, 618

\bibitem[\protect\citeauthoryear{Schaerer, Izotov, Nakajima, Worseck, Chisholm,
  Verhamme, Thuan  \& de Barros}{Schaerer
  et~al.}{2018}]{schaererIntenseIIIEnsuremath2018}
Schaerer D.,  Izotov Y.~I.,  Nakajima K.,  Worseck G.,  Chisholm J.,  Verhamme
  A.,  Thuan T.~X.,   de Barros S.,  2018, \mn@doi [\aap]
  {10.1051/0004-6361/201833823}, 616, L14

\bibitem[\protect\citeauthoryear{Schlafly \& Finkbeiner}{Schlafly \&
  Finkbeiner}{2011}]{schlaflyMeasuringReddeningSloan2011}
Schlafly E.~F.,  Finkbeiner D.~P.,  2011, \mn@doi [\apj]
  {10.1088/0004-637X/737/2/103}, 737, 103

\bibitem[\protect\citeauthoryear{Schmidt et~al.,}{Schmidt
  et~al.}{2017}]{schmidtGrismLensAmplifiedSurvey2017}
Schmidt K.~B.,  et~al., 2017, \mn@doi [\apj] {10.3847/1538-4357/aa68a3}, 839,
  17

\bibitem[\protect\citeauthoryear{Schneider et~al.,}{Schneider
  et~al.}{2014}]{schneiderAgesYoungStar2014}
Schneider F. R.~N.,  et~al., 2014, \mn@doi [\apj]
  {10.1088/0004-637X/780/2/117}, 780, 117

\bibitem[\protect\citeauthoryear{Schneider et~al.,}{Schneider
  et~al.}{2018}]{schneiderExcessMassiveStars2018}
Schneider F. R.~N.,  et~al., 2018, \mn@doi [Science] {10.1126/science.aan0106},
  359, 69

\bibitem[\protect\citeauthoryear{Senchyna et~al.,}{Senchyna
  et~al.}{2017}]{senchynaUltravioletSpectraExtreme2017}
Senchyna P.,  et~al., 2017, \mn@doi [\mnras] {10.1093/mnras/stx2059}, 472, 2608

\bibitem[\protect\citeauthoryear{Senchyna, Stark, Chevallard, Charlot, Jones
  \& {Vidal-Garc{\'i}a}}{Senchyna
  et~al.}{2019}]{senchynaExtremelyMetalpoorGalaxies2019}
Senchyna P.,  Stark D.~P.,  Chevallard J.,  Charlot S.,  Jones T.,
  {Vidal-Garc{\'i}a} A.,  2019, \mn@doi [\mnras] {10.1093/mnras/stz1907}, 488,
  3492

\bibitem[\protect\citeauthoryear{Shapley, Steidel, Pettini  \&
  Adelberger}{Shapley et~al.}{2003}]{shapleyRestFrameUltravioletSpectra2003}
Shapley A.~E.,  Steidel C.~C.,  Pettini M.,   Adelberger K.~L.,  2003, \mn@doi
  [\apj] {10.1086/373922}, 588, 65

\bibitem[\protect\citeauthoryear{Shirazi \& Brinchmann}{Shirazi \&
  Brinchmann}{2012}]{shiraziStronglyStarForming2012}
Shirazi M.,  Brinchmann J.,  2012, \mn@doi [\mnras]
  {10.1111/j.1365-2966.2012.20439.x}, 421, 1043

\bibitem[\protect\citeauthoryear{Smit et~al.,}{Smit
  et~al.}{2014}]{smitEvidenceUbiquitousHighequivalentwidth2014}
Smit R.,  et~al., 2014, \mn@doi [\apj] {10.1088/0004-637X/784/1/58}, 784, 58

\bibitem[\protect\citeauthoryear{Smith, Crowther, Calzetti  \& Sidoli}{Smith
  et~al.}{2016}]{smithVeryMassiveStar2016}
Smith L.~J.,  Crowther P.~A.,  Calzetti D.,   Sidoli F.,  2016, \mn@doi [\apj]
  {10.3847/0004-637X/823/1/38}, 823, 38

\bibitem[\protect\citeauthoryear{Soria, Wu  \& Hunstead}{Soria
  et~al.}{2000}]{soriaOpticalSpectroscopyGRO2000}
Soria R.,  Wu K.,   Hunstead R.~W.,  2000, \mn@doi [\apj] {10.1086/309194},
  539, 445

\bibitem[\protect\citeauthoryear{Stanway \& Eldridge}{Stanway \&
  Eldridge}{2019}]{stanwayInitialMassFunction2019}
Stanway E.~R.,  Eldridge J.~J.,  2019, \mn@doi [\aap]
  {10.1051/0004-6361/201834359}, 621, A105

\bibitem[\protect\citeauthoryear{Stanway, Chrimes, Eldridge  \&
  Stevance}{Stanway et~al.}{2020}]{stanwayEvaluatingImpactBinary2020}
Stanway E.~R.,  Chrimes A.~A.,  Eldridge J.~J.,   Stevance H.~F.,  2020,
  arXiv:2004.11913 [astro-ph]

\bibitem[\protect\citeauthoryear{Stark}{Stark}{2016}]{starkGalaxiesFirstBillion2016}
Stark D.~P.,  2016, \mn@doi [\araa] {10.1146/annurev-astro-081915-023417}, 54,
  761

\bibitem[\protect\citeauthoryear{Stark et~al.,}{Stark
  et~al.}{2015a}]{starkSpectroscopicDetectionsIII2015}
Stark D.~P.,  et~al., 2015a, \mn@doi [\mnras] {10.1093/mnras/stv688}, 450, 1846

\bibitem[\protect\citeauthoryear{Stark et~al.,}{Stark
  et~al.}{2015b}]{starkSpectroscopicDetectionIV2015}
Stark D.~P.,  et~al., 2015b, \mn@doi [\mnras] {10.1093/mnras/stv1907}, 454,
  1393

\bibitem[\protect\citeauthoryear{Steidel, Strom, Pettini, Rudie, Reddy  \&
  Trainor}{Steidel et~al.}{2016}]{steidelReconcilingStellarNebular2016}
Steidel C.~C.,  Strom A.~L.,  Pettini M.,  Rudie G.~C.,  Reddy N.~A.,   Trainor
  R.~F.,  2016, \mn@doi [\apj] {10.3847/0004-637X/826/2/159}, 826, 159

\bibitem[\protect\citeauthoryear{Storey \& Zeippen}{Storey \&
  Zeippen}{2000}]{storeyTheoreticalValuesOIII2000}
Storey P.~J.,  Zeippen C.~J.,  2000, \mn@doi [\mnras]
  {10.1046/j.1365-8711.2000.03184.x}, 312, 813

\bibitem[\protect\citeauthoryear{Storey, Sochi  \& Badnell}{Storey
  et~al.}{2014}]{storeyCollisionStrengthsNebular2014}
Storey P.~J.,  Sochi T.,   Badnell N.~R.,  2014, \mn@doi [\mnras]
  {10.1093/mnras/stu777}, 441, 3028

\bibitem[\protect\citeauthoryear{Strom, Steidel, Rudie, Trainor  \&
  Pettini}{Strom et~al.}{2018}]{stromMeasuringPhysicalConditions2018}
Strom A.~L.,  Steidel C.~C.,  Rudie G.~C.,  Trainor R.~F.,   Pettini M.,  2018,
  \mn@doi [\apj] {10.3847/1538-4357/aae1a5}, 868, 117

\bibitem[\protect\citeauthoryear{Tang, Stark, Chevallard, Charlot, Endsley  \&
  Congiu}{Tang et~al.}{2020}]{tangRestframeUVSpectroscopy2020}
Tang M.,  Stark D.,  Chevallard J.,  Charlot S.,  Endsley R.,   Congiu E.,
  2020, arXiv e-prints, 2007, arXiv:2007.12197

\bibitem[\protect\citeauthoryear{Tayal}{Tayal}{2007}]{tayalOscillatorStrengthsElectron2007}
Tayal S.~S.,  2007, \mn@doi [\apjs] {10.1086/513107}, 171, 331

\bibitem[\protect\citeauthoryear{Tayal \& Zatsarinny}{Tayal \&
  Zatsarinny}{2010}]{tayalBreitPauliTransitionProbabilities2010}
Tayal S.~S.,  Zatsarinny O.,  2010, \mn@doi [\apjs]
  {10.1088/0067-0049/188/1/32}, 188, 32

\bibitem[\protect\citeauthoryear{Todt, Sander, Hainich, Hamann, Quade  \&
  Shenar}{Todt et~al.}{2015}]{todtPotsdamWolfRayetModel2015}
Todt H.,  Sander A.,  Hainich R.,  Hamann W.-R.,  Quade M.,   Shenar T.,  2015,
  \mn@doi [\aap] {10.1051/0004-6361/201526253}, 579, A75

\bibitem[\protect\citeauthoryear{Tsamis, Barlow, Liu, Danziger  \&
  Storey}{Tsamis et~al.}{2003}]{tsamisHeavyElementsGalactic2003}
Tsamis Y.~G.,  Barlow M.~J.,  Liu X.-W.,  Danziger I.~J.,   Storey P.~J.,
  2003, \mn@doi [\mnras] {10.1046/j.1365-8711.2003.06081.x}, 338, 687

\bibitem[\protect\citeauthoryear{Val~Baker, Norton  \& Quaintrell}{Val~Baker
  et~al.}{2005}]{valbakerMassNeutronStar2005}
Val~Baker A. K.~F.,  Norton A.~J.,   Quaintrell H.,  2005, \mn@doi [\aap]
  {10.1051/0004-6361:20053074}, 441, 685

\bibitem[\protect\citeauthoryear{Vale~Asari, Stasi{\'n}ska, Morisset  \&
  Cid~Fernandes}{Vale~Asari et~al.}{2016}]{valeasariBONDBayesianOxygen2016}
Vale~Asari N.,  Stasi{\'n}ska G.,  Morisset C.,   Cid~Fernandes R.,  2016,
  \mn@doi [\mnras] {10.1093/mnras/stw971}, 460, 1739

\bibitem[\protect\citeauthoryear{Vidal-Garc\'{i}a, Charlot, Bruzual  \&
  Hubeny}{Vidal-Garc\'{i}a
  et~al.}{2017}]{vidal-garciaModellingUltravioletlineDiagnostics2017}
Vidal-Garc\'{i}a A.,  Charlot S.,  Bruzual G.,   Hubeny I.,  2017, \mn@doi
  [\mnras] {10.1093/mnras/stx1324}, 470, 3532

\bibitem[\protect\citeauthoryear{Vink, de Koter  \& Lamers}{Vink
  et~al.}{2001}]{vinkMasslossPredictionsStars2001}
Vink J.~S.,  de Koter A.,   Lamers H. J. G. L.~M.,  2001, \mn@doi [\aap]
  {10.1051/0004-6361:20010127}, 369, 574

\bibitem[\protect\citeauthoryear{Vink, Muijres, Anthonisse, {de Koter},
  Gr{\"a}fener  \& Langer}{Vink et~al.}{2011}]{vinkWindModellingVery2011}
Vink J.~S.,  Muijres L.~E.,  Anthonisse B.,  {de Koter} A.,  Gr{\"a}fener G.,
  Langer N.,  2011, \mn@doi [\aap] {10.1051/0004-6361/201116614}, 531, A132

\bibitem[\protect\citeauthoryear{Walborn}{Walborn}{1973}]{walbornSpaceDistributionStars1973}
Walborn N.~R.,  1973, \mn@doi [\aj] {10.1086/111509}, 78, 1067

\bibitem[\protect\citeauthoryear{Walborn et~al.,}{Walborn
  et~al.}{2010}]{walbornOnfpClassMagellanic2010}
Walborn N.~R.,  et~al., 2010, \mn@doi [\aj] {10.1088/0004-6256/139/3/1283},
  139, 1283

\bibitem[\protect\citeauthoryear{Wiese, Fuhr  \& Deters}{Wiese
  et~al.}{1996}]{wieseAtomicTransitionProbabilities1996}
Wiese W.~L.,  Fuhr J.~R.,   Deters T.~M.,  1996, Atomic transition
  probabilities of carbon, nitrogen, and oxygen : a critical data compilation.
  Edited by W.L. Wiese, J.R. Fuhr, and T.M. Deters. Washington, DC : American
  Chemical Society ... for the National Institute of Standards and Technology
  (NIST) c1996. QC 453 .W53 1996. Also Journal of Physical and Chemical
  Reference Data, Monograph 7. Melville, NY: AIP Press

\bibitem[\protect\citeauthoryear{Wofford, Leitherer, Chandar  \&
  Bouret}{Wofford et~al.}{2014}]{woffordRareEncounterVery2014}
Wofford A.,  Leitherer C.,  Chandar R.,   Bouret J.-C.,  2014, \mn@doi [\apj]
  {10.1088/0004-637X/781/2/122}, 781, 122

\bibitem[\protect\citeauthoryear{Wofford et~al.,}{Wofford
  et~al.}{2016}]{woffordComprehensiveComparativeTest2016}
Wofford A.,  et~al., 2016, \mn@doi [\mnras] {10.1093/mnras/stw150}, 457, 4296

\bibitem[\protect\citeauthoryear{Xiao, Stanway  \& Eldridge}{Xiao
  et~al.}{2018}]{xiaoEmissionlineDiagnosticsNearby2018}
Xiao L.,  Stanway E.~R.,   Eldridge J.~J.,  2018, \mn@doi [\mnras]
  {10.1093/mnras/sty646}, 477, 904

\bibitem[\protect\citeauthoryear{Yoon \& Langer}{Yoon \&
  Langer}{2005}]{yoonEvolutionRapidlyRotating2005}
Yoon S.-C.,  Langer N.,  2005, \mn@doi [\aap] {10.1051/0004-6361:20054030},
  443, 643

\bibitem[\protect\citeauthoryear{de Mink, Sana, Langer, Izzard  \&
  Schneider}{de~Mink et~al.}{2014}]{deminkIncidenceStellarMergers2014}
de Mink S.~E.,  Sana H.,  Langer N.,  Izzard R.~G.,   Schneider F. R.~N.,
  2014, \mn@doi [\apj] {10.1088/0004-637X/782/1/7}, 782, 7

\makeatother
\end{thebibliography}

\appendix

\section{Target star-forming regions in context}
\label{app:targetcontext}

Here we summarize the target objects in the context of their host galaxies and relevant previous studies in which they have been identified.

\textbf{SB 9} (SDSS JJ130432.27-033322.0, Plate-MJD-Fiber 339-51692-83) is a star-forming region embedded in the northern outskirts of the barred-spiral galaxy UGCA 322.
This system was also identified as WR 33 in the catalog of SDSS spectra with prominent optical Wolf-Rayet features presented by \citet{brinchmannGalaxiesWolfRayetSignatures2008}.

\textbf{SB 49} (SDSS J011533.82-005131.1, Plate-MJD-Fiber 695-52202-137, WR 354) is located in an eastern arm of the spiral galaxy NGC 450.

\textbf{SB 61} (SDSS J144805.38-011057.6, Plate-MJD-Fiber 308-51662-81\footnote{We use this SDSS spectrum over the other flux-matched spectrum 920-52411-575 as it is identified as the \texttt{sciencePrimary} deliverable by SDSS.}, WR 25/182) is an isolated compact galaxy at $z=0.0274$, placing it at the greatest estimated distance for our sample (117 Mpc).

\textbf{SB 119} (SDSS J143248.36+095257.1, Plate-MJD-Fiber 1709-53533-215, WR 380) is a star-forming region in the southeastern outskirts of the barred spiral NGC 5669.

\textbf{SB 125} (SDSS J113235.34+141129.8, Plate-MJD-Fiber 1754-53385-151, WR 384) is a lone compact star-forming region at the northeastern end of the tadpole-like spiral galaxy IC 2919.

\textbf{SB 126} (SDSS J115002.73+150123.4, Plate-MJD-Fiber 1761-53376-636, WR 385) is an isolated compact star-forming galaxy also identified as Mrk 750.

\textbf{SB 153} (SDSS J131447.36+345259.7, Plate-MJD-Fiber 2023-53851-263, WR 473) is the brightest star-forming region in the irregular galaxy Mrk 450.

\section{Case-by-case analysis of fits to the nebular lines and stellar wind features}
\label{sec:casebycase}
Our stellar population synthesis models generally struggle to simultaneously reproduce both the optical nebular line emission and UV stellar wind features detected in these systems.
To investigate this in more detail, it is instructive to consider the fits for each galaxy on a case-by-case basis:

{\bf SB 9} presents fairly average properties for our sample in the UV, including a \civ{} profile with absorption depth of -4.4 \AA{} and \heii{} emission at 2.2 \AA{} both near the median for our targets.
The gas-phase oxygen abundance inferred from the \beagle{} fits is $12+\log\mathrm{O/H}=8.39\pm0.04$, 0.09 dex higher than inferred from the direct-$T_e$ method above (Table~\ref{tab:optneb})\footnote{However, note that because SB 9 lacks a measurement of [\ion{O}{ii}] $\lambda 3727$, the \beagle{} fits do not include any \ion{O}{ii} lines.}.
The model \heii{} profile is in good agreement with that observed before any UV information is incorporated, though the \civ{} profile is somewhat underestimated in absorption depth and in emission.
Adding the stellar wind equivalent widths brings the absorption component of the \civ{} profile into better agreement with the data, though the emission is still underestimated.
In both fits, the optical WR features are well-modeled by the data, with good agreement to both the \heii{} $\lambda 4686$ and the weak \ion{N}{ii} and \ion{C}{iv} emission.
However, in order to match the \civ{} profile, the \beagle{} models prefer a higher metallicity.
The preferred gas-phase O/H is shifted 0.06 dex higher, while the stellar metallicity is shifted even higher, increasing by a factor of 1.4 from $\log_{10}(Z/Z_\odot)=-0.37$ to $-0.22$.
This larger boost to the stellar metallicity relative to the change in the gas-phase is accounted for by an increase in the amount of metals locked-up in dust, with $\xi_d$ nearly doubling from 0.29 to 0.46, in excess of the solar value $\xi_{d,\odot}=0.36$ \citep[and near the upper edge of the model range and our prior at 0.5;][]{gutkinModellingNebularEmission2016}.
This increase in depletion onto dust allows for an even higher total metallicity $Z$ at fixed O/H.
While most of the nebular lines are relatively unaffected by this change, this increase in the preferred model metallicity brings the [\ion{O}{iii}] $\lambda 4363$ line out of good agreement with the data, lowering its preferred value in the models to 0.8 times that observed.

The stellar wind properties of {\bf SB 49} prove to be even more challenging to reproduce in the models.
This system presents the deepest \civ{} P-Cygni absorption in our sample at $-6.51$ \AA{} and broad \heii{} emission at 3.20 \AA{}, in-excess of canonical stellar population synthesis predictions (Section~\ref{sec:ultraviolet}).
The \beagle{} fits to the nebular lines only are in reasonably good agreement with the observed stellar \heii{} profile, while the strength of the \civ{} P-Cygni absorption and emission is significantly under-predicted.
The retrieved gas-phase oxygen abundance for this fit is $12+\log\mathrm{O/H}=8.53\pm0.03$, 0.33 dex higher than found with the direct-$T_e$ method (Tables~\ref{tab:optneb} and \ref{tab:beagle_nebfit}) though with good agreement found for [\ion{O}{iii}] $\lambda 4363$.
However, incorporating the UV stellar wind equivalent widths in the fits leads to significant discord with these nebular line measurements.
In order to match the depth of the observed \civ{} P-Cygni profile, the models require a significantly higher metallicity; as for SB~9, this is accomplished by a simultaneous increase in both the total metallicity $Z$ and the dust-to-metal mass ratio $\xi_d$, this time bringing both quantities to the edge of their respective priors (at solar metallicity and $\xi_d\simeq 0.5$).
In addition, the model transitions to a preference for an essentially single-age stellar population, with both the start and end times for the current star formation epoch converging to 2 Myr ago.
This brings the stellar \civ{} profile into much better agreement with the data, though the models now slightly overestimate the depth of the absorption trough at intermediate velocities and the strength of emission near systemic (in the latter case, the model predicted emission appears to be in-part nebular, and this disagreement may be partially accounted for by interstellar absorption).
This process leaves \heii{} reasonably close to the observed strength (though slightly lowered from the prior estimate due to the general weakening of this feature in the models above $Z/Z_\odot \simeq 0.5$; Section~\ref{sec:ultraviolet}).
However, as for SB~9, this concordance models fail to match the strength of the temperature-sensitive [\ion{O}{iii}] $\lambda 4363$ line, underpredicting the observed flux by nearly a factor of two.
In addition, while the agreement with the optical WR features is very good with the nebular line information alone, the introduction of the UV stellar equivalent width information brings the models into substantial disagreement with the observed profiles.
In particular, both the broad \ion{He}{ii} $\lambda 4686$ and \ion{N}{iii} $\lambda 4640$ features are substantially overestimated by the models along with broad emission underlying \ion{He}{i} $\lambda 5876$; while the observed broad \ion{C}{iv} emission at 5808 \AA{} is now undetected in the models.
This suggests that the models prefer a solution to the UV stellar wind features relying on very prominent WN stars, which are too high in metallicity or present in too large of numbers to be consistent with the optical data.

The UV spectrum of {\bf SB~61} presents more modest stellar wind features, with both the \civ{} absorption and \heii{} emission near 2 \AA{}.
The nebular-only fit constrains the gas-phase metallicity to be $12+\log\mathrm{O/H}=8.25\pm 0.05$, only 0.14 dex in-excess of the direct-$T_e$ determination.
These optical-only results are already in good agreement with the depth of the \civ{} P-Cygni absorption, though they slightly underestimate the strength of the stellar emission component and predict a nebular emission component that we do not observe (though again, this comparison is complicated by ISM absorption).
In contrast, the stellar \heii{} profile predicted by the models is more prominent than observed.
Incorporating the stellar equivalent widths in the fits changes very little, with only a modest increase in the preferred metallicity from $\log_{10}Z/Z_\odot = -0.57$ to $-0.46$ and very little change to the predicted \civ{} profile.
Both model predictions show a narrow nebular emission component in the \heii{} profile, of which there is also some indication at low S/N in the observed spectrum.
This is the only system in which the model fits predict a clear nebular \heii{} component.
This is consistent with the fact that this system has the lowest preferred \beagle{} model metallicity in our sample ($Z/Z_\odot \simeq 0.3$), in the context of both observed and theoretical evidence towards enhancement in the $>54.4$ eV photons necessary to power gas recombination emission in this line at lower stellar metallicities \citep[e.g.][and references therein]{senchynaExtremelyMetalpoorGalaxies2019}.
Interestingly, the models incorporating the UV \heii{} equivalent width prefer a stronger broad stellar component and weaker nebular component than observed, as is also evident in the optical \heii{} $\lambda 4686$ profile.
This suggests that the models cannot reproduce the strength of the narrow nebular component of this line even at the moderately low metallicity of this system; though evaluating this confidently would require a direct fit to the \heii{} profile rather than just the total equivalent width.

{\bf SB~80} \citepalias[previously studied in][]{senchynaUltravioletSpectraExtreme2017} reveals a similar picture.
With only the nebular line constraints, we constrain the gas-phase oxygen abundance to be $12+\log\mathrm{O/H}=8.33\pm0.03$, 0.09 dex higher than the direct-$T_e$ measurement presented in \citetalias{senchynaUltravioletSpectraExtreme2017}.
Incorporating the stellar wind strengths in the fit changes little in the derived parameters, with a boost upwards in total metallicity of only 0.08 dex.
The fits with the UV stellar equivalent widths also show no significant improvement in the UV, and though \civ{} is fairly well-fit, \heii{} remains underestimated by the models (though continuum determination near \heii{} in the \hstcos{} data is somewhat uncertain).

The UV spectrum of {\bf SB~119} presents a more a more significant challenge to reproduce, with prominent stellar \heii{} emission at 4.3 \AA{}.
Fitting only the optical lines, we find a gas-phase $12+\log\mathrm{O/H}=8.37\pm0.05$ which exceeds the direct-$T_e$ measurement by $0.26$ dex.
However, this fit significantly underestimates the prominence of both the \civ{} and \heii{} wind features observed in the \hstcos{} G160M spectrum.
Incorporating the stellar wind equivalent widths in the fit brings the models into very good agreement with the full profiles of both features, without shifting the estimated gas-phase oxygen abundance at all.
Instead, the current epoch of star formation contracts, with a revised start point of 6.3 Myr rather than 20 and a slightly early truncation 1.3 Myr ago.
The inferred stellar metallicity increases by 0.13 dex, but purely through an increase in $\xi_d$ from $0.17$ to a significantly super-solar value of $0.46$ (near the edge of our prior).
The median model-predicted [\ion{O}{iii}] $\lambda 4363$ flux does decrease somewhat (to 93\% the observed flux), but the posterior distribution remains consistent with the observed value.
However, while the UV \heii{} feature is reasonably well-fit, the optical \heii{} $\lambda 4686$ profile is overestimated by the models incorporating the UV stellar equivalent width.
Given the relatively low signal-to-noise of the continuum in the UV, this disagreement may be due to an overestimate of the equivalent width of \heii{} $\lambda 1640$; or this may reflect disagreement in the model prediction of the stellar populations present.
In short, our model is able to simultaneously reproduce the set of nebular lines and stellar wind features provided for SB~119, even including its large stellar \heii{} $\lambda 1640$ equivalent width; but only with an unrealistically-high $\xi_d$ and with some disagreement in the optical WR features.

{\bf SB~125} reveals a \civ{} P-Cygni absorption equivalent width of $-4.6$ \AA{} and accompanying stellar \heii{} emission at an intermediate 2.7 \AA{}, both typical for our sample.
The nebular line-only fit supports a gas-phase abundance of $12+\log\mathrm{O/H}=8.38\pm0.03$, 0.19 dex higher than the direct-$T_e$ result.
The observed \civ{} profile is slightly underestimated by the models, while \heii{} is well-matched.
Addition of these stellar equivalent widths to the fit results in increases in both the preferred total metallicity and the corresponding gas-phase oxygen abundance by $\sim 0.2$ dex, producing another case in which the model [\ion{O}{iii}] $\lambda 4363$ flux distribution systematically underestimates the observed line flux.
In addition, while the agreement with \heii{} $\lambda 1640$ appears unchanged with the addition of the stellar equivalent width measurements to the fit, the optical stellar \ion{N}{iii} $\lambda 4640$ feature and \ion{He}{ii} $\lambda 4686$ feature are both over-estimated by the models after continuum normalization.
This suggests again that the models have resolved the disagreement with the UV stellar features by invoking WN stars at higher metallicity or incidence than supported by the data.
Intriguingly, the optical \heii{} profile shows a narrow nebular component in \heii{} which is not visible in the UV \heii{} $\lambda 1640$ profile, even though the contaminating MW \ion{Al}{ii} $\lambda 1670$ absorption line nearby does not reach the expected line center for nebular emission in \heii{} $\lambda 1640$.
However, we find that given the relatively high reddening measured towards this target from the Balmer decrement ($\mathrm{E(B-V)}=0.25$) and assuming an SMC extinction curve, the nondetection of \heii{} $\lambda 1640$ is consistent with the UV continuum noise and the flux of $\lambda 4686$ given a conservative flux ratio of 10:1 \citep{hummerRecombinationlineIntensitiesHydrogenic1987}.

{\bf SB~126} also presents fairly typical stellar wind properties for our sample, with \civ{} absorption at $-4.0$ \AA{} and $1.9$ \AA{} \heii{} emission.
The nebular line fits produce a gas-phase oxygen abundance estimate of $12+\log\mathrm{O/H}=8.28\pm0.03$, in-excess of the direct-$T_e$ measurement by 0.26 dex.
This fit appears in reasonable agreement with the observed broad emission in both the UV and optical \heii{} profiles, but significantly underestimates the strength of the \civ{} P-Cygni feature.
Adding the stellar equivalent widths brings the absorption component of the \civ{} profile into agreement with the data, but the model still underestimates the strength of the redshifted stellar emission.
As before, better matching the UV stellar wind features necessitates a move to higher metallicity, this time by 0.36 dex in $Z$ accompanied by a shift in $\xi_d$ to 0.49 at the edge of the model range.
In this case, [\ion{O}{iii}] $\lambda 4363$ is underestimated by a factor of 0.85 while in addition, the model flux in [\ion{O}{iii}] $\lambda 5007$ exceeds that observed by a factor of 1.22, leading to an even larger disagreement in the ratio of these lines.

Beginning with {\bf SB~153}, the remaining three targets are in the rarefied class of object with stellar \heii{} equivalent widths comparable to or exceeding 4 \AA{}, and represent a more significant challenge to reconcile.
The gas-phase metallicity inferred from the optical-only fit here is $12+\log\mathrm{O/H}=8.36\pm0.03$, 0.16 dex larger than estimated with the direct-$T_e$ method.
The posterior distribution of the models in the UV is broad with nebular lines only, and somewhat underestimates the strength of the \civ{} profile.
However, adding the stellar wind equivalent widths into the fits brings the models into very good agreement with both \civ{} and \heii{}.
Interestingly, with a relatively modest increase in the inferred metallicity (increasing from $\log_{10}(Z/Z_\odot)=-0.45\pm0.05$ to $-0.37\pm0.01$) and with a compensating increase in $\xi_d$ to $0.35\pm0.07$, all of the fitted nebular lines including $\lambda 4363$ are brought into good agreement with the model alongside the stellar wind features.
The optical WR features are generally in good agreement with the models, though the observed \ion{He}{ii} $\lambda 4686$ profile is somewhat lower in equivalent width than preferred by the best-fit model.

{\bf SB~179} presents the largest integrated \heii{} equivalent width in our sample at $4.67\pm 0.23$ \AA{}.
The nebular line fit yields a gas-phase metallicity of $12+\log\mathrm{O/H}=8.44\pm0.03$, only 0.09 dex larger than the direct-$T_e$ determination in \citetalias{senchynaUltravioletSpectraExtreme2017}.
This nebular line-only fit significantly underestimates the absorption and emission component of the \civ{} profile and does not match the strength of the observed \heii{} at all.
Adding these equivalent width constraints does bring the models into reasonable agreement with the \civ{} profile.
However, the modeled strength of \heii{} remains significantly below that observed, even with a significant reduction of the star-formation period to a brief burst sustained from 4.0 to 1.6 Myr ago.
In contrast, while the optical WR features are initially reasonably well-fit, the inclusion of the UV stellar equivalent widths in the fits increases the modeled strength of stellar \ion{N}{iii} $\lambda 4640$, \ion{He}{ii} $\lambda 4686$, and \ion{C}{iv} $\lambda 5808$ above their observed equivalent widths.
In addition, the increase in metallicity to nearly-solar ($\log Z/Z_\odot =-0.09\pm0.1$) again causes [\ion{O}{iii}] $\lambda 4363$ to fall out of alignment with the data (at a factor of 0.72 times the observed strength).

Finally, {\bf SB~191} reveals a similar challenge to the stellar models, with a total \heii{} equivalent width of $4.02\pm 0.16$ \AA{}.
With optical nebular lines only, we infer a gas-phase oxygen abundance $12+\log\mathrm{O/H}=8.51\pm0.02$, 0.21 dex higher than estimated with the direct-$T_e$ method.
The corresponding UV stellar wind line morphology is close to that observed for \civ{}, but significantly below the very prominent \heii{} emission.
We note that SB~179 displays an intriguing \heii{} emission shape, with both an apparent double-peak in the broad emission (also visible in SB~179) and narrow (likely nebular) emission superimposed -- we discuss this profile in more detail in Section~\ref{sec:peculiarheii}.
Fitting the stellar equivalent widths explicitly brings the models closer, but they remain well below the observed \heii{} emission peak.
In contrast, the optical WR features appear reasonably well-fit by the models, though with some indication that \ion{He}{ii} $\lambda 4686$ is over-estimated in the fits including the UV stellar lines.
In this case, in contrast to the other systems, we do not observe a significant increase in the preferred model metallicity when the stellar emission is incorporated; the median metallicity increases by only 0.04 dex in this case, likely because it is already near the peak of stellar \heii{} emission in the models at $Z/Z_\odot \simeq 0.5$ (Section~\ref{sec:ultraviolet}).

\label{lastpage}

\end{document}